\documentclass[11pt,twoside,a4paper]{report}
\usepackage{amssymb}

\usepackage{graphicx}
\usepackage{amsmath}

\newtheorem{theorem}{Theorem}[section]

\newtheorem{corollary}[theorem]{Corollary}

\newtheorem{definition}[theorem]{Definition}
\newtheorem{example}[theorem]{Example}

\newtheorem{lemma}[theorem]{Lemma}

\newtheorem{proposition}[theorem]{Proposition}
\newtheorem{remark}[theorem]{Remark}

\newenvironment{proof}[1][Proof]{\textbf{#1.} }{\ \rule{0.5em}{0.5em} \medskip}

\begin{document}

\begin{titlepage}
\begin{center}
{\Large Universit\`a degli Studi di Genova}\\
\vspace{0.2in}{Facolt\`a di Scienze Matematiche, Fisiche e Naturali}\\
{Anno Accademico 2004/2005}\\
\vspace{1.2in}{Tesi di Dottorato in Fisica}\\
\vspace{1.2in}{\bf{\LARGE Positive operator measures,}} \\
\vspace{0.1in}{\bf{\LARGE generalised imprimitivity theorem}}\\
\vspace{0.1in}{\bf{\LARGE and their applications}}\\
\vspace{1in}{Dottorando:}\\
{\large Alessandro Toigo}\\
\end{center}
\begin{center}
\vspace{1.2in}{\qquad\quad \ \ Relatore:}\hspace{2.4in}{Correlatore:}\\
{\large Prof. Gianni Cassinelli}\hspace{1.7in}{\large Prof. Pekka Lahti}\\
{\quad \ (Universit\`a di Genova)}\hspace{1.6in}{\ (Universit\`a di Turku)}\\
\end{center}
\end{titlepage}
\thispagestyle{empty}
\

\chapter*{Introduction}

In the common textbook presentation of quantum mechanics the observables of
a quantum system are represented by selfadjoint operators, or, equivalently,
by spectral measures. The origin of this point of view dates back to the very
beginning of quantum theory. Its rigorous mathematical formulation is mainly
due to von Neumann \cite{Von Neumann}, and, for a more recent and complete
review, we refer to the book of Varadarajan \cite{Var}.

But it is quite well known that, when particular quantum systems are
considered, this approach can not give a satisfactory description of some of
their physical properties. A famous example (dating back to Dirac \cite{ref.
a Dirac}) is provided by the phase of the electromagnetic field, which is a
well defined quantity in classical physics, but can not be described by any
selfadjoint operator in quantum mechanics \cite{QDET}, \cite{Holevo82}. This
drawback in the conventional formulation of quantum theory becomes even more
evident when one attempts to define a position observable for the photon. In
fact, a theorem of Wightman (\cite{Wightman},
\cite{Var}) asserts that there does not exists
any selfadjoint operator describing the localisation property of the photon.

Positive operator measures were introduced just to overcome difficulties of
this kind arising from the von Neumann formulation of quantum theory. Quite
soon after their introduction, it became also clear \cite{Davies76}, \cite
{QDET}, \cite{Holevo82}, \cite{FQMI} that the most general description of
the observables of quantum mechanics is provided by positive operator
measures rather than by spectral measures. In this extended setting, the
phase observable and the localisation observable for the photon are
associated to positive operator measures that are \emph{not} spectral maps,
and thus can not be represented by any selfadjoint operator.

It also turned out that joint measurements of quantities which are
incompatible in the von Neumann
framework can be described in terms of positive operator measures
(an example is provided in section \ref{Joint1}).

The aim of this thesis is to give a complete characterisation of an
important class of positive operator measures, namely the positive operator
measures that are covariant with respect to unitary representations of a
group. Indeed, in quantum physics the observables that describe a particular
physical quantity are defined by means of their property of covariance with
respect to a specific symmetry group. Thus, covariant positive operator
measures naturally acquire a privileged role.

Since the characterisation of covariant positive operator measures requires
a few deep results from abstract harmonic analysis and needs the elaboration
of some specific mathematical techniques, in general we will concentrate on
the mathematical point of view of the problem. Our examples and applications to
quantum mechanics are not intended to give a detailed exposition of the
particular physical framework to which they are addressed.

The thesis is organised as follows. Chapter \ref{Primo capitolo} is
of an introductory nature.
In section \ref{Sez. 1.1} we give a brief account of the
arising of covariant positive operator measures in quantum physics. In \S\ref
{subsec. 1.2} we fix the mathematical set-up that will be commonly used in
the subsequent chapters. In \S\ref{Sez. sugli stati coerenti} we will sketch
an application to the theory of coherent states (for more details on the
topics treated in this section, we refer to \cite{Ali}).

In chapters \ref{cap. 1} and \ref{cap. 2} we will achieve a complete
characterisation of covariant positive operator measures in the following
two cases:

\begin{enumerate}
\item  if $G$ is an abelian group, we will characterise the most general
positive operator measure based on a \emph{transitive} $G$-space and
covariant with respect to a unitary representation of $G$ (chapter \ref{cap.
1});

\item  if $G$ is a generic group and $\Omega $ is a \emph{transitive} $G$%
-space with \emph{compact stabiliser} in $G$, we will classify the positive
operator measures based on $\Omega $ and covariant with respect to an \emph{%
irreducible} projective unitary representation of $G$ (chapter \ref{cap. 2}).
\end{enumerate}

Although they are quite specific, actually these two cases cover almost all
the situations of physical interest. In particular, the first case will
enable us to give a complete classification of the phase observables for a
quantum electromagnetic field, while the second characterises an important
class of joint observables of position and momentum, namely the covariant
phase space observables.

In chapters \ref{Capitolo sulle Covariant position and momentum observables}
and \ref{Cap. sulla coesistenza} we will concentrate on positive operator
measures associated to
the position and momentum observables of nonrelativistic quantum mechanics. We will define 
such observables in terms of their properties of transformation under the action of the 
isochronous Galilei group. After that, we will give their characterisation and discuss their main 
properties. In
particular, in chapter \ref{Cap. sulla coesistenza} we will study the
problem of joint measurability of position and momentum observables. We will
find under which conditions a position and a momentum observable are jointly
measurable, showing that these conditions imply Heisenberg's
uncertainity relation. Here we remark that in our convention position and momentum 
observables are covariant under the Galilei group {\em by definition}. Nevertheless, in 
many situations of physical interest one actually needs to relax the covariance 
requirement in the definitions of such observables. This last topic is beyond the scope of 
the thesis. The interested reader is referred to \cite{Lahti95}, \cite{Werner04} and 
references therein for more details. 

The thesis is a re-elaboration of the results published in \cite{Articolo
con Bassano}, \cite{CHT04}, \cite{Nuovo articolo con Teiko}, \cite{CDT1}
and \cite{CDT2}, and in the paper \cite{Articolo con Teiko in preparazione}
in preparation.
\
\newpage
\

\tableofcontents

\chapter{Covariant positive operator measures in mathematical physics\label%
{Primo capitolo}}

\section{Positive operator measures in quantum mechanics\label{Sez. 1.1}}

The framework of positive operator measures naturally arises in quantum
physics as the mathematical foundation of the theory of measurements. In
this section, we will sketch a very brief account of the subject, with a
particular emphasis on the physical meaning of positive operator measures
which are covariant under symmetry transformations. For more details, we
refer to \cite{Lahti95}, \cite{Davies76}, \cite{QDET}, \cite{Holevo82}, \cite
{Holevo01}, \cite{FQMI}.

Quantum mechanics associates to each physical system $\mathcal{S}$ a corresponding
Hilbert space $\mathcal{H}$, identifying the \textbf{states} of $\mathcal{S}$ with the
elements of the
set $S \left( \mathcal{H}\right) $ of positive trace one operators
on $\mathcal{H}$. On the other hand, if $\left( \Omega ,\mathcal{A}\left(
\Omega \right) \right) $ is a measurable space\footnote{%
We recall that a measurable space is a pair $\left( \Omega ,\mathcal{A}%
\left( \Omega \right) \right) $ composed by a nonempty set $\Omega $ and a $%
\sigma $-algebra $\mathcal{A}\left( \Omega \right) $ of subsets
of $\Omega $.}, the 
\textbf{observables} of $\mathcal{S}$ taking values in $\Omega $ are represented by
positive operator measures based on $\Omega $ and acting in $\mathcal{H}$.
Here we recall that a \textbf{positive operator measure} (POM) based on $%
\Omega $ and acting in $\mathcal{H}$ is a map $E:\mathcal{A}\left( \Omega
\right) \longrightarrow \mathcal{L}\left( \mathcal{H}\right) $ ($\mathcal{L}%
\left( \mathcal{H}\right) $ = the set of bounded operators on $\mathcal{H}$)
such that\footnote{%
We recall that in the mathematical literature by \emph{positive operator
measure} based on $\Omega $ and acting in $\mathcal{H}$ one usually means a
map $E:\mathcal{A}\left( \Omega \right) \longrightarrow \mathcal{L}\left( 
\mathcal{H}\right) $ satisfying only conditions \ref{Positivita' POM} and 
\ref{sigma-additivita' POM}. If $E$ satisfies also condition \ref
{Normalizzazione POM}, then $E$ is called to be a \emph{positive normalised
operator measure}. Our slightly imprecise abbreviated notation is justified
by the fact that in the quantum theory of measurement only normalised
positive operator measures are considered.}

\begin{enumerate}
\item  \label{Positivita' POM}$E\left( X\right) $ is a positive operator for
all $X\in \mathcal{A}\left( \Omega \right) $;

\item  \label{Normalizzazione POM}$E\left( \Omega \right) =I$;

\item  \label{sigma-additivita' POM}$E\left( \bigcup_{n=1}^{\infty
}X_{n}\right) =\sum_{n=1}^{\infty }E\left( X_{n}\right) $ (in the weak
sense) if $X_{n}\in \mathcal{A}\left( \Omega \right) $ and $X_{n}\cap
X_{m}=\varnothing $ for $n\neq m$.
\end{enumerate}

If $E:\mathcal{A}\left( \Omega \right) \longrightarrow \mathcal{L}\left( 
\mathcal{H}\right) $ is an observable and $T\in S \left( \mathcal{H}%
\right) $, we define
\begin{equation}
p_{T}^{E}\left( X\right) :=\mathrm{tr}\,\left( TE\left( X\right) \right)
\quad \forall X\in \mathcal{A}\left( \Omega \right) \text{.}
\label{def. la prob. associata alla POM E}
\end{equation}
By properties 1, 2 and 3 the above equation defines a probability measure $%
p_{T}^{E}$ on $\Omega $. It is interpreted as the probability distribution
describing the outcomes of a measurement of $E$. More precisely, the number $%
p_{T}^{E}\left( X\right) $ is the probability that a measurement of the
observable $E$ performed on the system $\mathcal{S}$ prepared in the state $T$ yelds a
result in the subset $X\subset \Omega $.

If property \ref{Positivita' POM} in the definition of POM is replaced by
the stronger condition

\begin{enumerate}
\item[1'.]  $E\left( X\right) =E\left( X\right) ^{\ast }=E\left( X\right)
^{2}$ for all $X\in \mathcal{A}\left( \Omega \right) $,
\end{enumerate}

\noindent then the POM $E$ is a \textbf{projection valued measure}, and
the associated observable is said a \textbf{sharp observable}. If in
addition $\Omega =\mathbb{R}$ with its Borel $\sigma $-algebra, then $E$ is
actually a spectral map. We note that in this case the mean value of the
observable $E$ measured on a state $T$ is 
\begin{equation*}
\int \lambda \text{d}p_{T}^{E}\left( \lambda \right) =\mathrm{tr}\,\left(
T\int \lambda \text{d}E\left( \lambda \right) \right) =:\mathrm{tr}\,\left(
TA\right) \text{,}
\end{equation*}
where $A$ is the selfadjoint operator with spectral decomposition $A=\int
\lambda $d$E\left( \lambda \right) $ (here for simplicity we suppose that
the domain of $A$ is the whole space $\mathcal{H}$). We thus see that
observables described by selfadjoint operators are a strict subset of the
whole class of observables. For more details on the physical meaning of
condition 1', we again refer to \cite{Lahti95}.

%
We now enrich our setting, introducing the action of a group of transformations
(symmetry group), and studying how the quantum systrem $\mathcal{S}$ and its
observables change under the action of the group. This will enable us to
define a particular class of observables, whose properties of transformations
under the action of the group are in fact the natural ones.

Thus, suppose that $E:\mathcal{A}\left( \Omega \right)
\longrightarrow \mathcal{L}\left( \mathcal{H}\right) $ is an observable,
and that a transformation group $G$ acts both
on the space $\Omega $ on which $E$ takes its values and
on the quantum system $\mathcal{S}$. We denote by $S\left( 
\mathcal{H}\right) \ni T\longmapsto g\left[ T\right] \in S\left( \mathcal{H}%
\right) $ and by $\Omega \ni x\longmapsto g\left[ x\right] \in \Omega $ the
actions of an element $g\in G$ on $S\left( \mathcal{H}\right) $ and $\Omega $
respectively. If $G$ satisfies some general topological conditions, then
Wigner's theorem states that $G$ acts on the Hilbert space of the system by
means of a projective unitary representation $U$, i.e. 
\begin{equation}
g\left[ T\right] =U\left( g\right) TU\left( g\right) ^{-1}\quad \forall g\in
G  \label{Teo. di Wigner}
\end{equation}
for all states $T\in S\left( \mathcal{H}\right) $. We now require that, when the system 
undergoes
a transformation of $G$, the statistics of the measurement of $E$ is
affected by a corresponding variation. More precisely, we request that the
probability distribution of the outcomes obtained measuring $E$ satisfies
\begin{equation}
p_{g\left[ T\right] }^{E}\left( X\right) =p_{T}^{E}\left( g^{-1}\left[ X%
\right] \right) \quad \forall X\in \mathcal{A}\left( \Omega \right) 
\label{covarianza di p}
\end{equation}
for all states $T\in S\left( \mathcal{H}\right) $ and all transformations $%
g\in G$. Eqs.~(\ref{def. la prob. associata alla POM E}), (\ref{Teo. di
Wigner}) and (\ref{covarianza di p}) imply 
\begin{equation}
E\left( g\left[ X\right] \right) =U\left( g\right) E\left( X\right) U\left(
g\right) ^{-1}\quad \forall X\in \mathcal{A}\left( \Omega \right) 
\label{covarianza di E}
\end{equation}
for all $g\in G$. Eq.~(\ref{covarianza di E}) is a covariance condition
imposed upon the POM $E$. Such a
condition thus selects among all the possible
observables taking values in the space $\Omega $ those which actually
transform in a compatible way under the action of $G$.

A triple $\left( U,E,\mathcal{H}\right) $ formed by a unitary representation 
$U$ of $G$ in the Hilbert space $\mathcal{H}$ and by a POM $E$ on $\Omega $
acting in $\mathcal{H}$ which satisfies eq.~(\ref{covarianza di E}) is
called \textbf{system of covariance} for $G$ based on $\Omega $. A POM $E$
satisfying eq.~(\ref{covarianza di E}) is said to be
$U$\textbf{-covariant}.

As a simple and clarifying example, we can consider the position observables
for a quantum particle with one degree of freedom. Here the Hilbert space of
the system is $\mathcal{H}=L^{2}\left( \mathbb{R},\text{d}x\right) $ and the
outcome space is the real line $\Omega =\mathbb{R}$ with the Borel $\sigma $%
-algebra $\mathcal{B}\left( \mathbb{R}\right) $. We have the usual action of
the group of translations $G=\mathbb{R}$ on $\mathbb{R}$ itself. We require that the
statistics of the outcomes registered measuring a position observable $E$
satisfy 
\begin{equation*}
p_{x\left[ T\right] }^{E}\left( X\right) =p_{T}^{E}\left( X-x\right) \quad
\forall x\in \mathbb{R},\,X\in \mathcal{B}\left( \mathbb{R}\right) ,\,T\in
S\left( \mathcal{H}\right) \text{,}
\end{equation*}
where $x\left[ T\right] =e^{-ixP}Te^{ixP}$, $P$ being the selfadjoint generator
of translations. This implies 
\begin{equation}
e^{-ixP}E\left( X\right) e^{ixP}=E\left( X+x\right) \quad \forall x\in 
\mathbb{R},\,X\in \mathcal{B}\left( \mathbb{R}\right) \text{.}
\label{Cov. per trasl.}
\end{equation}
A possible solution of the above equation is the projection valued measure $%
\Pi $ given by 
\begin{equation*}
\left[ \Pi \left( X\right) f\right] \left( x\right) =\chi _{X}\left(
x\right) f\left( x\right) \quad \forall f\in L^{2}\left( \mathbb{R},\text{d}%
x\right) ,\,X\in \mathcal{B}\left( \mathbb{R}\right) \text{.}
\end{equation*}
We note that $\Pi $ is the spectral map associated to the selfadjoint
generator $Q$ of momentum boosts, with $\left( Qf\right) \left( x\right) =xf\left( 
x\right) $.
Neverthless, we shall see in the next chapter that there are many other
solutions of eq.~(\ref{Cov. per trasl.}), for which the condition 1' does
not hold in general. By the way, we note that the covariance under
translations is not sufficient to define a position observable; also
invariance under boosts is needed. We shall explore this with more details
in chapter \ref{Capitolo sulle Covariant position and momentum observables}.

In the following, we will be mainly concerned with the solution of eq.~(\ref
{covarianza di E}), i.e.~with the classification of the systems of
covariance for a group $G$ based on a $G$-space $\Omega $. We will always
assume that the action of $G$ on $\Omega $ is \emph{transitive}. Under this
essential hypothesis (and some general topological assumptions), we will
find the most general solution in the following two cases:

\begin{enumerate}
\item  $G$ is abelian and $U$ is an arbitrary unitary representation of $G$
(chapter \ref{cap. 1});

\item  $G$ is generic, $U$ is an irreducible projective unitary
representation of $G$, and the stabilizer of $\Omega $ in $G$ is compact
(chapter \ref{cap. 2}).
\end{enumerate}

\noindent The first case covers, for example, the class of position
observables just described, while the second characterises an important
class of joint observables of position and momentum, namely the covariant
phase space observables (see \S \ref{subsec. Two examples}).

The next section lays down the mathematical set-up which will be the
starting point for our solution of this classification problem.

\section{\label{subsec. 1.2}The induced representation and the generalised
imprimitivity theorem}

In the following, we will always be concerned with topological spaces
endowed with their Borel $\sigma $-algebra. If $\Omega $ is a topological
space, we denote by $\mathcal{B}\left( \Omega \right) $ the $\sigma $%
-algebra of its Borel subsets.

We will always assume that $G$ is a Hausdorff locally compact second
countable topological group acting continuously and transitively on a
Hausdorff locally compact space $\Omega $ (transitive $G$-space). We denote
by $e$ the identity element of $G$. Fixed a point $x\in \Omega $, $\Omega $
is canonically identified with the quotient space $G/H_{x}$, $H_{x}\subset G$
being the stability subgroup at $x$. By virtue of this identification, for
the rest of this section we shall assume that $\Omega =G/H$, with $H$ a
closed subgroup in $G$. The canonical projection $G\longrightarrow G/H$ is
denoted by $\pi $; the equivalence class of an element $g\in G$ in $G/H$ is
denoted alternatively by $\pi \left( g\right) $ or $\dot{g}$. The action of
an element $a\in G$ on a point$\ \dot{g}\in G/H$ is clearly $a\left[ \dot{g}%
\right] =\pi \left( ag\right) $.

We assume that $G/H$ \emph{admits an invariant measure} $\mu _{G/H}$. We
recall that this is equivalent to assume that the modular function $\Delta
_{G}$ of $G$ restricts on $H$ to the modular function $\Delta _{H}$ of $H$.
If the invariant measure $\mu _{G/H}$ exists, then it is unique up to a
constant. We let $\mu _{G}$ be a fixed left Haar measure of $G$. The
following fact will often be used: a set $X\in \mathcal{B}\left( G/H\right) $
is $\mu _{G/H}$-negligible if and only if $\pi ^{-1}\left( X\right) $ is $%
\mu _{G}$-negligible. For more details on these facts, we refer to \cite
{Foll}, \cite{Gaal}.

By a \textbf{unitary representation} (or simply a \textbf{representation})
we mean a weakly continuous unitary representation acting in a separable
Hilbert space. If $\mathcal{H}$ is a Hilbert space, we denote by $%
\left\langle \cdot ,\cdot \right\rangle _{\mathcal{H}}$ its scalar product,
linear in the second argument; when clear from the context, we drop the
subscript $\mathcal{H}$.

Let $\mathrm{rep}\,\left(H\right)$ and
$\mathrm{rep}\,\left(G\right)$ be the set of the unitary representations of $H$
and $G$ respectively. We now describe a canonical construction which will allow us to 
associate:
\begin{itemize}
\item to each unitary representation $\sigma $ of $H$ a unitary
representation of $G$, called the {\bf representation of $G$ induced by $\sigma$} and 
denoted by
${\rm ind}_H^G(\sigma)$;
\item to each map $A$ intertwining\footnote{We recall that, if $\sigma$
and $\sigma^{\prime}$ are representations of $H$ acting in the Hilbert spaces
$\mathcal{K}$ and $\mathcal{K}^{\prime}$, a bounded operator $A: \mathcal{K}
\longrightarrow \mathcal{K}^{\prime}$ {\bf intertwines} $\sigma$
and $\sigma^{\prime}$ if
\begin{equation*}
A\sigma (h) = \sigma^{\prime} (h) A \qquad \forall h\in H.
\end{equation*}
} $\sigma $ with $\sigma
^{\prime }$ a map $\tilde{A}$ intertwining ${\rm ind}_H^G(\sigma)$
with ${\rm ind}_H^G(\sigma^{\prime})$ in such a way that: a) $\widetilde{AB}
= \tilde{A} \tilde{B} $ if $A$ intertwines $\sigma $ with $\sigma
^{\prime }$ and $B$ intertwines $\sigma^{\prime} $ with $\sigma
^{\prime \prime}$; b) $\widetilde{A^{\ast}} = \tilde{A}^{\ast}$.
\end{itemize}

Suppose $\sigma $ is a representation of $H$ in the Hilbert space $\mathcal{K%
}$. Let $\mathcal{H}^{\sigma }$ be the space of functions $%
f:G\longrightarrow \mathcal{K}$ such that

\begin{enumerate}
\item  $f$ is weakly $\mu _{G}$-measurable;

\item  for all $h\in H$ and $g\in G$ 
\begin{equation*}
f\left( gh\right) =\sigma \left( h\right) ^{-1}f\left( g\right) \text{;}
\end{equation*}

\item  
\begin{equation*}
\int_{G/H}\left\| f\left( g\right) \right\| ^{2}\text{d}\mu _{G/H}\left( 
\dot{g}\right) <\infty \text{.}
\end{equation*}
\end{enumerate}

\noindent We identify functions in $\mathcal{H}^{\sigma }$ which are equal $%
\mu _{G}$-a.e.\footnote{%
Here and in the following `a.e.' is the acronym of `almost everywhere'.
Likewise, `a.a.' is the abbreviated form of `almost all'.}. Endowed with the
scalar product 
\begin{equation*}
\left\langle f_{1},f_{2}\right\rangle _{\mathcal{H}^{\sigma
}}=\int_{G/H}\left\langle f_{1}\left( g\right) ,f_{2}\left( g\right)
\right\rangle \text{d}\mu _{G/H}\left( \dot{g}\right)
\end{equation*}
$\mathcal{H}^{\sigma }$ becomes a separable Hilbert space (see \cite{Dieu2}%
). There is a natural unitary representation of $G$ in $\mathcal{H}^{\sigma
} $, in which $G$ acts on $\mathcal{H}^{\sigma }$ by left translations. We
denote it by $\lambda ^{\sigma }$: 
\begin{equation*}
\left[ \lambda ^{\sigma }\left( a\right) f\right] \left( g\right) =f\left(
a^{-1}g\right) \quad \text{for~}\mu _{G}\text{-a.a.~}g\in G\text{.}
\end{equation*}
for all $a\in G$.

We then define $\mathrm{ind}_{H}^{G}\left( \sigma \right) =\lambda ^{\sigma
} $. If $A$ intertwines $\sigma $ with $\sigma^{\prime}$, we let $\tilde{A}:
\mathcal{H}^{\sigma} \longrightarrow \mathcal{H}^{\sigma^{\prime}}$
be defined by
\begin{equation*}
\left(\tilde{A} f\right) (g) = Af (g).
\end{equation*}
It is immediately checked that $\tilde{A}$ satisfies the required properties.

\begin{remark}
\label{Remark sulla rappr. regolare}Note that if $H=\left\{ 1\right\} $ and $%
\sigma $ is the trivial one-dimensional representation of $H$, then $\mathrm{%
ind}_{H}^{G}\left( \sigma \right) $ is the left regular representation $%
\lambda $ of $G$. We recall that this representation acts by left
translations in the Hilbert space $L^{2}\left( G,\mu _{G}\right) $: 
\begin{equation*}
\left[ \lambda \left( a\right) f\right] \left( g\right) =f\left(
a^{-1}g\right) \quad \text{for~}\mu _{G}\text{-a.a.~}g\in G\text{.}
\end{equation*}
for all $f\in L^{2}\left( G,\mu _{G}\right) $, $a\in G$. In the following,
we will often drop the adjective `left' in referring to this representation.
\end{remark}

An equivalent realisation of $\mathrm{ind}_{H}^{G}$ is now given. Fix a
Borel section $s:G/H\longrightarrow G$. Define 
\begin{equation*}
\left( Vf\right) \left( x\right) =f\left( s\left( x\right) \right) \quad 
\text{for~}\mu _{G/H}\text{-a.a.~}x\in G/H
\end{equation*}
for all $f\in \mathcal{H}^{\sigma }$. Then, $Vf\in L^{2}\left( G/H,\mu
_{G/H};\mathcal{K}\right) $, and the linear operator $V:\mathcal{H}^{\sigma
}\longrightarrow L^{2}\left( G/H,\mu _{G/H};\mathcal{K}\right) $ is unitary.
By means of $V$, we transfer $\lambda ^{\sigma }$ to a unitary
representation $U^{\sigma }$ of $G$ in $L^{2}\left( G/H,\mu _{G/H};\mathcal{K%
}\right) $. Its action on $\phi \in L^{2}\left( G/H,\mu _{G/H};\mathcal{K}%
\right) $ is 
\begin{equation*}
\left[ U^{\sigma }\left( a\right) \phi \right] \left( x\right) =\sigma
\left( s\left( x\right) ^{-1}as\left( a^{-1}\left[ x\right] \right) \right)
\phi \left( a^{-1}\left[ x\right] \right)
\end{equation*}
for all $a\in G$.

In the following, we will mainly use the realisation of $\mathrm{ind}%
_{H}^{G}\left( \sigma \right) $ in $\mathcal{H}^{\sigma }$, although
sometimes we will refer also to the second construction.

In $\mathcal{H}^{\sigma }$ we define the projection valued measure $%
P^{\sigma }$ based on $G/H$ which maps each $X\in \mathcal{B}\left(
G/H\right) $ into the
multiplication by the corresponding characteristic
function $\chi _{X}$: 
\begin{equation}
\left[ P^{\sigma }\left( X\right) f\right] \left( g\right) =\chi _{X}\left( 
\dot{g}\right) f\left( g\right) \quad \forall f\in \mathcal{H}^{\sigma }%
\text{.}  \label{La PVM canonica}
\end{equation}
Clearly, 
\begin{equation*}
\lambda ^{\sigma }\left( g\right) P^{\sigma }\left( X\right) \lambda
^{\sigma }\left( g\right) ^{-1}=P^{\sigma }\left( g\left[ X\right] \right)
\quad \forall X\in \mathcal{B}\left( G/H\right) ,\,\forall g\in G\text{.}
\end{equation*}
A triple $\left( U,P,\mathcal{H}\right) $, in which $U$ is a unitary
representation of $G$ in the Hilbert space $\mathcal{H}$, $P:\mathcal{B}%
\left( G/H\right) \longrightarrow \mathcal{L}\left( \mathcal{H}\right) $ is
a projection valued measure, and 
\begin{equation*}
U\left( g\right) P\left( X\right) U\left( g\right) ^{-1}=P\left( g\left[ X%
\right] \right) \quad \forall X\in \mathcal{B}\left( G/H\right) ,\,\forall
g\in G\text{,}
\end{equation*}
is called \textbf{system of imprimitivity} for $G$ based on $G/H$. The
triple $\left( \lambda ^{\sigma },P^{\sigma },\mathcal{H}^{\sigma }\right) $
defined above is the \textbf{canonical system of imprimitivity} induced by
the representation $\sigma $ of $H$.

We can now state the fundamental theorem on which our classification of
covariant POM's will be based. It is a generalisation of the imprimitivity
theorem of Mackey (\cite{Mackey}, \cite{Foll}), and, in its first proof given by Cattaneo 
in \cite{Catt}, it is a consequence of Mackey's theorem and of the dilation
theorem of Naimark. A direct proof, which includes the proof of the Mackey
imprimitivity theorem, is given in \cite{Cass}. We recall from the previous
section that a system of covariance for $G$ based on $G/H$ is a triple $%
\left( U,E,\mathcal{H}\right) $ made up by a unitary representation $U$ of $%
G $ in the Hilbert space $\mathcal{H}$ and a POM $E$ on $G/H$ acting in $%
\mathcal{H}$ and satisfying the covariance condition of eq.~(\ref{covarianza
di E}).

\begin{theorem}[Generalised imprimitivity theorem]
\label{teo. GIT}Let $\left( U,E,\mathcal{H}\right) $ be a system of
covariance for $G$ based on $G/H$. There exists a representation $\sigma $
of $H$ such that

\begin{enumerate}
\item  \label{teo. GIT 1}there is an isometry $W:\mathcal{H}\longrightarrow 
\mathcal{H}^{\sigma }$, with $WU\left( g\right) =\lambda ^{\sigma }\left(
g\right) W$ $\forall g\in G$, such that 
\begin{equation*}
E\left( X\right) =W^{\ast }P^{\sigma }\left( X\right) W\quad \forall X\in 
\mathcal{B}\left( G/H\right) \text{;}
\end{equation*}

\item  \label{teo. GIT 2}the linear hull of the set 
\begin{equation*}
\left\{ P^{\sigma }\left( X\right) Wv\mid v\in \mathcal{H},\,X\in \mathcal{B}%
\left( G/H\right) \right\} 
\end{equation*}
is dense in $\mathcal{H}^{\sigma }$.
\end{enumerate}

\noindent Moreover, the representation $\sigma $ satisfying $\ref{teo. GIT 1}
$ and $\ref{teo. GIT 2}$ is unique up to equivalence.
\end{theorem}

Conversely, given a representation $U$ of $G$ on $\mathcal{H}$, suppose that
for some $\sigma \in \mathrm{rep}\,\left( H\right) $ there exists an
isometry $W:\mathcal{H}\longrightarrow \mathcal{H}^{\sigma }$ intertwining $%
U $ with $\lambda ^{\sigma }$. Then, it is immediately checked that 
\begin{equation*}
E\left( X\right) :=W^{\ast }P^{\sigma }\left( X\right) W\quad \forall X\in 
\mathcal{B}\left( G/H\right)
\end{equation*}
defines a system of covariance $\left( U,E,\mathcal{H}\right) $.

By virtue of the above theorem, the problem of classifying all the systems
of covariance for $G$ based on $G/H$ reduces to characterising the
representations of $G$ which are contained in representations induced from $%
H $. This is directly related to the problem of diagonalising the induced
representation, hence in particular to Plancherel theory. Thus, as we shall
see, a general solution is achievable only for particular choices of $H$,
such that a canonical method for diagonalising $\mathrm{ind}_{H}^{G}\left(
\sigma \right) $ ($\sigma \in \mathrm{rep}\,\left( H\right) $) can be worked
out.

\begin{remark}
Suppose $\left( U,E,\mathcal{H}\right) $ and $\left( U^{\prime },E^{\prime },%
\mathcal{H}^{\prime }\right) $ are two systems of covariance for $G$ based
on $G/H$. Then they are \textbf{equivalent} if there exists a unitary
operator $V:\mathcal{H}\longrightarrow \mathcal{H}^{\prime }$ such that $%
U^{\prime }\left( g\right) =VU\left( g\right) V^{-1}$ $\forall g\in G$ and $%
E^{\prime }\left( X\right) =VE\left( X\right) V^{-1}$ $\forall X\in \mathcal{%
B}\left( G/H\right) $.
\end{remark}

\begin{remark}
\label{Rem. sulla rappr. di E come funz. su Cc}Suppose $\Omega $ is a
locally compact second countable Hausdorff space, and let $C_{c}\left(
\Omega \right) $ be the linear space of continuous complex valued functions
on $\Omega $ with compact support. If $E:$ $\mathcal{B}\left( \Omega \right)
\longrightarrow \mathcal{L}\left( \mathcal{H}\right) $ is a POM, for $\varphi
\in C_{c}\left( \Omega \right) $ we define the following bounded operator on 
$\mathcal{H}$ 
\begin{equation*}
E\left( \varphi \right) :=\int_{\Omega }\varphi \left( x\right) \text{d}%
E\left( x\right) \text{,}
\end{equation*}
where the integral is understood in the weak sense. It is known that the map 
$\varphi \longmapsto E\left( \varphi \right) $ defines uniquely the POM $E$.
Indeed, if $u,v\in \mathcal{H}$, let $\mu _{u,v}$ be the complex measure
given by 
\begin{equation}
\mu _{u,v}\left( X\right) =\left\langle u,E\left( X\right) v\right\rangle
\quad \forall X\in \mathcal{B}\left( \Omega \right) \text{.}
\label{Def. di mu u,v}
\end{equation}
By the Riesz representation theorem, the complex measure $\mu _{u,v}$ can be
recovered from the bounded linear functional it induces on $C_{c}\left(
\Omega \right) $, i.e.~from the mapping 
\begin{equation*}
C_{c}\left( \Omega \right) \ni \varphi \longmapsto \int_{\Omega}
\varphi \left(
x\right) \text{d}\mu _{u,v}\left( x\right) \equiv \left\langle u,
E\left(\varphi \right) v\right\rangle 
\end{equation*}
(here we used the definition of $E\left( \varphi \right) $). Hence, the
mapping $\varphi \longmapsto E\left( \varphi \right) $ determines all the
complex measures $\mu _{u,v}$ ($u,v\in \mathcal{H}$), and these in turn
determine the POM $E$ by means of (\ref{Def. di mu u,v}). For more details,
we refer to $\cite{NST}$. In the following, we will often use this
alternative description of $E$.

If $\Omega =G/H$, denoting by $\varphi ^{g}\left( x\right) :=\varphi \left(
g^{-1}\left[ x\right] \right) $ the action of an element $g\in G$ on a
function $\varphi \in C_{c}\left( G/H\right) $, covariance condition $(\ref
{covarianza di E})$ is equivalent to the following 
\begin{equation*}
E\left( \varphi ^{g}\right) =U\left( g\right) E\left( \varphi \right) U\left(
g\right) ^{-1}\quad \forall \varphi \in C_{c}\left( G/H\right) ,\,g\in G\text{%
.}
\end{equation*}
If $P^{\sigma }$ is the projection valued measure of eq.~$(\ref{La PVM
canonica})$, we have, for $\varphi \in C_{c}\left( G/H\right) $, 
\begin{equation*}
\left[ P^{\sigma }\left( \varphi \right) f\right] \left( g\right) =\varphi
\left( \dot{g}\right) f\left( g\right) \quad \forall f\in \mathcal{H}%
^{\sigma }\text{.}
\end{equation*}
\end{remark}

\begin{remark}
\label{Rem. sulla densita' di H0sigma}Let $\sigma $ be a representation of $H
$ in the Hilbert space $\mathcal{K}$. For $\varphi \in C_{c}\left( G\right) $
and $v\in \mathcal{K}$, let 
\begin{equation*}
f_{\varphi v}\left( g\right) :=\int_{H}\varphi \left( gh\right) \sigma
\left( h\right) v\text{d}\mu _{H}\left( h\right) \qquad \forall g\in G
\end{equation*}
(here $\mu _{H}$ is a Haar measure of $H$). It is easily checked that $%
f_{\varphi v}$ is a continuous function in $\mathcal{H}^{\sigma }$, whose
support is compact modulo $H$. Moreover, if $\mathcal{D}\subset \mathcal{K}$
is a subset which is total in $\mathcal{K}$, it is known that the set 
\begin{equation*}
\left\{ f_{\varphi v}\mid \varphi \in C_{c}\left( G\right) ,\,v\in \mathcal{D%
}\right\} 
\end{equation*}
is total in $\mathcal{H}^{\sigma }$ (see $\cite{Foll}$).
\end{remark}

\begin{remark}
\label{Rem. sulla rappr. proiettiva}In chapter $\ref{cap. 2}$, we will deal
also with \textbf{projective unitary representations} of the group $G$
(sometimes abbreviated in `projective representations'). We recall that such
a representation acting in a Hilbert space $\mathcal{H}$ is a weakly
measurable map $G\ni g\longmapsto U\left( g\right) \in \mathcal{U}\left( 
\mathcal{H}\right) $, $\mathcal{U}\left( \mathcal{H}\right) $ being the
unitary group of $\mathcal{H}$, such that $U\left( e\right) =1$ and $U\left(
g_{1}g_{2}\right) =m\left( g_{1},g_{2}\right) U\left( g_{1}\right) U\left(
g_{2}\right) $. The measurable
map $m:G\times G\longrightarrow \mathbb{T}$, $\mathbb{T}$
being the set of complex numbers with modulus one, is the \textbf{multiplier}
of $U$ (see $\cite{Var}$ for more details; in particular, if the multiplier $m$
is trivial, there exists a measurable map $a:G\longrightarrow \mathbb{T}$ such
that $G\longmapsto a(g)U(g)$ is a strongly continuous unitary representation of $G$).
As in the unitary case, a POM $E:%
\mathcal{B}\left( G/H\right) \longrightarrow \mathcal{H}$ is $U$-covariant
if $U\left( g\right) E\left( X\right) U\left( g\right) ^{-1}=E\left( g\left[
X\right] \right) $ for all $X\in \mathcal{B}\left( G/H\right) $ and $g\in G$.
\end{remark}

\section{\label{Sez. sugli stati coerenti}Systems of covariance and coherent
states}

Let $G$, $H$ and $G/H$ be as in the previous section. In addition, suppose $%
U\in \mathrm{rep}\,\left( G\right) $ is fixed. Let $\mathcal{H}$ be the
Hilbert space of $U$. We are now interested in the $U$-covariant POM's that
are expressible by means of an operator density with respect to some measure
on $G/H$. In other words, we are looking for those $U$-covariant POM's $E$
based on $G/H$ such that there exist a positive Borel measure $\nu $ on $G/H$
and a weakly $\nu $-measurable map $G/H\ni x\longmapsto E\left( x\right) \in 
\mathcal{L}\left( \mathcal{H}\right) $ satisfying 
\begin{equation}
E\left( X\right) =\int_{X}E\left( x\right) \text{d}\nu \left( x\right) \quad
\forall X\in \mathcal{B}\left( G/H\right).  \label{(E)}
\end{equation}
Here and in the following, our operator valued integrals are always understood in
the weak sense.
If eq.~(\ref{(E)})
holds, then we have the following resolution of the identity 
\begin{equation}
I=\int_{G/H}E\left( x\right) \text{d}\nu \left( x\right) \text{,}
\label{ris}
\end{equation}
so that $E\left( x\right) $, $x\in G/H$, define a family of \textbf{%
generalised coherent states }in the sense of \cite{Ali}. This terminology is
justified by the following fact: if there exists a measurable map $G/H\ni
x\longmapsto \psi \left( x\right) \in \mathcal{H}$, with $\left\| \psi
\left( x\right) \right\| =1$ a.e., such that $E\left( x\right) $ is the
orthogonal projection along $\psi \left( x\right) $, then eq.~(\ref{ris})
reads\footnote{%
We recall that our inner product is linear in the second argument. Eq.~(\ref
{ris}) is also equivalent to 
\begin{equation*}
\left\langle u,v\right\rangle =\int \left\langle u,\psi \left( x\right)
\right\rangle \left\langle \psi \left( x\right) ,v\right\rangle \text{d}\nu
\left( x\right) \quad \forall u,v\in \mathcal{H}\text{.}
\end{equation*}
} 
\begin{equation*}
I=\int_{G/H}\left\langle \psi \left( x\right),\cdot  \right\rangle \psi
\left( x\right) \text{d}\nu \left( x\right) \text{,}
\end{equation*}
thus showing the connection with classical coherent states \cite{Davies76}.

Theorem \ref{Teo. sugli stati coerenti}
below is a consequence of the result found by Cattaneo in \cite{Catt2}.
Here we give a different and simplified proof, which, up to our
knowledge, is new. It is based on the following lemma.

\begin{lemma}
\label{Lemma 1.3.1}If eq.~$(\ref{(E)})$ holds for the $U$-covariant POM $E$,
then there exists a weakly continuous map $G/H\ni x\longmapsto \widetilde{E}%
\left( x\right) \in \mathcal{L}\left( \mathcal{H}\right) $ such that

\begin{enumerate}
\item  
\begin{equation*}
U\left( g\right) \widetilde{E}\left( x\right) U\left( g\right) ^{-1}=%
\widetilde{E}\left( g\left[ x\right] \right) \quad \forall g\in G,\,\forall
x\in G/H\text{;}
\end{equation*}

\item  
\begin{equation*}
E\left( X\right) =\int_{X}\widetilde{E}\left( x\right) \text{d}\mu
_{G/H}\left( x\right) \quad \forall X\in \mathcal{B}\left( G/H\right) \text{.%
}
\end{equation*}
\end{enumerate}
\end{lemma}

\begin{proof}
We first show that in eq.~(\ref{(E)}) $E\left( x\right) $ is a positive
operator for $\nu $-a.a.~$x$, and $\nu $ can always be chosen to be the
invariant measure $\mu _{G/H}$. First of all, possibly redefining $\nu $, we
can assume $E\left( x\right) \neq 0$ for $\nu $-a.a.~$x$. Let $\left(
v_{n}\right) _{n\geq 1}$ be a countable sequence of vectors which is dense
in $\mathcal{H}$. Since, for all $n\geq 1$, $\left\langle E\left( X\right)
v_{n},v_{n}\right\rangle \geq 0$ for all $X\in \mathcal{B}\left( G/H\right) $%
, it follows from eq.~(\ref{(E)}) that there is a $\nu $-null set $N\in 
\mathcal{B}\left( G/H\right) $ such that $\left\langle E\left( x\right)
v_{n},v_{n}\right\rangle \geq 0$ for all $n\geq 1$ and $x\notin N$. By
continuity, $E\left( x\right) \geq 0$ for all $x\notin N$. For each $n$, we
define the bounded measures 
\begin{equation*}
\mu _{n}\left( X\right) =\left\| v_{n}\right\| ^{-2}\left\langle 
v_{n} , E\left(X\right) v_{n}
\right\rangle \quad \forall X\in \mathcal{B}\left(
G/H\right)
\end{equation*}
and 
\begin{equation*}
\mu =\sum_{n}2^{-n}\mu _{n}\text{.}
\end{equation*}
We then have the equivalence 
\begin{equation*}
\mu \left( X\right) =0\Longleftrightarrow E\left( X\right) =0\text{,}
\end{equation*}
and, since $E\left( g\left[ X\right] \right) =U\left( g\right) E\left(
X\right) U\left( g\right) ^{-1}$, 
\begin{equation*}
\mu \left( X\right) =0\Longleftrightarrow \mu \left( g\left[ X\right]
\right) =0\text{.}
\end{equation*}
It follows that $\mu $ is equivalent to the invariant measure $\mu _{G/H}$
(see \cite{Foll}). On the other hand, by eq.~(\ref{(E)}) and monotone
convergence theorem, $\mu $ has density 
\begin{equation*}
x\longmapsto \sigma \left( x\right) :=\sum_{n}2^{-n}\left\| v_{n}\right\|
^{-2}\left\langle v_{n}, E\left( x\right) v_{n}\right\rangle
\end{equation*}
with respect to $\nu $, and, since $\sigma \left( x\right) >0$ for $\nu $%
-a.a.~$x$, $\mu $ and $\nu $ are equivalent. Let $\rho $ be the density of $%
\nu $ with respect to $\mu _{G/H}$, and define $\widetilde{E}\left( x\right)
=\rho \left( x\right) E\left( x\right) $. We then have 
\begin{equation*}
E\left( X\right) =\int_{X}\widetilde{E}\left( x\right) \text{d}\mu
_{G/H}\left( x\right) \quad \forall X\in \mathcal{B}\left( G/H\right)
\end{equation*}
as claimed.

The covariance condition on $E$ easily implies that for all $g\in G$ there
is a $\mu _{G/H}$-null set $N_{g}$ such that 
\begin{equation}
U\left( g\right) \widetilde{E}\left( x\right) U\left( g\right) ^{-1}=%
\widetilde{E}\left( g\left[ x\right] \right) \quad \forall x\notin N_{g}%
\text{.}  \label{cov. di E(x)}
\end{equation}
We now show that $N_{g}$ can be chosen independent on $g\in G$. This will
complete the proof of the lemma. Note that, since 
\begin{equation*}
\exp \left[ \pm i\widetilde{E}\left( x\right) \right] =\sum_{k=0}^{\infty }%
\frac{\left( \pm i\right) ^{k}}{k!}\widetilde{E}\left( x\right) ^{k}\text{,}
\end{equation*}
the convergence being in the uniform norm of $\mathcal{L}\left( \mathcal{H}%
\right) $, the maps $x\longmapsto \exp \left( \pm i\widetilde{E}\left(
x\right) \right) $ are weakly $\mu _{G/H}$-measurable. In fact, it suffices
to show that each map $x\longmapsto \widetilde{E}\left( x\right) ^{k}$ is
weakly $\mu _{G/H}$-measurable, and this is seen by induction on $k$: fixed
an orthonormal basis $\left( \phi _{n}\right) _{n\geq 1}$ of $\mathcal{H}$,
we have 
\begin{equation*}
\left\langle u, \widetilde{E}\left( x\right)^{k} v\right\rangle
=\sum_{n}\left\langle u, \widetilde{E}\left( x\right) \phi _{n}\right\rangle
\phi _{n}, \left\langle \widetilde{E}\left( x\right) ^{k-1} v\right\rangle
\quad \forall x\in G/H
\end{equation*}
and by the inductive hypothesis the right hand side is the sum of $\mu
_{G/H} $-measurable functions, hence is $\mu _{G/H}$-measurable. Inserting
again $I=\sum_{n}\left\langle \cdot ,\phi _{n}\right\rangle \phi _{n}$
between the composed operators, we see that the maps 
\begin{eqnarray*}
g &\longmapsto &2I+U\left( g\right) ^{-1}\exp \left[ i\widetilde{E}\left( 
\dot{g}\right) \right] U\left( g\right) +\text{h.c.}=:B^{+}\left( g\right) \\
g &\longmapsto &2I+iU\left( g\right) ^{-1}\exp \left[ i\widetilde{E}\left( 
\dot{g}\right) \right] U\left( g\right) +\text{h.c.}=:B^{-}\left( g\right)
\end{eqnarray*}
are weakly $\mu _{G}$-measurable. Moreover, $B^{\pm }\left( g\right) \geq 0$
and $\left\| B^{\pm }\left( g\right) \right\| \leq 4$ for all $g\in G$. For $%
n\geq 1$, we can thus define the positive Borel measures $\alpha _{n}^{\pm }$
on $G$, given by 
\begin{equation*}
\int_{G}f\left( g\right) \text{d}\alpha _{n}^{\pm }\left( g\right)
=\int_{G}f\left( g\right) \left\langle 
v_{n},B^{\pm }\left( g\right) v_{n}\right\rangle \text{d}\mu _{G}\left( g\right) \quad 
\forall f\in
C_{c}\left( G\right) \text{.}
\end{equation*}
For all $a\in G$, by eq.~(\ref{cov. di E(x)}) we have 
\begin{equation*}
B^{\pm }\left( ag\right) =B^{\pm }\left( g\right) \quad \text{for~a.a.~}g\in
G\text{,}
\end{equation*}
and so $\alpha _{n}^{\pm }$ are invariant measures on $G$. This implies $%
\alpha _{n}^{\pm }=c_{n}^{\pm }\mu _{G}$ for some constants $c_{n}^{\pm }$,
hence $\left\langle B^{\pm }\left( g\right) v_{n},v_{n}\right\rangle $ is a
constant for a.a.~$g$. It follows that there is a $\mu _{G}$-null set $Z$
and operators $C^{\pm }$ such that 
\begin{equation*}
B^{\pm }\left( g\right) =C^{\pm }\quad \forall g\notin Z\text{.}
\end{equation*}
Since 
\begin{equation*}
U\left( g\right) ^{-1}\exp \left[ i\widetilde{E}\left( \dot{g}\right) \right]
U\left( g\right) =\frac{1}{2i}\left( C^{+}-iC^{-}\right) +\left( i-1\right)
I\quad \text{for~a.a.~}g\in G\text{,}
\end{equation*}
by spectral theorem $g\longmapsto U\left( g\right) ^{-1}\widetilde{E}\left( 
\dot{g}\right) U\left( g\right) $ is constant almost everywhere. Our claim
is thus proved.
\end{proof}

The next theorem characterises those $U$-covariant POM's which define a
family of generalised coherent states.

\begin{theorem}
\label{Teo. sugli stati coerenti}Suppose $E$ is a $U$-covariant POM based on 
$G/H$. Let $\sigma $ and $W$ be as in Theorem $\ref{teo. GIT}$. Then, $E$
admits the representation of eq.~$(\ref{(E)})$ if and only there exists a
bounded operator $A:\mathcal{H}\longrightarrow \mathcal{K}$, $\mathcal{K}$
being the Hilbert space of $\sigma $, such that

\begin{enumerate}
\item  
\begin{equation}
AU\left( h\right) =\sigma \left( h\right) A\quad \forall h\in H\text{;}
\label{wav. 1}
\end{equation}

\item  
\begin{equation}
\left( Wv\right) \left( g\right) =AU\left( g\right) ^{-1}v\quad \text{for
a.a.~}g\in G\text{.}  \label{wav. 2}
\end{equation}
\end{enumerate}
\end{theorem}

\begin{proof}
If $A$ satisfies the conditions in the statement, then 
\begin{eqnarray*}
\left\langle u,E\left( X\right) v\right\rangle _{\mathcal{H}} & = & \left\langle
 Wu,P^{\sigma }\left( X\right) Wv\right\rangle _{\mathcal{H}^{\sigma
}} \\
& = & \int_{X}\left\langle AU\left( g\right) ^{-1}u,AU\left( g\right)
^{-1}v\right\rangle _{\mathcal{K}}\text{d}\mu _{G/H}\left( \dot{g}\right)
\end{eqnarray*}
and eq.~(\ref{(E)}) follows with 
\begin{equation*}
E\left( \dot{g}\right) :=U\left( g\right) A^{\ast }AU\left( g\right)
^{-1}\quad \forall g\in G\text{.}
\end{equation*}

Conversely, if eq.~(\ref{(E)}) holds, then by the previous lemma we can take 
$\nu =\mu _{G/H}$ and $E\left( \dot{g}\right) =U\left( g\right) TU\left(
g\right) ^{-1}$ for some fixed positive operator $T$, where $T$ commutes
with the representation $\sigma ^{\prime }$ of $H$ obtained restricting $U$
to $H$. For all $v\in \mathcal{H}$, define 
\begin{equation*}
\left( W^{\prime }v\right) \left( g\right) =T^{1/2}U\left( g\right)
^{-1}v\quad \forall g\in G\text{.}
\end{equation*}
By eq.~(\ref{ris}), $W^{\prime }$ is an isometry $\mathcal{H}\longrightarrow 
\mathcal{H}^{\sigma ^{\prime }}$. It clearly intertwines $U$ with $\lambda
^{\sigma ^{\prime }}$. Let $P$ be the orthogonal projection onto the closed
subspace $\overline{\mathrm{span}}\,\{P^{\sigma ^{\prime }}\left( X\right)
W^{\prime }v\mid v\in \mathcal{H},\,X\in \mathcal{B}\left( G/H\right) \}$.
Then, as $P$ commutes with $\lambda ^{\sigma ^{\prime }}$ and $P^{\sigma
^{\prime }}$, the imprimitivity theorem of Mackey implies that there exists
an orthogonal projection $P_{\mathcal{H}}$ of $\mathcal{H}$ commuting with $%
\sigma ^{\prime }$ and such that 
\begin{equation*}
\left( Pf\right) \left( g\right) =P_{\mathcal{H}}f\left( g\right) \quad
\forall f\in \mathcal{H}^{\sigma ^{\prime }}\text{.}
\end{equation*}
By the uniqueness statement in Theorem \ref{teo. GIT}, we can assume $\sigma
=\left. \sigma ^{\prime }\right| _{P_{\mathcal{H}}\mathcal{H}}$, the
restriction of $\sigma ^{\prime }$ to the subspace $\mathcal{K}=P_{\mathcal{H%
}}\mathcal{H}$, and $W=W^{\prime }$. Since $PW^{\prime }=W^{\prime }$, we
have $P_{\mathcal{H}}T^{1/2}v=\left( PW^{\prime }v\right) \left( e\right)
=\left( W^{\prime }v\right) \left( e\right) =T^{1/2}v$ for all $v\in 
\mathcal{H}$. Thus, $\mathrm{ran}\,T^{1/2}\subset \mathcal{K}$. If we set $%
A=T^{1/2}$, then $A$ satisfies eqs.~(\ref{wav. 1}) and (\ref{wav. 2}), and
the proof of the theorem is complete.
\end{proof}

An operator $A:\mathcal{H}\longrightarrow \mathcal{K}$ as in the statement
of the theorem above is called \textbf{generalised wavelet operator} (see 
\cite{Ali}). Note that an intertwining isometry $W:\mathcal{H}%
\longrightarrow \mathcal{H}^{\sigma }$ is expressible in the form of eq.~(%
\ref{wav. 2}), with $A:\mathcal{H}\longrightarrow \mathcal{K}$ bounded and
satisfying eq.~(\ref{wav. 1}), if and only if the subspace $W\mathcal{H}%
\subset \mathcal{H}^{\sigma }$ is a reproducing kernel Hilbert space of
continuous functions on $G$ (recall that a reproducing kernel Hilbert space
on $G$ is a Hilbert space of functions in which the evaluation maps at each
point $g\in G$ are continuous functionals). We thus find in a different way
the result found by Cattaneo in (\cite{Catt2}).
\newpage
\
\newpage

\chapter{The abelian case\label{cap. 1}}

\section{Introduction}

\noindent Throughout all this chapter, $G$ will be a Hausdorff locally
compact \emph{abelian} group satisfying the second countability axiom, and $%
H $ will be a closed subgroup in $G$.

In the following, we describe all the systems of covariance for $G$ based on 
$G/H$. First of all, given $U\in \mathrm{rep}\,\left( G\right) $, we will
find a necessary and sufficient condition in order that $U$ admits covariant
POM's based on $G/H$. To give a brief explanation of this point, let $%
\widehat{G}$ be the group of unitary characters of $G$, and $[M(\widehat{G}%
)] $ be the partially ordered set of equivalence classes of positive Borel
measures on $\widehat{G}$ (here $\left[ \rho \right] \leq \left[ \nu \right] 
$ iff $\rho $ has density with respect to $\nu $). As we shall explain with
more details in section \ref{subsec. 2.4}, the
Stone-Naimark-Ambrose-Godement (SNAG) theorem canonically associates to $U$
a unique measure class $\mathcal{C}_{U}\in \lbrack M(\widehat{G})]$. We will
construct a canonical order preserving map $\Phi _{H}:[M(\widehat{G}%
)]\longrightarrow \lbrack M(\widehat{G})]$ such that the following holds: $U$
admits covariant POM's based on $G/H$ if and only if $\left[ \mathcal{C}_{U}%
\right] \leq \Phi _{H}\left( \left[ \mathcal{C}_{U}\right] \right) $. If
this condition is satisfied, the next step consists in characterising the
most general $U$-covariant POM based on $G/H$. We will see that such a POM
is described in terms of a family $W_{x}:E_{x}\rightarrow E$ of isometries,
where the index $x$ runs over the dual group $\widehat{G}$, $\dim E_{x}$
equals the multiplicity of the character $x$ in $U$ and $E$ is a fixed
(infinite dimensional) Hilbert space. Since our classification of covariant
POM's is based on the generalised imprimitivity theorem, an essential point
is the definition of a unitary transform $\Sigma $ which diagonalises $%
\mathrm{ind}_{H}^{G}\left( \sigma \right) $ for $\sigma \in \mathrm{rep}%
\,\left( H\right) $. This sort of ``generalised Fourier transform'' is the
subject of \S \ref{subsec. 2.3}. In \S \ref{subsec. 2.4} we will state the
basic results of this chapter. Finally, as an application in \S \ref{subsec.
2.5} we give three examples:

\begin{enumerate}
\item  the \emph{regular} representation of the real line, where the
positive operator measures describe a class of \emph{position observables}
in one dimension (\S \ref{Esempio 2.5.1});

\item  the \emph{number}-representation of the torus, where the positive
operator measures describe the \emph{phase observables} for a single mode
bosonic field (\S \ref{Esempio 2.5.2});

\item  the tensor product of two \emph{number}-representations of the torus,
where the positive operator valued measures describe the \emph{phase
difference observables} (\S \ref{Esempio 2.5.3}).
\end{enumerate}

In the literature, the problem of classifying the $U$-covariant POM's for
the abelian group $G$ had been solved by Holevo in the special case $%
H=\left\{ e\right\} $ \cite{holevo1}. Our results are then a generalisation
of the results of Holevo. In addition, in the case $H=\left\{ e\right\} $
our approach gives a description of covariant POM's which is more manageable
than the already known one.

The material in this chapter is taken from \cite{CDT2}.

\section{Notations}

All the groups considered in this chapter are Hausdorff, locally compact,
second countable and abelian. If $\Omega $ is a locally compact second
countable Hausdorff space, by a \emph{measure} on $\Omega $ we always mean a
positive measure defined on the Borel $\sigma $-algebra $\mathcal{B}\left(
\Omega \right) $ of $\Omega $ which is finite on compact sets. If $\mathcal{H%
}$ is a Hilbert space, we denote by $C_{c}\left( \Omega ;\mathcal{H}\right) $
the space of functions $\varphi :\Omega \longrightarrow \mathcal{H}$ which
are continuous and with compact support. If $\mathcal{H}=\mathbb{C}$, we use
the standard abbreviated notation $C_{c}\left( \Omega \right) $ for $%
C_{c}\left( \Omega ;\mathbb{C}\right) $.

In the sequel we shall use rather freely basic results of harmonic analysis
on abelian groups, as exposed, for example, in refs.~\cite{Dieu2} and~\cite
{Foll}.

We fix a group $G$ and a closed subgroup $H$. We denote by $\widehat{G}$ and 
$\widehat{H}$ the corresponding dual groups and by $\left\langle
x,g\right\rangle $ the canonical pairing.

We recall that 
\begin{equation*}
\pi :G\longrightarrow G/H\text{,}\qquad \pi \left( g\right) =\dot{g}
\end{equation*}
is the canonical projection onto the quotient group $G/H$. If $a\in G$ and $%
\dot{g}\in G/H$, the action of $a$ on the point $\dot{g}$ is $a\left[ \dot{g}%
\right] =\dot{a}\dot{g}$.

Let $H^{\perp }$ be the annihilator of $H$ in $\widehat{G}$, that is 
\begin{equation*}
H^{\perp }=\left\{ y\in \widehat{G}\mid \left\langle y,h\right\rangle =1%
\text{ }\forall h\in H\right\} \text{.}
\end{equation*}
The group $H^{\perp }$ is a closed subgroup of $\widehat{G}$ and $\widehat{%
G/H}$ can be identified (and we will do) with $H^{\perp }$ by means of 
\begin{equation*}
\left\langle y,\dot{g}\right\rangle :=\left\langle y,g\right\rangle \quad
\forall y\in H^{\perp },\,\forall \dot{g}\in G/H\text{.}
\end{equation*}

Since $H^{\perp }$ is closed, we can consider the quotient group $\widehat{G}%
/H^{\perp }$. We denote by 
\begin{equation*}
q:\widehat{G}\longrightarrow \widehat{G}/H^{\perp }\text{,}\qquad q\left(
x\right) =\dot{x}
\end{equation*}
the canonical projection. The group $\widehat{H}$ can be identified (and we
will do) with the quotient group $\widehat{G}/H^{\perp }$ by means of 
\begin{equation*}
\left\langle \dot{x},h\right\rangle :=\left\langle x,h\right\rangle \quad
\forall \dot{x}\in \widehat{G}/H^{\perp }\text{, }\forall h\in H\text{.}
\end{equation*}

Let $\mu _{G}$, $\mu _{H}$ and $\mu _{G/H}$ be fixed Haar measures on $G$, $%
H $ and $G/H$, respectively.

We denote by $\mu _{H^{\perp }}$\ the Haar measure on $H^{\perp }$ such that
the Fourier-Plancherel cotransform $\overline{\mathcal{F}}_{G/H}$ is a unitary
operator from $L^{2}\left( G/H,\mu _{G/H}\right) $ onto $L^{2}\left( H^{\perp },\mu
_{H^{\perp }}\right) $, where $\overline{\mathcal{F}}_{G/H}$ is given by 
\begin{equation*}
\left( \overline{\mathcal{F}}_{G/H}f\right) \left( y\right)
=\int_{G/H}\left\langle y,\dot{g}\right\rangle f\left( \dot{g}\right) \text{d%
}\mu _{G/H}\left( \dot{g}\right) \quad \text{for a.a.~}y\in H^{\perp }
\end{equation*}
for all $f\in \left( L^{1}\cap L^{2}\right) \left( G/H,\mu _{G/H}\right) $.

Given $\varphi \in C_{c}\left( \widehat{G}\right) $, let 
\begin{equation*}
\widetilde{\varphi }\left( \dot{x}\right) :=\int_{H^{\perp }}\varphi \left(
xy\right) \text{d}\mu _{H^{\perp }}\left( y\right) \quad \forall \dot{x}\in 
\widehat{G}/H^{\perp }\text{.}
\end{equation*}
It is well known that $\widetilde{\varphi }$ is in $C_{c}\left( \widehat{G}%
/H^{\perp }\right) $ and that $\widetilde{\varphi }\geq 0$ if $\varphi \geq
0 $. Given a measure $\nu $ on $\widehat{G}/H^{\perp }$, the map 
\begin{equation}
C_{c}\left( \widehat{G}\right) \ni \varphi \longmapsto \int_{\widehat{G}%
/H^{\perp }}\widetilde{\varphi }\left( \dot{x}\right) \text{d}\nu \left( 
\dot{x}\right) \in \mathbb{C}  \label{def. nu tilde}
\end{equation}
is linear and positive. Hence, by Riesz-Markov theorem, there is a unique
measure $\widetilde{\nu }$ on $\widehat{G}$ such that 
\begin{equation*}
\int_{\widehat{G}}\phi \left( x\right) \text{d}\widetilde{\nu }\left(
x\right) =\int_{\widehat{G}/H^{\perp }}\text{d}\nu \left( \dot{x}\right)
\int_{H^{\perp }}\phi \left( xy\right) \text{d}\mu _{H^{\perp }}\left(
y\right)
\end{equation*}
for all $\phi \in L^{1}\left( \widehat{G},\widetilde{\nu }\right) $. One can
check that the correspondence $\nu \longmapsto \widetilde{\nu }$ preserves
equivalence and orthogonality of measures.

Given a finite measure $\mu $ on $\widehat{G}$, we denote by $\mu ^{q}$ the
image measure of $\mu $ with respect to $q$, i.e.~the measure on $\widehat{G}%
/H^{\perp }$ given by 
\begin{equation*}
\mu ^{q}\left( A\right) =\mu \left( q^{-1}\left( A\right) \right) \quad
\forall A\in \mathcal{B}\left( \widehat{G}/H^{\perp }\right) \text{.}
\end{equation*}

The following well known theorem of harmonic analisys characterises the most
general unitary representation of the abelian group $G$. It is stated in
this form for example in \cite[Theorem 7.40]{Foll} (see also
\cite[Theorem 22.15.1]{Dieu2}). We recall that if $\rho $ is a measure on $%
\widehat{G}$ and $F$ is a Hilbert space, the \textbf{diagonal representation}
$U^{\rho ,F}$ of $G$ in the space $L^{2}\left( \widehat{G},\rho ;F\right) $
is defined by 
\begin{equation*}
\left( U^{\rho ,F}\phi \right) \left( x\right) =\left\langle
x,g\right\rangle \phi \left( x\right) \quad \text{for }\rho \text{-a.a~}x\in 
\widehat{G}
\end{equation*}
for all $\phi \in L^{2}\left( \widehat{G},\rho ;F\right) $ and $g\in G$.

\begin{theorem}[Stone, Naimark, Ambrose, Godement]
\label{Teo. SNAG}Let $U$ be a unitary representation of the abelian group $G$%
. Then there exists a family of disjoint measures $\left( \rho _{j}\right)
_{j\in \mathbb{Z}_{+}\cup \left\{ \infty \right\} }$ on $\widehat{G}$ and a
family of Hilbert spaces $(F_{k})_{j\in \mathbb{Z}_{+}\cup \left\{ \infty
\right\} }$ with $\dim F_{k}=k$ such that $U$ is unitarily equivalent to the
direct sum $\bigoplus_{j\in \mathbb{Z}_{+}\cup \left\{ \infty \right\}
}U^{\rho _{j},F_{j}}$.

If $\left( \rho _{j}^{\prime }\right) _{j\in \mathbb{Z}_{+}\cup \left\{
\infty \right\} }$ is another family of disjoint measures on $\widehat{G}$
such that $U$ is unitarily equivalent to the direct sum $\bigoplus_{j\in 
\mathbb{Z}_{+}\cup \left\{ \infty \right\} }U^{\rho _{j}^{\prime
},F_{j}^{\prime }}$, then $\rho _{j}$ and $\rho _{j}^{\prime }$ are
equivalent measures for all $j\in \mathbb{Z}_{+}\cup \left\{ \infty \right\} 
$.
\end{theorem}

\noindent We say that $U\in \mathrm{rep}\,\left( G\right) $ has \textbf{%
uniform multiplicity} $m\in \mathbb{Z}_{+}\cup \left\{ \infty \right\} $ if,
with the notations of the above theorem, we have $\rho =0$ for all $j\neq m$.

We now fix a representation $U$ of $G$ acting on a Hilbert space $\mathcal{H}
$. Our aim is to describe all the positive operator measures covariant with
respect to $U$. From the generalised imprimitivity theorem the following
fact immediately follows.

\begin{theorem}
\label{Teo. GIT abeliano}A POM $E$ based on $G/H$ and acting on $\mathcal{H}$
is covariant with respect to $U$ if and only if there exists a
representation $\sigma $ of $H$ and an isometry $W$ intertwining $U$ with $%
\mathrm{ind}_{H}^{G}\left( \sigma \right) $ such that 
\begin{equation*}
E\left( \omega \right) =W^{\ast }P^{\sigma }\left( \omega \right) W
\end{equation*}
for all $\omega \in C_{c}\left( G/H\right) $.
\end{theorem}

Note that here and in the rest of this chapter we define the POM $E$ by
means of its representation as a linear form on $C_{c}\left( G/H\right) $,
as explained in Remark \ref{Rem. sulla rappr. di E come funz. su Cc}. This
is done in order to avoid some technical difficulties (like problems when in
the following some order of integration is changed). If $\sigma ^{\prime }$
is another representation of $H$ such that $\sigma $ is contained (as
subrepresentation) in $\sigma $, then the induced imprimitivity system $%
\left( \lambda ^{\sigma },P^{\sigma },\mathcal{H}^{\sigma }\right) $ is
contained in $\left( \lambda ^{\sigma ^{\prime }},P^{\sigma ^{\prime }},%
\mathcal{H}^{\sigma ^{\prime }}\right) $. Hence, we can always assume that $%
\sigma $ in the previous theorem has uniform infinite multiplicity.

With this assumption, and by the equivalence $\widehat{H}\simeq \widehat{G}%
/H^{\perp }$, there exist a measure $\nu $ on $\widehat{G}/H^{\perp }$ and
an infinite dimensional Hilbert space $\mathcal{M}$ such that, up to a
unitary equivalence, $\sigma $ acts diagonally on $L^{2}\left( \widehat{G}%
/H^{\perp },\nu ;\mathcal{M}\right) $. The first step of our construction is
to diagonalise the representation $\mathrm{ind}_{H}^{G}\left( \sigma \right) 
$.

\section{\label{subsec. 2.3}Diagonalisation of $\mathrm{ind}_{H}^{G}\left( 
\protect\sigma \right) $}

In this section, given a representation of $H$ with uniform multiplicity, we
diagonalise the corresponding induced representation.

Let $\nu $ be a measure on $\widehat{G}/H^{\perp }$ and $\mathcal{M}$ be a
Hilbert space. Let $\sigma ^{\nu }$ be the diagonal representation of $H$
acting on the space $L^{2}\left( \widehat{G}/H^{\perp },\nu ;\mathcal{M}%
\right) $, that is

\begin{equation*}
\left( \sigma ^{\nu }\left( h\right) \xi \right) \left( \dot{x}\right)
=\left\langle \dot{x},h\right\rangle \xi \left( \dot{x}\right) {\text{,}}
\end{equation*}
for all $\xi \in L^{2}\left( \widehat{G}/H^{\perp },\nu ;\mathcal{M}\right) $
and $h\in H$.

We denote by $\mathcal{H}^{\nu }$ the space of functions $f:G\times \widehat{%
G}/H^{\perp }\longrightarrow \mathcal{M}$ such that

\begin{enumerate}
\item  $f$ is weakly $\left( \mu _{G}\otimes \nu \right) $-measurable;

\item  for all $h\in H$ and $g\in G$ 
\begin{equation}
f\left( gh,\dot{x}\right) =\overline{\left\langle \dot{x},h\right\rangle }%
f\left( g,\dot{x}\right) \quad \text{for }\nu \text{-a.a.~}\dot{x}\in 
\widehat{G}/H^{\perp }\text{;}  \label{inv}
\end{equation}

\item  
\begin{equation*}
\int_{G/H\times \widehat{G}/H^{\perp }}\left\| f\left( g,\dot{x}\right)
\right\| ^{2}\,\text{d}\left( \mu _{G/H}\otimes \nu \right) \left( \dot{g},%
\dot{x}\right) <+\infty \text{.}
\end{equation*}
\end{enumerate}

\noindent We identify functions in $\mathcal{H}^{\nu }$ that are equal $%
\left( \mu _{G}\otimes \nu \right) $-a.e.. Let $G$ act on $\mathcal{H}^{\nu
} $ as 
\begin{equation*}
\left( \lambda ^{\nu }\left( a\right) f\right) \left( g,\dot{x}\right)
:=f\left( a^{-1}g,\dot{x}\right)
\end{equation*}
for all $a\in G$. Define 
\begin{equation*}
\left( P^{\nu }\left( \omega \right) f\right) \left( g,\dot{x}\right)
:=\omega \left( \dot{g}\right) f\left( g,\dot{x}\right)
\end{equation*}
for all $f\in \mathcal{H}^{\nu }$, $\omega \in C_{c}\left( G/H\right) $.

The following proposition is stated in \cite{CDT2} without a proof.

\begin{proposition}
The space $\mathcal{H}^{\nu }$ is a Hilbert space with respect to the inner
product 
\begin{equation*}
\left\langle f_{1},f_{2}\right\rangle _{\mathcal{H}^{\nu }}=\int_{G/H\times 
\widehat{G}/H^{\perp }}\left\langle f_{1}\left( g,\dot{x}\right)
,f_{2}\left( g,\dot{x}\right) \right\rangle \text{d}\left( \mu _{G/H}\otimes
\nu \right) \left( \dot{g},\dot{x}\right) \text{.}
\end{equation*}
If $\varphi \in C_{c}\left( G\times \widehat{G}/H^{\perp };\mathcal{M}%
\right) $, let 
\begin{equation*}
f_{\varphi }\left( g,\dot{x}\right) :=\int_{H}\left\langle \dot{x}%
,h\right\rangle \varphi \left( gh,\dot{x}\right) \text{d}\mu _{H}\left(
h\right) \qquad \forall \left( g,\dot{x}\right) \in G\times \widehat{G}%
/H^{\perp }\text{.}
\end{equation*}
Then $f_{\varphi }$ is a continuous function in $\mathcal{H}^{\nu }$ such
that $\left( \pi \times \mathrm{id}_{\widehat{G}/H^{\perp }}\right) \left( 
\mathrm{supp}\,f_{\varphi }\right) $ is compact, and the set 
\begin{equation*}
\mathcal{H}_{0}^{\nu }=\left\{ f_{\varphi }\mid \varphi \in C_{c}\left(
G\times \widehat{G}/H^{\perp };\mathcal{M}\right) \right\}
\end{equation*}
is a dense subspace of $\mathcal{H}^{\nu }$. The triple $\left( \lambda
^{\nu },P^{\nu },\mathcal{H}^{\nu }\right) $ is the canonical imprimitivity
system induced by $\sigma ^{\nu }$ from $H$ to $G$.
\end{proposition}

\begin{proof}
Recall the definitions of $s$ and $V$ given after Remark \ref{Remark sulla
rappr. regolare} in \S \ref{subsec. 1.2}. We define a unitary operator $%
V^{\prime }:\mathcal{H}^{\nu }\longrightarrow L^{2}\left( G/H\times \widehat{%
G}/H^{\perp },\mu _{G/H}\otimes \nu ;\mathcal{M}\right) $ by 
\begin{equation*}
\left( V^{\prime }f\right) \left( \dot{g},\dot{x}\right) =f\left( s\left( 
\dot{g}\right) ,\dot{x}\right) \text{.}
\end{equation*}
We denote by $J$ the canonical identification of $L^{2}\left( G/H\times 
\widehat{G}/H^{\perp },\mu _{G/H}\otimes \nu ;\mathcal{M}\right) $ with $%
L^{2}\left( G/H,\mu _{G/H};L^{2}\left( \widehat{G}/H^{\perp },\nu ;\mathcal{M%
}\right) \right) $, where 
\begin{equation*}
\left[ J\phi \left( \dot{g}\right) \right] \left( \dot{x}\right) =\phi
\left( \dot{g},\dot{x}\right) \quad \text{for a.a.~}\dot{x}\in \widehat{G}%
/H^{\perp }
\end{equation*}
for a.a.~$\dot{g}\in G/H$. It is then easy to check that the unitary map $%
f\longmapsto \hat{f}:=V^{-1}JV^{\prime }f$ intertwines the imprimitivity
systems $\left( \lambda ^{\nu },P^{\nu },\mathcal{H}^{\nu }\right) $ and $%
\left( \lambda ^{\sigma ^{\nu }},P^{\sigma ^{\nu }},\mathcal{H}^{\sigma
^{\nu }}\right) $, since 
\begin{equation*}
\left[ \hat{f}\left( g\right) \right] \left( \dot{x}\right) =f\left( g,\dot{x%
}\right) \quad \text{for }\nu \text{-a.a.~}\dot{x}\in \widehat{G}/H^{\perp }
\end{equation*}
for $\mu _{G}$-a.a.~$g\in G$.

For the second statement in the theorem, the compactness of the set $\left( \pi
\times \mathrm{id}_{\widehat{G}/H^{\perp }}\right) \left( \mathrm{supp}%
\,f_{\varphi }\right) $ is clear from the definition of $f_{\varphi }$.
Also, the continuity of $f_{\varphi }$ is a standard consequence of
dominated convergence theorem. It only remains to prove the density of $%
\mathcal{H}_{0}^{\nu }$. Let $\varphi _{1}\in C_{c}\left( G\right) $, $%
\varphi _{2}\in C_{c}\left( \widehat{G}/H^{\perp };\mathcal{M}\right) $, and 
$\psi \left( g,x\right) =\varphi _{1}\left( g\right) \varphi _{2}\left( \dot{%
x}\right) $. For $f\in \mathcal{H}^{\nu }$, we have 
\begin{eqnarray*}
\left\langle f, f_{\psi } \right\rangle _{\mathcal{H}^{\nu }} & = & \int_{G/H}\text{d}
\mu _{G/H}\left( \dot{g}\right) \int_{\widehat{G}/H^{\perp }}\text{d}\nu
\left( \dot{x}\right) \int_{H}\text{d}\mu _{H}\left( h\right) \varphi
_{1}\left( gh\right) \\
&& \times \left\langle \left\langle \dot{x},h\right\rangle 
f\left( g,\dot{x}\right),
\varphi _{2}\left( 
\dot{x}\right) \right\rangle _{\mathcal{M}} \\
& = & \int_{G/H}\text{d}\mu _{G/H}\left( \dot{g}\right) \int_{H}\text{d}\mu
_{H}\left( h\right) \varphi _{1}\left( gh\right) \\
&& \times \int_{\widehat{G}/H^{\perp }}\text{d}\nu \left( \dot{x}\right)
\left\langle \left[ \hat{f}\left( g\right) \right] \left( \dot{x}\right) ,
\left( \sigma ^{\nu }\left( h\right) \varphi _{2}\right) \left( 
\dot{x}\right)
\right\rangle _{\mathcal{M}} \\
& = & \int_{G/H}\text{d}\mu _{G/H}\left( \dot{g}\right) \int_{H}\text{d}\mu
_{H}\left( h\right) \varphi _{1}\left( gh\right) \\
&& \times \left\langle \hat{f}\left( g\right) ,
\sigma ^{\nu
}\left( h\right) \varphi _{2}
\right\rangle_{L^{2}\left( \widehat{G}/H^{\perp },\nu ;\mathcal{M}\right) } \\
& = &\left\langle \hat{f}, f_{\varphi _{1}\varphi _{2}}
\right\rangle _{\mathcal{H}%
^{\sigma ^{\nu }}}\text{.}
\end{eqnarray*}
Since the set 
\begin{equation*}
\left\{ f_{\varphi _{1}\varphi _{2}}\mid \varphi _{1}\in C_{c}\left(
G\right) ,\,\varphi _{2}\in C_{c}\left( \widehat{G}/H^{\perp };\mathcal{M}%
\right) \right\}
\end{equation*}
is total in $\mathcal{H}^{\sigma ^{\nu }}$ (see Remark \ref{Rem. sulla
densita' di H0sigma}), the density of $\mathcal{H}_{0}^{\nu }$ follows.
\end{proof}

We now diagonalise the representation $\lambda ^{\nu }$. First of all, we
let $\widetilde{\nu }$ be the measure defined in $\widehat{G}$ by eq.~(\ref
{def. nu tilde}). Let $\Lambda ^{\nu }$ be the diagonal representation of $G$
acting on $L^{2}\left( \widehat{G},\widetilde{\nu };\mathcal{M}\right) $ as 
\begin{equation*}
\left( \Lambda ^{\nu }\left( g\right) \phi \right) \left( x\right)
=\left\langle x,g\right\rangle \phi \left( x\right) \quad \text{for a.a.~}%
x\in \widehat{G}
\end{equation*}
for all $g\in G$.

Moreover, given $\phi :\widehat{G}\longrightarrow \mathcal{M}$ and fixed $%
x\in \widehat{G}$, define $\phi _{x}$ from $H^{\perp }$ to $\mathcal{M}$ as 
\begin{equation*}
\phi _{x}\left( y\right) :=\phi \left( xy\right) \quad \forall y\in H^{\perp
}\text{.}
\end{equation*}

\begin{theorem}
There is a unique unitary operator $\Sigma $ from $\mathcal{H}^{\nu }$ onto $%
L^{2}\left( \widehat{G},\widetilde{\nu };\mathcal{M}\right) $ such that, for
all $f\in \mathcal{H}_{0}^{\nu }$, 
\begin{equation}
\left( \Sigma f\right) \left( x\right) =\int_{G/H}\left\langle
x,g\right\rangle f\left( g,\dot{x}\right) \text{\textnormal{d}}\mu
_{G/H}\left( \dot{g}\right) \quad \text{for a.a.~}x\in \widehat{G}\text{.}
\label{Formula di sigma}
\end{equation}
The operator $\Sigma $ intertwines $\lambda ^{\nu }$ with $\Lambda ^{\nu }$.
Moreover, 
\begin{equation}
\left( \Sigma ^{\ast }\varphi \right) \left( g,\dot{x}\right)
=\int_{H^{\perp }}\overline{\left\langle xy,g\right\rangle }\varphi \left(
xy\right) \text{\textnormal{d}}\mu _{H^{\perp }}\left( y\right) \quad \text{%
for a.a.~}\left( g,\dot{x}\right) \in G\times \widehat{H}  \label{star}
\end{equation}
for all $\varphi \in C_{c}\left( \widehat{G};\mathcal{M}\right) $.
\end{theorem}

\begin{proof}
We first define $\Sigma $ on $\mathcal{H}_{0}^{\nu }$. Let $f\in \mathcal{H}%
_{0}^{\nu }$. Fixed $x\in \widehat{G}$, by virtue of eq.~(\ref{inv}) the
function 
\begin{equation*}
g\longmapsto \left\langle x,g\right\rangle f\left( g,\dot{x}\right)
\end{equation*}
depends only on the equivalence class $\dot{g}$ of $g$ and we let $f^{x}$ be
the corresponding map on $G/H$. Due to the properties of $f$, $f^{x}$ is
continuous and has compact support, so it is $\mu _{G/H}$-integrable and we
define $\Sigma f$ by means of eq.~(\ref{Formula di sigma}).

We claim that $\Sigma f$ is in $L^{2}\left( \widehat{G},\widetilde{\nu };%
\mathcal{M}\right) $ and $\left\| \Sigma f\right\| _{L^{2}\left( \widehat{G},%
\widetilde{\nu };\mathcal{M}\right) }=\left\| f\right\| _{\mathcal{H}^{\nu
}} $. Since the map 
\begin{equation*}
\left( x,\dot{g}\right) \longmapsto f^{x}\left( \dot{g}\right)
\end{equation*}
is continuous from $\widehat{G}\times G/H$ to $\mathcal{M}$ and has compact
support, by a standard argument $\Sigma f$ is continuous. Moreover, if $x\in 
\widehat{G}$ and $y\in H^{\perp }$, 
\begin{eqnarray*}
\left( \Sigma f\right) \left( xy\right) & = & \int_{G/H}\left\langle
xy,g\right\rangle f\left( g,\dot{x}\right) \text{\textnormal{d}}\mu
_{G/H}\left( \dot{g}\right) \\
& = & \int_{G/H}\left\langle y,\dot{g}\right\rangle \left\langle x,g\right\rangle
f\left( g,\dot{x}\right) \text{\textnormal{d}}\mu _{G/H}\left( \dot{g}\right)
\\
& = & \overline{\mathcal{F}}_{G/H}\left( f^{x}\right) \left( y\right) \text{.}
\end{eqnarray*}
We then have 
\begin{eqnarray*}
\left\| \Sigma f\right\| _{L^{2}\left( \widehat{G},\widetilde{\nu };\mathcal{%
M}\right) }^{2} & = & \int_{\widehat{G}}\left\| \left( \Sigma f\right) \left(
x\right) \right\| ^{2}\text{d}\widetilde{\nu }\left( x\right) \\
& = & \int_{\widehat{G}/H^{\perp }}\text{d}\nu \left( \dot{x}\right)
\int_{H^{\perp }}\left\| \left( \Sigma f\right) \left( xy\right) \right\|
^{2}\text{d}\mu _{H^{\perp }}\left( y\right) \\
& =& \int_{\widehat{G}/H^{\perp }}\text{d}\nu \left( \dot{x}\right)
\int_{H^{\perp }}\left\| \overline{\mathcal{F}}_{G/H}\left( f^{x}\right)
\left( y\right) \right\| ^{2}\text{d}\mu _{H^{\perp }}\left( y\right) \\
&& (\text{unitarity~of~}\overline{\mathcal{F}}_{G/H}) \\
& = & \int_{\widehat{G}/H^{\perp }}\text{d}\nu \left( \dot{x}\right)
\int_{G/H}\left\| f^{x}\left( \dot{g}\right) \right\| ^{2}\text{d}\mu
_{G/H}\left( \dot{g}\right) \\
& = & \int_{\widehat{G}/H^{\perp }}\text{d}\nu \left( \dot{x}\right)
\int_{G/H}\left\| f\left( g,\dot{x}\right) \right\| ^{2}\text{d}\mu
_{G/H}\left( \dot{g}\right) \\
& = & \int_{G/H\times \widehat{G}/H^{\perp }}\left\| f(g,\dot{x})\right\| ^{2}%
\text{d}\left( \mu _{G/H}\otimes \nu \right) \left( \dot{g},\dot{x}\right) \\
& = & \Vert f\Vert _{\mathcal{H}^{\nu }}^{2}\text{.}
\end{eqnarray*}
By density, $\Sigma $ extends to an isometry from $\mathcal{H}^{\nu }$ to $%
L^{2}\left( \widehat{G},\widetilde{\nu };\mathcal{M}\right) $. Clearly, eq.~(%
\ref{Formula di sigma}) holds and it defines uniquely $\Sigma $.

The second step is computing the adjoint of $\Sigma $. Let $\varphi \in
C_{c}\left( \widehat{G};\mathcal{M}\right) $, by standard arguments the
right hand side of eq.~(\ref{star}) is a continuous function of $\left( g,%
\dot{x}\right) $. Moreover, it satisfies eq.~(\ref{inv}). We have 
\begin{equation*}
\int_{H^{\perp }}\overline{\left\langle xy,g\right\rangle }\varphi \left(
xy\right) \text{\textnormal{d}}\mu _{H^{\perp }}\left( y\right) =\overline{%
\left\langle x,g\right\rangle }\overline{\mathcal{F}}_{G/H}^{\ast }\left(
\varphi _{x}\right) \left( \dot{g}\right) \text{.}
\end{equation*}
First of all, we show that the above function of $\left( g,\dot{x}\right) $
is in $\mathcal{H}^{\nu }$. Indeed, {\setlength\arraycolsep{0pt} 
\begin{eqnarray}
&& \int_{\widehat{G}/H^{\perp }}\text{d}\nu \left( \dot{x}\right)
\int_{G/H}\left\| \overline{\left\langle x,g\right\rangle }\overline{%
\mathcal{F}}_{G/H}^{\ast }\left( \varphi _{x}\right) \left( \dot{g}\right)
\right\| ^{2}\text{d}\mu _{G/H}\left( \dot{g}\right)  \notag \\
&&\quad \quad \quad =\int_{\widehat{G}/H^{\perp }}\text{d}\nu \left( \dot{x}%
\right) \int_{G/H}\left\| \overline{\mathcal{F}}_{G/H}^{\ast }\left( \varphi
_{x}\right) \left( \dot{g}\right) \right\| ^{2}\text{d}\mu _{G/H}\left( \dot{%
g}\right)  \notag \\
&& \quad \quad \quad \quad (\text{unitarity~of~}\overline{\mathcal{F}}_{G/H}) 
\notag \\
&& \quad \quad \quad =\int_{\widehat{G}/H^{\perp }}\text{d}\nu \left( \dot{x}%
\right) \int_{H^{\bot }}\left\| \varphi _{x}\left( y\right) \right\| ^{2}%
\text{d}\mu _{H^{\bot }}\left( y\right)  \notag \\
&& \quad \quad \quad =\int_{\widehat{G}/H^{\perp }}\text{d}\nu \left( \dot{x}%
\right) \int_{H^{\bot }}\left\| \varphi \left( xy\right) \right\| ^{2}\text{d%
}\mu _{H^{\bot }}\left( y\right)  \notag \\
&& \quad \quad \quad =\left\| \varphi \right\| _{L^{2}\left( \widehat{G},%
\widetilde{\nu };\mathcal{M}\right) }^{2}\text{.}  \label{modulo quadro}
\end{eqnarray}
}Moreover, for all $f\in \mathcal{H}_{0}^{\nu }$, we have {%
\setlength\arraycolsep{0pt} 
\begin{eqnarray*}
\left\langle f, \Sigma ^{\ast }\varphi \right\rangle _{\mathcal{H}^{\nu
}} & = & \left\langle \Sigma f , \varphi \right\rangle _{L^{2}\left( \widehat{G},%
\widetilde{\nu };\mathcal{M}\right) } \\
& = & \int_{\widehat{G}/H^{\perp }}\text{d}\nu \left( \dot{x}\right)
\int_{H^{\perp }}\left\langle \left( \Sigma
f\right) \left( xy\right),
\varphi \left( xy\right) \right\rangle \text{d}\mu _{H^{\perp }}\left(
y\right) \\
& = & \int_{\widehat{G}/H^{\perp }}\text{d}\nu \left( \dot{x}\right)
\int_{H^{\perp }}\left\langle \overline{%
\mathcal{F}}_{G/H}\left( f^{x}\right) \left( y\right) ,
\varphi _{x}\left( y\right) \right\rangle \text{ d}%
\mu _{H^{\perp }}\left( y\right) \\
&& \text{(unitarity of }\overline{\mathcal{F}}_{G/H}\text{) } \\
& = & \int_{\widehat{G}/H^{\perp }}\text{d}\nu \left( \dot{x}\right)
\int_{G/H}\left\langle f^{x}\left( \dot{g}\right) ,
\overline{\mathcal{F}}_{G/H}^{\ast }\left( \varphi
_{x}\right) \left( \dot{g}\right) \right\rangle 
\text{d}\mu _{G/H}\left( \dot{g}\right) \\
& = & \int_{\widehat{G}/H^{\perp }}\text{d}\nu \left( 
\dot{x}\right) \int_{G/H}\left\langle \left\langle
x,g\right\rangle f\left( g,\dot{x}\right) ,
\overline{\mathcal{F}}_{G/H}^{\ast
}\left( \varphi _{x}\right) \left( \dot{g}\right) \right\rangle \text{d}\mu
_{G/H}\left( \dot{g}\right) \\
& = & \int_{\widehat{G}/H^{\perp }}\text{d}\nu \left( \dot{x}\right)
\int_{G/H}\left\langle 
f\left( g,\dot{x}\right) ,
\overline{\left\langle x,g\right\rangle }\overline{%
\mathcal{F}}_{G/H}^{\ast }\left( \varphi _{x}\right) \left( \dot{g}\right) 
\right\rangle \text{d}\mu _{G/H}\left( \dot{g}%
\right) \\
& = & \int_{G/H\times \widehat{G}/H^{\perp }}\left\langle f\left( g,\dot{x}\right) ,
\overline{\left\langle
x,g\right\rangle }\overline{\mathcal{F}}_{G/H}^{\ast }\left( \varphi
_{x}\right) \left( \dot{g}\right) \right\rangle 
\text{d}\left( \mu _{G/H}\otimes \nu \right) \left( \dot{g},\dot{x}\right) 
\text{.}
\end{eqnarray*}
}Since $\mathcal{H}_{0}^{\nu }$ is dense, eq.~(\ref{star}) follows. By eq.~(%
\ref{modulo quadro}) $\Sigma ^{\ast }$ is isometric, hence $\Sigma $ is
unitary.

Finally, we show the intertwining property. Let $a\in G$ and $f\in \mathcal{H%
}_{0}^{\nu }$. Then $\lambda ^{\nu }\left( a\right) f\in \mathcal{H}%
_{0}^{\nu }$, and so one has 
\begin{eqnarray*}
\left( \Sigma \lambda ^{\nu }\left( a\right) f\right) \left( x\right)
& = & \int_{G/H}\left\langle x,g\right\rangle f\left( a^{-1}g,\dot{x}\right) 
\text{d}\mu _{G/H}\left( \dot{g}\right) \\
& = & \left\langle x,a\right\rangle \int_{G/H}f^{x}\left( a^{-1}[\dot{g}]\right) 
\text{d}\mu _{G/H}\left( \dot{g}\right) \\
&& \text{(}\dot{g}\longrightarrow a\left[ \dot{g}\right] \text{)} \\
& = & \left\langle x,a\right\rangle \int_{G/H}\left\langle x,g\right\rangle
f\left( g,\dot{x}\right) \text{d}\mu _{G/H}\left( \dot{g}\right) \\
& = & \left( \Lambda ^{\nu }\left( a\right) \Sigma f\right) \left( x\right) \text{%
.}
\end{eqnarray*}
By density of $\mathcal{H}_{0}^{\nu }$, it follows that $\Sigma \lambda
^{\nu }\left( a\right) =\Lambda ^{\nu }\left( a\right) \Sigma $.
\end{proof}

Given $\omega \in C_{c}\left( G/H\right) $, let $\widetilde{P^{\nu }}\left(
\omega \right) =\Sigma P^{\nu }\left( \omega \right) \Sigma ^{\ast }$. Then

\begin{proposition}
For all $\omega \in C_{c}\left( G/H\right) $ and $\phi \in L^{2}\left( 
\widehat{G},\widetilde{\nu };\mathcal{M}\right) $, 
\begin{equation}
\left( \widetilde{P^{\nu }}\left( \omega \right) \phi \right) \left(
x\right) =\int_{H^{\perp }}\overline{\mathcal{F}}_{G/H}\left( \omega \right)
\left( y\right) \phi \left( xy^{-1}\right) \text{d}\mu _{H^{\perp }}\left(
y\right) \quad \text{for a.a.~}x\in \widehat{G}\text{.}  \label{La PVM !}
\end{equation}
\end{proposition}

\begin{proof}
Let $\omega \in C_{c}\left( G/H\right) $. We compute the action of $%
\widetilde{P^{\nu }}\left( \omega \right) $ on $C_{c}\left( \widehat{G};%
\mathcal{M}\right) $. If $\varphi \in C_{c}\left( \widehat{G};\mathcal{M}%
\right) $, let 
\begin{equation*}
\xi \left( x\right) :=\int_{H^{\perp }}\overline{\mathcal{F}}_{G/H}\left(
\omega \right) \left( y\right) \varphi \left( xy^{-1}\right) \text{d}\mu
_{H^{\perp }}\left( y\right) \qquad \forall x\in \widehat{G}\text{,}
\end{equation*}
which is well defined and continuous. Moreover, for all $x\in \widehat{G}$
and $y\in H^{\perp }$, 
\begin{eqnarray}
\xi \left( xy\right) & = & \int_{H^{\perp }}\overline{\mathcal{F}}_{G/H}\left(
\omega \right) \left( y^{\prime }\right) \varphi \left( xyy^{\prime
-1}\right) \text{d}\mu _{H^{\perp }}\left( y^{\prime }\right)  \notag \\
& = & \int_{H^{\perp }}\overline{\mathcal{F}}_{G/H}\left( \omega \right) \left(
y^{\prime }\right) \varphi _{x}\left( yy^{\prime -1}\right) \text{d}\mu
_{H^{\perp }}\left( y^{\prime }\right)  \notag \\
& = & \left( \overline{\mathcal{F}}_{G/H}\left( \omega \right) \ast \varphi
_{x}\right) \left( y\right) \text{.}  \label{Eq. 1 in POVM}
\end{eqnarray}
Here and in the following, convolutions are always taken in $H^{\perp }$. If 
$\varphi ,\psi \in C_{c}\left( \widehat{G};\mathcal{M}\right) $,{%
\setlength
\arraycolsep{0pt} 
\begin{eqnarray*}
&& \left\langle \psi ,\widetilde{P^{\nu }}\left( \omega \right) \varphi 
\right\rangle _{L^{2}\left( \widehat{G},\widetilde{\nu };\mathcal{M}\right)
}=\left\langle \Sigma ^{\ast }\psi ,P^{\nu }\left( \omega \right) \Sigma
^{\ast }\varphi \right\rangle _{\mathcal{H}^{\nu }} \\
&& \quad \quad =\int_{\widehat{G}/H^{\perp }}\text{d}\nu \left( \dot{x}\right)
\int_{G/H}\text{d}\mu _{G/H}\left( \dot{g}\right) \Big\langle \left( 
\dot{g}\right) \overline{\left\langle x,g\right\rangle }\overline{\mathcal{F}%
}_{G/H}^{\ast }\left( \psi_{x}\right) \left( \dot{g}\right) ,\omega  \\
&& \quad \quad \quad \times \overline{%
\left\langle x,g\right\rangle } \overline{\mathcal{F}}_{G/H}^{\ast }\left( \varphi
_{x}\right) \left( \dot{g}\right) \Big\rangle \\
&& \quad \quad =\int_{\widehat{G}/H^{\perp }}\text{d}\nu \left( \dot{x}\right)
\int_{G/H}\text{d}\mu _{G/H}\left( \dot{g}\right) \left\langle \left( 
\dot{g}\right) \overline{\mathcal{F}}_{G/H}^{\ast }\left( \psi
_{x}\right) \left( \dot{g}\right) ,\omega \overline{\mathcal{F}}_{G/H}^{\ast
}\left( \varphi_{x}\right) \left( \dot{g}\right) \right\rangle \\
&& \quad \quad \quad \text{(unitarity of }\overline{\mathcal{F}}_{G/H}\text{
and properties of convolution)} \\
&& \quad \quad =\int_{\widehat{G}/H^{\perp }}\text{d}\nu \left( \dot{x}\right)
\int_{H^{\perp }}\text{d}\mu _{H^{\perp }}\left( y\right) \left\langle
\psi_{x}\left( y\right) ,
\left( \overline{\mathcal{F}}_{G/H}\left( \omega \right) \ast \varphi 
_{x}\right) \left( y\right) \right\rangle \\
&& \quad \quad =\int_{\widehat{G}/H^{\perp }}\text{d}\nu \left( \dot{x}\right)
\int_{H^{\perp }}\text{d}\mu _{H^{\perp }}\left( y\right) \left\langle 
\psi \left( xy\right) ,
\xi \left( xy\right) \right\rangle \text{,}
\end{eqnarray*}
} hence eq.~(\ref{La PVM !}) holds on $C_{c}\left( \widehat{G};\mathcal{M}%
\right) $.

Let now $\phi \in L^{2}\left( \widehat{G},\widetilde{\nu };\mathcal{M}%
\right) $. Since 
\begin{equation*}
\left\| \phi \right\| _{L^{2}\left( \widehat{G},\widetilde{\nu };\mathcal{M}%
\right) }^{2}=\int_{\widehat{G}/H^{\perp }}\text{d}\nu \left( \dot{x}\right)
\int_{H^{\perp }}\Vert \phi (xy)\Vert ^{2}\text{d}\mu _{H^{\perp }}\left(
y\right) <+\infty \text{,}
\end{equation*}
by virtue of Fubini's theorem there is a $\nu $-negligible set $X_{1}\subset 
\widehat{G}/H^{\perp }$ such that, for all $x\in \widehat{G}$ with $\dot{x}%
\not\in X_{1}$, $\phi _{x}\in L^{2}\left( H^{\perp },\mu _{H^{\perp }};%
\mathcal{M}\right) $. Moreover, using the definition of $\widetilde{\nu }$,
one can check that $q^{-1}\left( X_{1}\right) $ is $\widetilde{\nu }$%
-negligible. Then, for $\widetilde{\nu }$-almost all $x\in \widehat{G}$, $%
\phi _{x}$ is in $L^{2}\left( H^{\perp },\mu _{H^{\perp }};\mathcal{M}%
\right) $. We observe that the map 
\begin{equation*}
\dot{g}\longmapsto \omega \left( \dot{g}\right) \left( \overline{\mathcal{F}}%
_{G/H}^{\ast }\left( \phi _{x}\right) \right) \left( \dot{g}\right)
\end{equation*}
is then in $\left( L^{1}\cap L^{2}\right) \left( G/H,\mu _{G/H};\mathcal{M}%
\right) $ for $\widetilde{\nu }$-almost all $x\in \widehat{G}$, hence its
Fourier cotransform is continuous, and we have 
\begin{eqnarray}
\overline{\mathcal{F}}_{G/H}\left( \omega \overline{\mathcal{F}}_{G/H}^{\ast
}\left( \phi _{x}\right) \right) \left( e\right) & = & \left( \overline{\mathcal{F%
}}_{G/H}\left( \omega \right) \ast \phi _{x}\right) \left( e\right)  \notag
\\
& = & \int_{H^{\perp }}\overline{\mathcal{F}}_{G/H}\left( \omega \right) \left(
y\right) \phi \left( xy^{-1}\right) \text{d}\mu _{H^{\perp }}\left( y\right) 
\text{.}  \label{Eq. 2 in POVM}
\end{eqnarray}
Now, we let $\left( \varphi _{k}\right) _{k\geq 1}$ be a sequence in $%
C_{c}\left( \widehat{G};\mathcal{M}\right) $ converging to $\phi $ in $%
L^{2}\left( \widehat{G},\widetilde{\nu };\mathcal{M}\right) $. Then 
\begin{equation*}
\int_{\widehat{G}/H^{\perp }}\text{d}\nu \left( \dot{x}\right)
\int_{H^{\perp }}\left\| \left( \varphi _{k}\right) _{x}\left( y\right)
-\phi _{x}\left( y\right) \right\| ^{2}\text{d}\mu _{H^{\perp }}\left(
y\right) \longrightarrow 0
\end{equation*}
and so, possibly passing to a subsequence, there is a $\nu $-negligible set $%
X_{2}\subset \widehat{G}/H^{\perp }$ such that 
\begin{equation*}
\int_{H^{\perp }}\left\| \left( \varphi _{k}\right) _{x}\left( y\right)
-\phi _{x}\left( y\right) \right\| ^{2}\text{d}\mu _{H^{\perp }}\left(
y\right) \longrightarrow 0
\end{equation*}
for all $x\in \widehat{G}$ with $\dot{x}\not\in X_{2}$. This fact means
that, for $\widetilde{\nu }$-almost all $x\in \widehat{G}$, 
\begin{equation*}
\left( \varphi _{k}\right) _{x}\longrightarrow \phi _{x}
\end{equation*}
in $L^{2}\left( H^{\perp },\mu _{H^{\perp }};\mathcal{M}\right) $. It
follows that 
\begin{equation*}
\omega \overline{\mathcal{F}}_{G/H}^{\ast }\left( \left( \varphi _{k}\right)
_{x}\right) \longrightarrow \omega \overline{\mathcal{F}}_{G/H}^{\ast
}\left( \phi _{x}\right)
\end{equation*}
in $L^{1}\left( G/H,\mu _{G/H};\mathcal{M}\right) $. Then, for $\widetilde{%
\nu }$-almost all $x\in \widehat{G}$, 
\begin{equation*}
\overline{\mathcal{F}}_{G/H}\left( \omega \overline{\mathcal{F}}_{G/H}^{\ast
}\left( \left( \varphi _{k}\right) _{x}\right) \right) \longrightarrow 
\overline{\mathcal{F}}_{G/H}\left( \omega \overline{\mathcal{F}}_{G/H}^{\ast
}\left( \phi _{x}\right) \right)
\end{equation*}
uniformly, and, using eqs.~(\ref{Eq. 1 in POVM}), (\ref{Eq. 2 in POVM}), {%
\setlength\arraycolsep{0pt} 
\begin{eqnarray*}
&& \left( \widetilde{P^{\nu }}\left( \omega \right) \varphi _{k}\right) \left(
x\right) =\overline{\mathcal{F}}_{G/H}\left( \omega \overline{\mathcal{F}}%
_{G/H}^{\ast }\left( \left( \varphi _{k}\right) _{x}\right) \right) \left(
e\right) \longrightarrow \\
&& \quad \quad \quad \longrightarrow \overline{\mathcal{F}}_{G/H}\left( \omega 
\overline{\mathcal{F}}_{G/H}^{\ast }\left( \phi _{x}\right) \right) \left(
e\right) =\int_{H^{\perp }}\overline{\mathcal{F}}_{G/H}\left( \omega \right)
\left( y\right) \phi \left( xy^{-1}\right) \text{d}\mu _{H^{\perp }}\left(
y\right) \text{.}
\end{eqnarray*}
}Since $\widetilde{P^{\nu }}\left( \omega \right) \varphi _{k}$ converges to 
$\widetilde{P^{\nu }}\left( \omega \right) \phi $ in $L^{2}\left( \widehat{G}%
,\widetilde{\nu };\mathcal{M}\right) $, eq.~(\ref{La PVM !}) follows from
uniqueness of the limit.
\end{proof}

\section{Characterisation of covariant POM's\label{subsec. 2.4}}

We fix in the following an \emph{infinite dimensional} Hilbert space $%
\mathcal{M}$. According to the results of the previous sections, theorem \ref
{Teo. GIT abeliano} can be stated in the following way.

\begin{theorem}
\label{GIT} A POM $E$ based on $G/H$ and acting on $\mathcal{H}$ is
covariant with respect to $U$ if and only if there exist a measure $\nu $ on 
$\widehat{G}/H^{\perp }$ and an isometry $W$ intertwining $U$ with $\Lambda
^{\nu }$ such that 
\begin{equation*}
E\left( \omega \right) =W^{\ast }\widetilde{P^{\nu }}\left( \omega \right) W
\end{equation*}
for all $\omega \in C_{c}\left( G/H\right) $.
\end{theorem}

To get an explicit form of $W$, we assume that $U$ acts diagonally on $%
\mathcal{H}$. This means that $\mathcal{H}$ is the orthogonal sum of
invariant subspaces 
\begin{equation}
\mathcal{H}=\bigoplus_{k\in I}L^{2}\left( \widehat{G},\rho _{k};F_{k}\right) 
\text{,}  \label{decomp. di U,H 2}
\end{equation}
where $I$ is a denumerable set, $\left( \rho _{k}\right) _{k\in I}$ is a
family of \emph{pairwise disjoint} measures on $\widehat{G}$, $\left(
F_{k}\right) _{k\in I}$ is a family of Hilbert spaces, and the action of $U$
is given by 
\begin{equation*}
\left( U\left( g\right) \phi _{k}\right) \left( x\right) =\left\langle
x,g\right\rangle \phi _{k}\left( x\right) \qquad x\in \widehat{G}\text{,}
\end{equation*}
where $\phi _{k}\in L^{2}\left( \widehat{G},\rho _{k};F_{k}\right) $ and $%
g\in G$. We will denote by $P_{k}$ the orthogonal projector onto the
invariant subspace $L^{2}\left( \widehat{G},\rho _{k};F_{k}\right) $.

The assumption~(\ref{decomp. di U,H 2}) is not restrictive. Indeed, by
Theorem \ref{Teo. SNAG} there are a family of disjoint measures $(\rho
_{k})_{k\in \mathbb{Z}_{+}\cup \{\infty \}}$ and a family of Hilbert spaces $%
(F_{k})_{k\in \mathbb{Z}_{+}\cup \{\infty \}}$ such that $\dim F_{k}=k$ and, up
to unitary equivalence, eq.~(\ref{decomp. di U,H 2}) holds.

Given the decomposition~(\ref{decomp. di U,H 2}), let $\rho $ be a measure
on $\widehat{G}$ such that 
\begin{equation}
\rho (N)=0\Longleftrightarrow \rho _{k}(N)=0\ \ \forall k\in I\text{.}
\label{implic.}
\end{equation}
We recall that the equivalence class of $\rho $ is uniquely defined by the
family $(\rho _{k})_{k\in I}$.

The following proposition was stated in \cite{CDT2} without a proof.

\begin{proposition}
The equivalence class of the measure $\rho $ defined in eq.~(\ref{implic.})
is independent of the choice of decomposition (\ref{decomp. di U,H 2}).
\end{proposition}

\begin{proof}
Fix a decomposition of $\mathcal{H}$ as in eq.~(\ref{decomp. di U,H 2}). For 
$j\in \mathbb{Z}_{+}\cup \left\{ \infty \right\} $, let $I_{j}=\left\{ k\in
I\mid \dim F_{k}=j\right\} $. Let $\rho _{j}^{\prime }$ be a measure on $%
\widehat{G}$ such that 
\begin{equation*}
\rho _{j}^{\prime }(N)=0\Longleftrightarrow \rho _{k}(N)=0\ \ \forall k\in
I_{j}\text{.}
\end{equation*}
Fix Hilbert spaces $\left( F_{j}^{\prime }\right) _{j\in \mathbb{Z}_{+}\cup
\left\{ \infty \right\} }$ such that $\dim F_{j}^{\prime }=j$. Then we can
estabilish a unitary equivalence $V$ between $U$ and the diagonal
representation $U^{\prime }$ acting in $\mathcal{H}^{\prime
}=\bigoplus_{j\in \mathbb{Z}_{+}\cup \left\{ \infty \right\} }L^{2}\left( 
\widehat{G},\rho _{j}^{\prime };F_{j}^{\prime }\right) $. Namely, for $k\in
I_{j}$, let $\gamma _{k,j}$ be the density of $\rho _{k}$ with respect to $%
\rho _{j}^{\prime }$, and $V_{k}:F_{k}\longrightarrow F_{j}^{\prime }$ be a
fixed unitary map. Then we set 
\begin{equation*}
\left( P_{j}^{\prime }VP_{k}\phi \right) \left( x\right) =\sqrt{\gamma
_{k,j}\left( x\right) }V_{k}\left( P_{k}\phi \right) \left( x\right) \text{,}
\end{equation*}
where $P_{j}^{\prime }$ is the orthogonal projector onto the invariant
subspace $L^{2}\left( \widehat{G},\rho _{j}^{\prime };F_{j}^{\prime }\right) 
$ of $\mathcal{H}^{\prime }$. By the definitions we have 
\begin{equation*}
\rho (N)=0\Longleftrightarrow \rho _{j}^{\prime }(N)=0\ \ \forall j\in 
\mathbb{Z}_{+}\cup \left\{ \infty \right\} \text{.}
\end{equation*}
Since by Theorem \ref{Teo. SNAG} the equivalence class of each $\rho
_{j}^{\prime }$ is uniquely determined by the representation $U$, the
equivalence class of $\rho $ is independent of the choice of decomposition (%
\ref{decomp. di U,H 2}).
\end{proof}

By the above proposition, the representation $U$ defines uniquely an
equivalence class $\mathcal{C}_{U}$ of measures $\rho $ such that relation~(%
\ref{implic.}) holds. Choosing in this equivalence class a \emph{finite}
measure $\rho $, we denote by $\mathcal{C}_{U}^{q}$ the equivalence class of
the image measure $\rho ^{q}$ on $\widehat{G}/H^{\perp }$. Clearly $\mathcal{%
C}_{U}^{q}$ depends only on $\mathcal{C}_{U}$.

We now give the central result of this chapter (and of \cite{CDT2}).

\begin{theorem}
\label{Prop. centr.} Let $U$ be a representation of $G$ acting diagonally on
the space $\mathcal{H}$ of eq.~(\ref{decomp. di U,H 2}). Given $\nu _{U}\in 
\mathcal{C}_{U}^{q}$, let $\widetilde{\nu }_{U}$ be the measure given by eq.~%
$(\ref{def. nu tilde})$. The representation $U$ admits covariant positive
operator valued measures based on $G/H$ if and only if, for all $k\in I$, $%
\rho _{k}$ has density with respect to $\widetilde{\nu }_{U}$. In this case,
for every $k\in I$, let $\alpha _{k}$ be the density of $\rho _{k}$ with
respect to $\widetilde{\nu }_{U}$.

Let $\mathcal{M}$ be a fixed infinite dimensional Hilbert space. For each $%
k\in I$, let 
\begin{equation*}
\widehat{G}\ni x\longmapsto W_{k}\left( x\right) \in \mathcal{L}\left( F_{k};%
\mathcal{M}\right)
\end{equation*}
be a weakly measurable map such that $W_{k}\left( x\right) $ are isometries
for $\rho _{k}$-almost all $x\in \widehat{G}$. For $\omega \in C_{c}\left(
G/H\right) $, let $E\left( \omega \right) $ be the operator whose action on $%
\phi \in \mathcal{H}$ is given by 
\begin{eqnarray}
\left( P_{j}E\left( \omega \right) P_{k}\phi \right) \left( x\right)
&= & \int_{H^{\perp }}\text{d}\mu _{H^{\perp }}\left( y\right) \overline{%
\mathcal{F}}_{G/H}\left( \omega \right) \left( y\right) \sqrt{\frac{\alpha
_{k}\left( xy^{-1}\right) }{\alpha _{j}\left( x\right) }}  \notag \\
&& \times W_{j}\left( x\right) ^{\ast }W_{k}\left( xy^{-1}\right) \left(
P_{k}\phi \right) \left( xy^{-1}\right)  \label{eq. di M buona}
\end{eqnarray}
for $\rho _{j}$-almost all~$x\in \widehat{G}$ and $k,j\in I$. Then, $E$ is a
POM covariant with respect to $U$.

Conversely, any POM based on $G/H$ and covariant with respect to $U$ is of
the form given by eq.~$(\ref{eq. di M buona})$.
\end{theorem}

We add some comments before the proof of the theorem.

\begin{remark}
We observe that eq.~$(\ref{eq. di M buona})$ is invariant with respect to
the choice of the measure $\nu _{U}\in \mathcal{C}_{U}^{q}$. Indeed, let $%
\nu _{U}^{\prime }\in \mathcal{C}_{U}^{q}$, and $\beta >0$ be the density of 
$\nu _{U}$ with respect to $\nu _{U}^{\prime }$. Clearly 
\begin{equation*}
\widetilde{\nu _{U}}=\left( \beta \circ q\right) \widetilde{\nu _{U}^{\prime
}}\text{,}
\end{equation*}
so that the density $\alpha _{k}^{\prime }$ of $\rho _{k}$ with respect to 
$\widetilde{\nu _{U}^{\prime }}$ is 
\begin{equation*}
\alpha _{k}^{\prime }=\left( \beta \circ q\right) \alpha _{k}\text{.}
\end{equation*}
It follows that eq.~$(\ref{eq. di M buona})$ does not depend on the choice
of $\nu _{U}\in \mathcal{C}_{U}^{q}$.
\end{remark}

\begin{corollary}
\label{banale}Let $H$ be the trivial subgroup $\{e\}$. The representation $U$
admits covariant positive operator valued measures based on $G$ if and only
if the measures $\rho _{k}$ have density with respect to the Haar measure $%
\mu _{\widehat{G}}$. In this case, the functions $\alpha _{k}$ in eq.~$(\ref
{eq. di M buona})$ are the densities of $\rho _{k}$ with respect to $\mu _{%
\widehat{G}}$.
\end{corollary}

\begin{remark}
The content of the previous corollary was first shown by Holevo in ref.~$
\cite{holevo1}$ for non-normalised POM. In order to compare the two results
observe that, if $\phi \in \left( L^{1}\cap L^{2}\right) \left( \widehat{G}%
,\rho _{k};F_{k}\right) $ and $\psi \in \left( L^{1}\cap L^{2}\right) \left( 
\widehat{G},\rho _{j};F_{j}\right) $, eq.~$(\ref{eq. di M buona})$ becomes {%
\setlength\arraycolsep{0pt} 
\begin{eqnarray*}
&& \left\langle \phi ,E\left( \omega \right) \psi \right\rangle _{\mathcal{H}%
} = \int_{G}\text{d}\mu _{G}\left( g\right) \omega \left( g\right) \int_{%
\widehat{G}\times \widehat{G}}\left\langle y,g\right\rangle \overline{%
\left\langle x,g\right\rangle }\sqrt{\alpha _{k}\left( y\right) \alpha
_{j}\left( x\right) } \\
&& \quad \quad \quad \quad \times \left\langle W_{j}\left( x\right) ^{\ast
}W_{k}\left( y\right) \phi \left( y\right) ,\psi \left( x\right)
\right\rangle \text{d}\left( \mu _{\widehat{G}}\otimes \mu _{\widehat{G}%
}\right) \left( x,y\right) \\
&& \quad \quad \quad =\int_{G}\text{d}\mu _{G}\left( g\right) \omega \left(
g\right) \int_{\widehat{G}\times \widehat{G}}K_{U\left( g^{-1}\right) \psi
,U\left( g^{-1}\right) \phi }\left( x,y\right) \text{d}\left( \mu _{\widehat{%
G}}\otimes \mu _{\widehat{G}}\right) \left( x,y\right) \text{,}
\end{eqnarray*}
}where 
\begin{equation*}
K_{\psi ,\phi }\left( x,y\right) =\sqrt{\alpha _{k}\left( y\right) \alpha
_{j}\left( x\right) }\left\langle W_{k}\left( y\right) \phi \left( y\right)
,W_{j}\left( x\right) \psi \left( x\right) \right\rangle
\end{equation*}
is a bounded positive definite measurable field of forms (compare with eqs.~$%
(4.2)$ and~$(4.3)$ in ref.~$\cite{holevo1}$).
\end{remark}

In order to prove Theorem~\ref{Prop. centr.}, we need the following lemma.

\begin{lemma}
\label{prop. 2.2}Let $\rho $ be a finite measure on $\widehat{G}$. Assume
that there is a measure $\nu $ on $\widehat{G}/H^{\perp }$ such that $\rho $
has density with respect to $\widetilde{\nu }$. Then $\rho $ has density
with respect to $\widetilde{\rho ^{q}}$. In this case, $\nu $ uniquely
decomposes as 
\begin{equation*}
\nu =\nu _{1}+\nu _{2}\text{,}
\end{equation*}
where $\nu _{1}$ is equivalent to $\rho ^{q}$ and $\nu _{2}\perp \rho ^{q}$.
\end{lemma}

\begin{proof}
Suppose that $\nu $ is a measure on $\widehat{G}/H^{\perp }$ such that $\rho
=\alpha \widetilde{\nu }$, where $\alpha $ is a non-negative $\widetilde{\nu 
}$-integrable function on $\widehat{G}$. Then, for all $\varphi \in
C_{c}\left( \widehat{G}/H^{\perp }\right) $, 
\begin{eqnarray*}
\rho ^{q}\left( \varphi \right)& = &\int_{\widehat{G}}\varphi \left( q\left(
x\right) \right) \text{d}\rho \left( \dot{x}\right) \\
& = & \int_{\widehat{G}/H^{\perp }}\text{d}\nu \left( \dot{x}\right)
\int_{H^{\perp }}\varphi \left( \dot{x}\right) \alpha \left( xy\right) \text{%
d}\mu _{H^{\perp }}\left( y\right) \\
& = & \int_{\widehat{G}/H^{\perp }}\varphi \left( \dot{x}\right) \alpha ^{\prime
}\left( \dot{x}\right) \text{d}\nu \left( \dot{x}\right) \text{,}
\end{eqnarray*}
where the function 
\begin{equation*}
\alpha ^{\prime }\left( \dot{x}\right) :=\int_{H^{\perp }}\alpha \left(
xy\right) \text{d}\mu _{H^{\perp }}\left( y\right) \geq 0
\end{equation*}
is $\nu $-integrable by virtue of Fubini theorem. It follows that 
\begin{equation}
\rho ^{q}=\alpha ^{\prime }\nu \text{.}  \label{Mostra}
\end{equation}
Using Lebesgue theorem, we can uniquely decompose 
\begin{equation*}
\nu =\nu _{1}+\nu _{2}\text{,}
\end{equation*}
where $\nu _{1}$ has base $\rho ^{q}$ and $\nu _{2}\perp \rho ^{q}$. From
eq.~(\ref{Mostra}), it follows that $\nu _{1}$ and $\rho ^{q}$ are
equivalent, and this proves the second statement of the lemma. If $A,B\in 
\mathcal{B}\left( \widehat{G}/H^{\perp }\right) $ are disjoint sets such
that $\nu _{2}$ is concentrated in $A$ and $\nu _{1}$ is concentrated in $B$%
, then $\widetilde{\nu _{2}}$ and $\widetilde{\nu _{1}}$ are respectively
concentrated in the disjoint sets $\widetilde{A}=q^{-1}\left( A\right) $ and 
$\widetilde{B}=q^{-1}\left( B\right) $. By definition of $\rho ^{q}$, we
also have 
\begin{equation*}
\rho \left( \widetilde{A}\right) =\rho ^{q}\left( A\right) =0\text{.}
\end{equation*}
Since $\rho $ has density with respect to $\widetilde{\nu }=\widetilde{\nu
_{1}}+\widetilde{\nu _{2}}$ and $\widetilde{\nu _{2}}$ is concentrated in $%
\widetilde{A}$, it follows that $\rho $ has density with respect to $%
\widetilde{\nu _{1}}\cong \widetilde{\rho ^{q}}$. The claim is now clear.
\end{proof}

\begin{proof}[Proof of Theorem~$\ref{Prop. centr.}$]
Let $\rho $ be a finite measure in $\mathcal{C}_{U}$. By virtue of Theorem 
\ref{GIT}, $U$ admits a covariant POM $\Longleftrightarrow $ there exists a
measure $\nu $ in $\widehat{G}/H^{\perp }$ such that $U$ is a
subrepresentation of $\Lambda ^{\nu }\Longleftrightarrow $ each measure $%
\rho _{k}$ has density with respect to $\widetilde{\nu }$ $%
\Longleftrightarrow $ $\rho $ has density with respect to $\widetilde{\nu }$%
. From Lemma \ref{prop. 2.2}, $U$ admits a covariant POM if and only if $%
\rho $ has density with respect to $\widetilde{\rho ^{q}}$. Since $\rho
^{q}\in \mathcal{C}_{U}^{q}$, the first claim follows.

Let now $E$ be a covariant POM. By Theorem \ref{GIT}, there is a measure $%
\nu $ on $\widehat{G}/H^{\perp }$ and an isometry $W$ intertwining $U$ with $%
\Lambda ^{\nu }$ such that 
\begin{equation*}
E\left( \omega \right) =W^{\ast }\widetilde{P^{\nu }}\left( \omega \right)
W\qquad \forall \omega \in C_{c}\left( G/H\right) \text{.}
\end{equation*}
Using Lemma \ref{prop. 2.2}, we (uniquely) decompose 
\begin{equation*}
\nu =\nu _{1}+\nu _{2}\text{,}
\end{equation*}
where $\nu _{1}$ is equivalent to $\nu _{U}$ and $\nu _{2}\perp \nu _{U}$.
Then we have 
\begin{equation*}
\sigma ^{\nu }\simeq \sigma ^{\nu _{U}}\oplus \sigma ^{\nu _{2}}\text{,}
\end{equation*}
from which the decomposition of the corresponding induced imprimitivity
systems follows: 
\begin{eqnarray*}
\left( \Lambda ^{\nu },\widetilde{P^{\nu }},L^{2}\left( \widehat{G},%
\widetilde{\nu };\mathcal{M}\right) \right) & \simeq &\left( \Lambda ^{\nu
_{U}},\widetilde{P^{\nu _{U}}},L^{2}\left( \widehat{G},\widetilde{\nu _{U}};%
\mathcal{M}\right) \right) \\
&&\oplus \left( \Lambda ^{\nu _{2}},\widetilde{P^{\nu _{2}}},L^{2}\left( 
\widehat{G},\widetilde{\nu _{2}};\mathcal{M}\right) \right) \text{.}
\end{eqnarray*}
Moreover, since each $\rho _{k}$ has density with respect to $\widetilde{\nu
_{U}}$ and $\widetilde{\nu _{U}}$ is disjoint from $\widetilde{\nu _{2}}$,
it follows that $W\mathcal{H}\subset L^{2}\left( \widehat{G},\widetilde{\nu
_{U}};\mathcal{M}\right) $. Thus, we can always assume that the measure $\nu 
$ on $\widehat{G}/H^{\perp }$ which occurs in Theorem \ref{GIT} is $\nu _{U}$%
.

We now characherise the form of $W$. For $k\in I$, we can always fix an
isometry $T_{k}:F_{k}\longrightarrow \mathcal{M}$ such that $T_{k}\left(
F_{k}\right) $ are mutually orthogonal subspaces of $\mathcal{M}$. Hence, if
we define, for $\phi _{k}\in L^{2}\left( \widehat{G},\rho _{k};F_{k}\right) $%
, 
\begin{equation*}
\left( T\phi _{k}\right) \left( x\right) :=\sqrt{\alpha _{k}\left( x\right) }%
T_{k}\phi _{k}\left( x\right) \text{,}
\end{equation*}
$T$ is an isometry intertwining $U$ with $\Lambda ^{\nu _{U}}$. We define $%
W_{k}=WP_{k}$. The operator $V=WT^{\ast }$ is a partial isometry commuting
with $\Lambda ^{\nu _{U}}$, hence there exists a weakly measurable
correspondence $\widehat{G}\ni x\longmapsto V\left( x\right) \in \mathcal{L}%
\left( \mathcal{M}\right) $ such that $V\left( x\right) $ are partial
isometries for $\widetilde{\nu _{U}}$-almost all $x\in \widehat{G}$ and 
\begin{equation*}
\left( V\phi \right) \left( x\right) =V\left( x\right) \phi \left( x\right)
\quad \text{for a.a.~}x\in \widehat{G}\text{,}
\end{equation*}
where $\phi \in L^{2}\left( \widehat{G},\widetilde{\nu _{U}};\mathcal{M}%
\right) $. We have $W=WT^{\ast }T=VT$, then 
\begin{eqnarray}
\left( W_{k}\phi _{k}\right) \left( x\right) & =& \sqrt{\alpha _{k}\left(
x\right) }V\left( x\right) T_{k}\phi _{k}\left( x\right)  \notag \\
& = & \sqrt{\alpha _{k}\left( x\right) }W_{k}\left( x\right) \phi _{k}\left(
x\right) \text{,}  \label{Eq. di W}
\end{eqnarray}
where we set 
\begin{equation*}
W_{k}\left( x\right) =V\left( x\right) T_{k}\quad \forall x\in \widehat{G}%
\text{.}
\end{equation*}
Since $W$ is isometric, then $W_{k}^{\ast }W_{k}$ is the identity operator
on $L^{2}\left( \widehat{G},\rho _{k};F_{k}\right) $, hence 
\begin{equation*}
T_{k}^{\ast }V\left( x\right) ^{\ast }V\left( x\right) T_{k}=I_{k}
\end{equation*}
for $\rho _{k}$-almost all $x\in \widehat{G}$, where $I_{k}$ is the identity
operator on $F_{k}$. Since $T_{k}$ is isometric and $V\left( x\right) $ is a
partial isometry for $\widetilde{\nu _{U}}$-almost every $x\in \widehat{G}$
(that is for $\rho _{k}$-almost every $x\in \widehat{G}$), it follows that $%
V\left( x\right) ^{\ast }V\left( x\right) $ is the identity on $\mathrm{ran}%
\,T_{k}$ and that $W_{k}\left( x\right) $ is isometric, for $\rho _{k}$%
-almost every $x\in \widehat{G}$. Weak measurability of the maps $%
x\longmapsto W_{k}\left( x\right) $ is immediate.

The explicit form of $E$ is then given by 
\begin{eqnarray*}
\left( P_{j}E\left( \omega \right) P_{k}\phi \right) \left( x\right) & = & \left(
W_{j}^{\ast }\widetilde{P^{\nu }}\left( \omega \right) W_{k}\phi \right)
\left( x\right) \\
& = & \frac{1}{\sqrt{\alpha _{j}\left( x\right) }}W_{j}\left( x\right) ^{\ast
}\int_{H^{\perp }}\overline{\mathcal{F}}_{G/H}\left( \omega \right) \left(
y\right) \\
&& \times \sqrt{\alpha _{k}\left( xy^{-1}\right) }W_{k}\left( xy^{-1}\right)
\left( P_{k}\phi \right) \left( xy^{-1}\right) \text{d}\mu _{H^{\perp
}}\left( y\right) \text{,}
\end{eqnarray*}
where $\phi \in \mathcal{H}$, $\omega \in C_{c}\left( G/H\right) $.

Conversely, let $\widehat{G}\ni x\longmapsto W_{k}\left( x\right) \in 
\mathcal{L}\left( F_{k};\mathcal{M}\right) $ be a weakly measurable map such
that $W_{k}\left( x\right) $ are isometries for $\rho _{k}$-almost every $%
x\in \widehat{G}$ and for all $k\in I$. We define, for $\phi _{k}\in
L^{2}\left( \widehat{G},\rho _{k};F_{k}\right) $, 
\begin{equation*}
\left( W\phi _{k}\right) \left( x\right) :=\sqrt{\alpha _{k}\left( x\right) }%
W_{k}\left( x\right) \phi _{k}\left( x\right) \text{,}
\end{equation*}
then $W$ is clearly an intertwining isometry between $U$ and $\Lambda ^{\nu
_{U}}$ and eq.~(\ref{eq. di M buona}) defines a covariant POM.
\end{proof}

We now study the problem of equivalence of covariant POM's.

Let $E$ and $E^{\prime }$ be two covariant positive operator valued measures
that are equivalent, i.e.~there exists an unitary operator $S:\mathcal{H}%
\longrightarrow \mathcal{H}$ such that 
\begin{eqnarray}
SU\left( g\right) &=&U\left( g\right) S\quad \forall g\in G\text{,}
\label{condiz. 1} \\
SE\left( \omega \right) &=&E^{\prime }\left( \omega \right) S\quad \forall
\omega \in C_{c}(G/H)\text{.}  \label{condiz. 2}
\end{eqnarray}
We have the following result.

\begin{proposition}
Let $\left( W_{j}\right) _{j\in I}$ and $\left( W_{j}^{\prime }\right)
_{j\in I}$ be families of maps such that eq.~$(\ref{eq. di M buona})$ holds
for $E$ and $E^{\prime }$, respectively.

The POM's $E$ and $E^{\prime }$ are equivalent if and only if, for each $%
k\in I$, there exists a weakly measurable map $x\longmapsto S_{k}\left(
x\right) \in \mathcal{L}\left( F_{k}\right) $ such that $S_{k}\left(
x\right) $ are unitary operators for $\rho _{k}$-almost all $x$ and 
\begin{equation}
\sqrt{\alpha _{k}\left( xy\right) }W_{j}\left( x\right) ^{\ast }W_{k}\left(
xy\right) =\sqrt{\alpha _{k}\left( xy\right) }S_{j}\left( x\right) ^{\ast
}W_{j}^{\prime }\left( x\right) ^{\ast }W_{k}^{\prime }\left( xy\right)
S_{k}\left( xy\right)  \label{Equivalenza}
\end{equation}
for $\left( \rho _{j}\otimes \mu _{H^{\perp }}\right) $-almost all $\left(
x,y\right) $.
\end{proposition}

\begin{proof}
By virtue of condition (\ref{condiz. 1}) and orthogonality of the measures $%
\rho _{k}$, $S$ preserves decomposition (\ref{decomp. di U,H 2}). Moreover,
for each $k\in I$, there exists a weakly measurable map $x\longmapsto
S_{k}\left( x\right) \in \mathcal{L}\left( F_{k}\right) $ such that $%
S_{k}\left( x\right) $ is unitary for $\rho _{k}$-almost all $x$ and, if $%
\phi _{k}\in L^{2}\left( \widehat{G},\rho _{k};F_{k}\right) $, 
\begin{equation*}
\left( S\phi _{k}\right) \left( x\right) =S_{k}\left( x\right) \phi
_{k}\left( x\right) \quad \text{for a.a.~}x\in \widehat{G}\text{.}
\end{equation*}
Condition (\ref{condiz. 2})\ is equivalent to 
\begin{equation*}
P_{j}E\left( \omega \right) P_{k}\phi =P_{j}S^{\ast }E^{\prime }\left(
\omega \right) SP_{k}\phi
\end{equation*}
for all $\phi \in \mathcal{H}$, $\omega \in C_{c}\left( G/H\right) $ and $%
j,k\in I$. It is not restrictive to assume that the densities $\alpha _{k}$
are measurable functions. Let 
\begin{equation*}
\Omega _{j,k}\left( x,x^{\prime }\right) =\sqrt{\frac{\alpha _{k}\left(
x^{\prime }\right) }{\alpha _{j}\left( x\right) }}\left( W_{j}\left(
x\right) ^{\ast }W_{k}\left( x^{\prime }\right) -S_{j}\left( x\right) ^{\ast
}W_{j}^{\prime }\left( x\right) ^{\ast }W_{k}^{\prime }\left( x^{\prime
}\right) S_{k}\left( x^{\prime }\right) \right) ,
\end{equation*}
using eq.~(\ref{eq. di M buona}), the previous condition becomes 
\begin{equation}
\int_{H^{\perp }}\mathcal{F}_{G/H}\left( \omega \right) \left( y\right)
\Omega _{j,k}\left( x,xy^{-1}\right) \left( P_{k}\phi \right) \left(
xy^{-1}\right) \text{d}\mu _{H^{\perp }}\left( y\right) =0  \label{sopra}
\end{equation}
$\rho _{j}$-almost everywhere for all $\phi \in \mathcal{H}$, $\omega \in
C_{c}\left( G/H\right) $ and $j,k\in I$.

Let $K$ be a compact set of $\widehat{G}$ and $v\in F_{k}$. In eq.~(\ref
{sopra}) we choose 
\begin{equation*}
\phi =\chi _{K}\,v\in L^{2}\left( \widehat{G},\rho _{k};F_{k}\right)
\end{equation*}
and $\omega \in C_{c}\left( G/H\right) $ running over a denumerable subset
dense in $L^{2}\left( G/H,\mu _{H^{\perp }}\right) $. It follows that there
exists a $\rho _{j}$-null set $N\subset \widehat{G}$ such that, for all $%
x\notin N$, 
\begin{equation*}
\chi _{K}\left( xy^{-1}\right) \Omega _{j,k}\left( x,xy^{-1}\right) v=0
\end{equation*}
for $\mu _{H^{\perp }}$-almost all $y\in H^{\perp }$. Since $\Omega _{j,k}$
is weakly measurable, the last equation holds in a measurable subset $%
X\subset \widehat{G}\times H^{\perp }$ whose complement is a $\left( \rho
_{j}\otimes \mu _{H^{\perp }}\right) $-null set. Define 
\begin{equation*}
m\left( x,y\right) =xy^{-1}\qquad \forall \left( x,y\right) \in \widehat{G}%
\times H^{\perp }\text{.}
\end{equation*}
For all $\left( x,y\right) \in X\cap m^{-1}\left( K\right) $ we then have 
\begin{equation*}
\Omega _{j,k}\left( x,xy^{-1}\right) v=0\text{.}
\end{equation*}
Since $F_{k}$ is separable and $\widehat{G}$ is $\sigma $-compact, we get 
\begin{equation*}
\Omega _{j,k}\left( x,xy\right) =0
\end{equation*}
for $\left( \rho _{j}\otimes \mu _{H^{\perp }}\right) $-almost all $\left(
x,y\right) \in \widehat{G}\times H^{\perp }$, that is, 
\begin{equation*}
\sqrt{\alpha _{k}\left( xy\right) }W_{j}\left( x\right) ^{\ast }W_{k}\left(
xy\right) =\sqrt{\alpha _{k}\left( xy\right) }S_{j}\left( x\right) ^{\ast
}W_{j}^{\prime }\left( x\right) ^{\ast }W_{k}^{\prime }\left( xy\right)
S_{k}\left( xy\right)
\end{equation*}
for $\left( \rho _{j}\otimes \mu _{H^{\perp }}\right) $-almost all $\left(
x,y\right) $.

Conversely, if condition (\ref{Equivalenza}) is satisfied for all $j,k\in I$%
, then clearly $E$ is equivalent to $E^{\prime }$.
\end{proof}

\section{Examples\label{subsec. 2.5}}

\subsection{\label{Esempio 2.5.1}Translation covariant observables}

Let $\mathcal{H}=L^{2}\left( \mathbb{R},\text{d}x\right) $, where d$x$ is
the Lebesgue measure on $\mathbb{R}$. We consider the representation $U$ of
the group $\mathbb{R}$ acting on $\mathcal{H}$ as 
\begin{equation*}
\left( U(a)\phi \right) \left( x\right) =e^{iax}\phi \left( x\right) \qquad
x\in \mathbb{R}
\end{equation*}
for all $a\in \mathbb{R}$. By means of Fourier transform $\overline{\mathcal{%
F}}_{\mathbb{R}}^{\ast }$, $U$ is clearly equivalent to the regular
representation of $\mathbb{R}$. We classify the POM's based on $\mathbb{R}$
and covariant with respect to $U$. With the notations of the previous
sections, we have 
\begin{equation*}
G=\mathbb{R}\text{,\quad }H=\left\{ 0\right\} \text{,\quad }G/H=\mathbb{R}%
\text{,\quad }\widehat{G}=H^{\perp }=\mathbb{R}\text{,\quad }\widehat{G}%
/H^{\perp }=\left\{ 0\right\} \text{.}
\end{equation*}
We choose $\mu _{G/H}=\frac{1}{2\pi }$d$x$, so that $\mu _{H^{\perp }}=$ d$x$%
, and $\mathcal{M}=\mathcal{H}$.

The representation $U$ is already diagonal with multiplicity equal to $1$,
so that in the decomposition (\ref{decomp. di U,H 2}) we can set $I=\left\{
1\right\} $, $\rho _{1}=$ d$x$, $F_{1}=\mathbb{C}$. Hence, by Corollary~\ref
{banale}, $U$ admits covariant POM's based on $\mathbb{R}$ and $\alpha
_{1}=1 $.

According to Theorem \ref{Prop. centr.}, any covariant POM $E$ is defined in
terms of a weakly measurable map $x\longmapsto W_{1}\left( x\right) $ such
that $W_{1}\left( x\right) :\mathbb{\ C}\longrightarrow \mathcal{H}$ is an
isometry for every $x\in \mathbb{R}$. This is equivalent to selecting a
weakly measurable map $x\longmapsto h_{x}\in \mathcal{H}$, with $\left\|
h_{x}\right\| _{\mathcal{H}}=1$ $\forall x\in \mathbb{R}$, such that $%
W_{1}\left( x\right) =h_{x}$ $\forall x\in \mathbb{R}$. Explicitly, if $\phi
\in L^{2}\left( \mathbb{R},\text{d}x\right) $,
\begin{eqnarray}
\left( E\left( \omega \right) \phi \right) \left( y\right) &=&\int_{\mathbb{R%
}}\overline{\mathcal{F}}_{\mathbb{R}}\left( \omega \right) \left( x\right)
\left\langle h_{y} , h_{y-x}\right\rangle \phi \left( y-x\right) \text{d}x 
\notag \\
&&\text{(by unitarity of }\overline{\mathcal{F}}_{\mathbb{R}}\text{)}  \notag
\\
&=&\int_{\mathbb{R}}\omega \left( x\right) \overline{\mathcal{F}}_{\mathbb{R}%
}\left[ \left\langle h_{y} , h_{y-\cdot } \right\rangle \phi \left( y-\cdot
\right) \right] \left( x\right) \text{d}x  \notag \\
&=&\int_{\mathbb{R}}\omega \left( x\right) e^{iyx}\left\langle h_{y} ,
\overline{%
\mathcal{F}}_{\mathbb{R}}\left[ \phi \left( \cdot \right) h_{\cdot }\right]
\left( -x\right)
\right\rangle \text{d}x\text{.}
\label{Oss. di localizz. su R}
\end{eqnarray}
The corresponding translation covariant observable is thus $E^{\prime
}\left( \omega \right) =\overline{\mathcal{F}}_{\mathbb{R}}^{\ast }E\left(
\omega \right) \overline{\mathcal{F}}_{\mathbb{R}}$ for all $\omega \in
C_{c}\left( \mathbb{R}\right) $.

If $X\in \mathcal{B}\left( \mathbb{R}\right) $ has finite Lebesgue measure,
one can explicitly write down the action of $E\left( X\right) $ on a
function $\phi \in L^{2}\left( \mathbb{R},\text{d}x\right) $. If $\psi \in
L^{1}\cap L^{2}$, with the notations of Remark \ref{Rem. sulla rappr. di E
come funz. su Cc} we have 
\begin{eqnarray*}
\int_{\mathbb{R}}\omega \left( x\right) \mu _{\phi ,\psi }\left( x\right)
&=&\left\langle  \phi ,E\left( \omega \right) \psi \right\rangle \\
&=&\int_{\mathbb{R}}\text{d}x\omega \left( x\right) \int_{\mathbb{R}%
}e^{iyx}\left\langle \phi \left( y\right)
h_{y} ,
\overline{\mathcal{F}}_{\mathbb{R}}\left[ \psi \left(
\cdot \right) h_{\cdot }\right] \left( -x\right) 
\right\rangle \text{d}y\text{.}
\end{eqnarray*}
It follows that the complex measure $\mu _{\phi ,\psi }$ has density 
\begin{equation*}
x\longmapsto \int_{\mathbb{R}}e^{iyx}\left\langle \phi \left( y\right) h_{y} ,
\overline{\mathcal{F}}_{%
\mathbb{R}}\left[ \psi \left( \cdot \right) h_{\cdot }\right] \left(
-x\right) \right\rangle \text{d}y
\end{equation*}
with respect to the Lebesgue measure. We then have 
\begin{eqnarray*}
\left\langle \phi ,E\left( X\right) \psi \right\rangle &=&\int_{\mathbb{R}}%
\text{d}x\chi _{X}\left( x\right) \int_{\mathbb{R}}e^{iyx}
\left\langle \phi \left( y\right) h_{y} ,
\overline{\mathcal{F}}_{%
\mathbb{R}}\left[ \psi \left( \cdot \right) h_{\cdot }\right] \left(
-x\right) \right\rangle
\text{d}y
\\
&=&\int_{\mathbb{R}}\text{d}y\overline{\phi \left( y\right) }\int_{\mathbb{R}%
}\chi_{X}\left( x\right) e^{iyx}
\left\langle h_{y} ,
\overline{\mathcal{F}}_{%
\mathbb{R}}\left[ \psi \left( \cdot \right) h_{\cdot }\right] \left(
-x\right) \right\rangle
\text{d}x\text{,}
\end{eqnarray*}
where we used $\chi _{X}\in L^{2}\left( \mathbb{R},\text{d}x\right) $ to
change the order of integration. By density of $L^{1}\cap L^{2}$ it follows 
\begin{equation*}
\left( E\left( X\right) \psi \right) \left( y\right)
=\int_{X}e^{iyx}\left\langle h_{y} ,
\overline{\mathcal{F}}_{\mathbb{R}}\left[ \psi
\left( \cdot \right) h_{\cdot }\right] \left( -x\right) \right\rangle 
\text{d}x\text{.}
\end{equation*}

If $h_{x}=h$ is constant for almost all $x$, eq.~(\ref{Oss. di localizz. su
R}) gives 
\begin{equation*}
\left( E\left( \omega \right) \phi \right) \left( y\right) =\left[ \overline{%
\mathcal{F}}_{\mathbb{R}}\left( \omega \right) \ast \phi \right] \left(
y\right) =\overline{\mathcal{F}}_{\mathbb{R}}\left[ \omega \overline{%
\mathcal{F}}_{\mathbb{R}}^{-1}\left( \phi \right) \right] \left( y\right) 
\text{,}
\end{equation*}
and so the corresponding translation covariant observable is 
\begin{equation*}
\left( E^{\prime }\left( \omega \right) \phi \right) \left( y\right) =\omega
\left( y\right) \phi \left( y\right) \text{,}
\end{equation*}
i.e.~$E^{\prime }$ is the spectral map associated to the canonical position
operator $Q$.

In \cite{holevo1}, the results summarised in this subsection were obtained
with a different method.

\subsection{\label{Esempio 2.5.2}Covariant phase observables}

We give a complete characterisation of the covariance systems based on the
one dimensional torus 
\begin{equation*}
\mathbb{T}=\left\{ z\in \mathbb{C}\mid \left| z\right| =1\right\} =\left\{
e^{i\theta }\mid \theta \in \lbrack 0,2\pi ]\right\} \text{.}
\end{equation*}
We have 
\begin{gather*}
G=\mathbb{T}\text{,\quad }H=\left\{ 1\right\} \text{,\quad }G/H=\mathbb{T}%
\text{,} \\
\widehat{G}=H^{\perp }=\left\{ \left( \mathbb{T}\ni z\longmapsto z^{n}\in 
\mathbb{C}\right) \mid n\in \mathbb{Z}\right\} \cong \mathbb{Z}\text{,} \\
\widehat{G}/H^{\perp }=\left\{ 1\right\} \text{.}
\end{gather*}
We choose $\mu _{G/H}=\frac{1}{2\pi }$d$\theta =:\mu _{\mathbb{T}}$, so that 
$\mu _{H^{\perp }}$ is the counting measure $\mu _{\mathbb{Z}}$ on $\mathbb{Z%
}$.

Let $U$ be a representation of $\mathbb{T}$. Since $\mathbb{T}$ is compact,
we can always assume that $U$ acts diagonally on 
\begin{equation*}
\mathcal{H}=\bigoplus_{k\in I} F_k \text{,}
\end{equation*}
where $I\subset \mathbb{Z}$, and $F_{k}$ are Hilbert spaces such that $\dim
F_{k}$ is the multiplicity of the representation $k\in \mathbb{Z}$ in $U$.
Explicitly, 
\begin{equation*}
\left( U\left( z\right) \phi _{k}\right) =z^{k}\phi_{k}
\end{equation*}
for all $z\in \mathbb{T}$ and $\phi _{k}\in F_k$.

In order to use eq.~(\ref{decomp. di U,H 2}), we notice that $%
F_{k}=L^{2}\left( \mathbb{Z},\delta _{k};F_{k}\right) $ (where $\delta _{k}$
is the Dirac measure at $k$), so that $\rho _{k}=\delta _{k}$, . By
Corollary \ref{banale}, one has that $U$ admits covariant POM's based on $%
\mathbb{T}$ and that $\alpha _{k}(j)=\delta _{k,j}$ (where $\delta _{k,j}$
is the Kronecker delta).

Choose an infinite dimensional Hilbert space $\mathcal{M}$ and, for each $%
k\in I$, fix an isometry $W_{k}$ from $F_{k}$ to $\mathcal{M}$. The
corresponding covariant POM is given by
\begin{eqnarray*}
P_{j}E\left( \omega \right) P_{k}\phi &=&\overline{\mathcal{F}}_{\mathbb{T}%
}\left( \omega \right) \left( j-k\right) W_{j}^{\ast }W_{k}P_{k}\phi \\
&=&\frac{1}{2\pi }\int_{0}^{2\pi }\omega (e^{i\theta })e^{i(j-k)\theta
}\,W_{j}^{\ast }W_{k}P_{k}\phi \ \text{d}\theta \text{,}
\end{eqnarray*}
where $\phi \in \mathcal{H}$ and $\omega \in \mathcal{C}(\mathbb{T})$.

Reasoning as in \ref{Esempio 2.5.1}, we can find the explicit form of $%
E\left( X\right) $ for every $X\in \mathcal{B}\left( \mathbb{T}\right) $. We
have

\begin{equation*}
P_{j}E\left( X\right) P_{k}\phi =\frac{1}{2\pi }\int_{0}^{2\pi }\chi
_{X}(e^{i\theta })e^{i(j-k)\theta }\,W_{j}^{\ast }W_{k}P_{k}\phi \ \text{d}%
\theta \text{.}
\end{equation*}

If $I=\mathbb{N}$ and $\dim F_{k}=1$ $\forall k\in \mathbb{N}$, $U$ is the 
\textbf{number representation} for a quantum harmonic oscillator. Each
subspace $F_{k}$ is the one dimensional eigenspace of the number operator $N$
corresponding to the eigenvalue $k$. We have $U\left( e^{i\theta }\right)
=e^{i\theta N}$. A transformation by $U\left( e^{i\theta }\right) $
corresponds to a phase shift by $\theta $. Identifying $\mathbb{T}$ with the
additive group of the real numbers $\mathrm{mod}\,2\pi $, the covariane
condition on $E$ thus defines the \textbf{covariant phase observable} for
the quantum harmonic oscillator. For more details, we refer to refs.~\cite
{Lahti95}, \cite{Torus}, \cite{Holevo82}. We only remark that there do not
exist sharp covariant phase observables. In fact, if we assume the contrary,
an easy application of Mackey imprimitivity theorem implies that $U$ is
equivalent to the regular representation of $\mathbb{T}$. But this is
absurd, since $U$ is strictly contained in the regular representation.

\subsection{\label{Esempio 2.5.3}Covariant phase difference observables}

As in the previous section, let $\mu _{\mathbb{T}}$ be the normalised Haar
measure on the one dimensional torus $\mathbb{T}$. We consider the following
representation $U$ of the direct product $G=\mathbb{T}\times \mathbb{T}$
acting on the space $\mathcal{H}=L^{2}\left( \mathbb{T}\times \mathbb{T},\mu
_{\mathbb{T}}\otimes \mu _{\mathbb{T}}\right) $ as 
\begin{equation*}
\left( U\left( a,b\right) f\right) \left( z_{1},z_{2}\right) =f\left(
az_{1},b^{-1}z_{2}\right) \qquad \left( z_{1},z_{2}\right) \in \mathbb{\
T\times T}
\end{equation*}
for all $\left( a,b\right) \in \mathbb{T\times T}$.

Let $H$ be the closed subgroup 
\begin{equation*}
H=\left\{ \left( a,b\right) \in \mathbb{T}\times \mathbb{T}\mid b=a\right\}
\cong \mathbb{T}.
\end{equation*}
We classify all the POM's based on $G/H$ and covariant with respect to $U$.
These describe the phase difference observables for two single mode bosonic
fields (for a more detailed account about the physical meaning of such
observables and for a different approach to the same problem, see ref.~\cite
{Teiko}).

We have 
\begin{gather*}
G=\mathbb{T\times T}\text{,}\quad G/H\cong \mathbb{T}\text{,\quad }\widehat{%
G }=\widehat{\mathbb{T}}\times \widehat{\mathbb{T}}\cong \mathbb{Z}\times 
\mathbb{Z}\text{,} \\
H^{\perp }=\left\{ \left( j,k\right) \in \mathbb{Z\times Z}\mid k=-j\right\}
\cong \mathbb{Z}\text{,} \\
\widehat{G}/H^{\perp }\cong \mathbb{Z}\text{.}
\end{gather*}
We fix $\mu _{G/H}=\mu _{\mathbb{T}}$, so that $\mu _{H^{\perp }}=\mu _{%
\mathbb{Z}}$.

We choose the following orthonormal basis $\left( e_{i,j}\right) _{i,j\in 
\mathbb{Z}}$ of $\mathcal{H}$ 
\begin{equation*}
e_{i,j}\left( z_{1},z_{2}\right) =z_{1}^{i}z_{2}^{-j}\qquad \left(
z_{1},z_{2}\right) \in \mathbb{T}\times \mathbb{T}\text{,}
\end{equation*}
so that 
\begin{equation*}
U\left( a,b\right) e_{i,j}=a^{i}b^{j}e_{i,j}\qquad \forall \left( a,b\right)
\in \mathbb{T}\times \mathbb{T}\text{.}
\end{equation*}
Let $F_{i,j}=\mathbb{C}e_{i,j}$, then $U$ acts diagonally on $F_{i,j}$ as
the character $(i,j)\in \mathbb{Z}\times \mathbb{Z}$. Then, one can choose
as decomposition~(\ref{decomp. di U,H 2}) 
\begin{equation*}
\mathcal{H}=\bigoplus_{i,j\in \mathbb{Z}}F_{i,j}\cong \bigoplus_{i,j\in 
\mathbb{Z}}L^{2}(\mathbb{Z}\times \mathbb{Z},\delta _{i}\otimes \delta
_{j};F_{i,j})
\end{equation*}
With the notations of Section \ref{subsec. 2.4}, we have $I=\mathbb{Z}\times 
\mathbb{Z}$ and $\rho _{i,j}=\delta _{i}\otimes \delta _{j}$. It follows
that $\mathcal{C}_{U}^{q}$ is the equivalence class of $\mu _{\mathbb{Z}}$.
With the choice $\nu _{U}=\mu _{\mathbb{Z}}$, it follows that $\widetilde{%
\nu }=\mu _{\mathbb{Z}}\otimes \mu _{\mathbb{Z}}$. According to Theorem~(\ref
{Prop. centr.}), $U$ admits covariant POM's and $\alpha _{i,j}(n,m)=\delta
_{n,i}\delta _{m,j}$.

With the choice $\mathcal{M}=\mathcal{H}$, we select a map $\left(
i,j\right) \longmapsto W_{i,j}$, where $W_{i,j}$ is an isometry from $%
F_{i,j} $ to $\mathcal{H}$. Since $F_{i,j}$ are one dimensional, there
exists a family of vectors $\left( h_{i,j}\right) _{i,j\in \mathbb{Z}}$ in $%
\mathcal{H}$, with $\left\| h_{i,j}\right\| _{\mathcal{H}}=1$ $\forall
\left( i,j\right) \in \mathbb{Z\times Z}$, such that 
\begin{equation*}
W_{i,j}e_{i,j}=h_{i,j}\qquad \forall \left( i,j\right) \in \mathbb{Z\times Z}%
\text{.}
\end{equation*}
The corresponding covariant POM $E$ is given, for every $\phi \in \mathcal{H}
$, by 
\begin{eqnarray*}
P_{l,m}E\left( \omega \right) P_{i,j}\phi &=&\sum_{h\in \mathbb{Z}}\mathcal{F%
}_{\mathbb{T}}\left( \omega \right) \left( h\right) \delta _{l-h,i}\delta
_{m+h,j}\left\langle h_{l,m}, h_{i,j}\right\rangle \ \left\langle 
e_{i,j}, \phi \right\rangle e_{l,m} \\
&=&\delta _{l+m,i+j}\ \mathcal{F}_{\mathbb{T}}\left( \omega \right) \left(
j-m\right) \left\langle h_{l,m}, h_{i,j} \right\rangle \ \left\langle e_{i,j} ,
\phi \right\rangle e_{l,m}\text{.}
\end{eqnarray*}
In particular, if $l+m=i+j$, we have 
\begin{eqnarray*}
\left\langle e_{l,m} , E\left( \omega \right) e_{i,j} \right\rangle &=&%
\overline{\mathcal{F}}_{\mathbb{T}}(\omega )(j-m)\left\langle
h_{l,m} , h_{i,j} \right\rangle \\
&=&\frac{1}{2\pi }\int_{0}^{2\pi }\omega (e^{i\theta })e^{i(j-m)\theta
}\left\langle h_{l,m} , h_{i,j} \right\rangle \text{d}\theta .
\end{eqnarray*}
If $l+m\neq i+j$, one has 
\begin{equation*}
\left\langle e_{l,m} , E\left( \omega \right) e_{i,j} \right\rangle =0\text{.}
\end{equation*}

If $X\in \mathcal{B}\left( \mathbb{T}\times \mathbb{T}\right) $, reasoning
as in \ref{Esempio 2.5.1} we can obtain the explicit form of $E\left(
X\right) $ simply substituing $\omega $ with the characteristic function $%
\chi _{X}$ in the two previous equations.
\newpage
\
\newpage

\chapter{The case of an irreducible representation\label{cap. 2}}

\section{Introduction}

It is well known~(\cite{Isot}, \cite{Holevo82}, \cite{scud}) that, given an
irreducible square-integrable representation $U$ of a unimodular group $G$
and a trace class, trace one positive operator $T$, the family of operators 
\begin{equation}
E(X)=\int_{X}U(g)TU(g^{-1})\text{d}\mu _{G}(g)\quad X\in \mathcal{B}\left(
G\right)  \label{eq. introduttiva}
\end{equation}
defines a positive operator valued measure based on $G$ and covariant with
respect to $U$ ($\mu _{G}$ is a Haar measure on $G$). In this chapter, we
prove that also the converse holds. More precisely, a POM $E$ based on $G$
is covariant with respect to $U$ iff $E$ is expressible in the form of eq.~(%
\ref{eq. introduttiva}) for some positive trace one operator $T$. We will
extend this result to non-unimodular groups and to POM's based on the
quotient space $G/H$, where $H$ is a compact subgroup. Moreover, we prove
that square-integrability of $U$ is not only a sufficient condition, but it
is also necessary in order that $U$ admits covariant POM's based on $G/H$.
Finally, we extend this result to the case of $U$ being an irreducible
projective unitary representation of $G$ (see Remark \ref{Rem. sulla rappr.
proiettiva}).

The result presented here are a rielaboration of refs.~\cite{Articolo con
Bassano} and \cite{CDT1}.

We start with the case in which $U$ is a unitary representation and $%
H=\left\{ e\right\} $. The more general case in which $U$ is projective and $%
H$ is compact is discussed in \S \ref{Caso H/Z compatto generico}. As usual,
we assume that $G$ is a Hausdorff locally compact second countable
topological group.

We fix a left Haar measure $\mu _{G}$ on the group $G$. We denote by $\Delta 
$ the modular function of $G$.

We recall some basic properties of square integrable representations (see
\cite[Theorem 2 and 3]{df}). Recall the definition of the (left) regular representation 
given
in Remark \ref{Remark sulla rappr. regolare}.

\begin{proposition}
\label{rivaldo}Let $U$ be an irreducible representation of $G$ in the
Hilbert space $\mathcal{H}$. The following facts are equivalent:

\begin{enumerate}
\item  there exists a vector $u\in \mathcal{H}$ such that 
\begin{equation}
0<\int_{G}\left| \left\langle U\left( g\right) u, u\right\rangle \right| ^{2}%
\text{d}\mu _{G}\left( g\right) <\infty \text{;}  \label{galliani}
\end{equation}

\item  $U$ is a subrepresentation of the regular representation $\lambda $
of $G$.
\end{enumerate}

If either of the above conditions is satisfied, there exists a selfadjoint
injective positive operator $C$ with $U$-invariant domain and dense range such that 
\begin{equation}
U\left( g\right) C=\Delta \left( g\right) ^{-\frac{1}{2}}CU\left( g\right)
\qquad \forall g\in G\text{,} \label{cheppalle}
\end{equation}
and an isometry $\Sigma :\mathcal{H}\otimes \mathcal{H}^{\ast
}\longrightarrow L^{2}\left( G,\mu _{G}\right) $ such that

\begin{enumerate}
\item  for all $u\in \mathcal{H}$ and $v\in \mathrm{dom}\,C$ 
\begin{equation*}
\Sigma (u\otimes v^{\ast })(g)=\left\langle U\left( g\right)
Cv , u\right\rangle \qquad g\in G\text{,}
\end{equation*}

\item  for all $g\in G$ 
\begin{equation*}
\Sigma (U(g)\otimes I_{\mathcal{H}^{\ast }})=\lambda (g)\Sigma \text{,}
\end{equation*}

\item  the range of $\Sigma $ is the isotypic\footnote{%
If $U$ and $U^{\prime }$ are representations of $G$ in the Hilbert spaces $%
\mathcal{H}$ and $\mathcal{H}^{\prime }$, and $U$ is irreducible, the 
\textbf{isotypic space} of $U$ in $\mathcal{H}^{\prime }$ is the maximal invariant
subspace $\mathcal{K}\subset \mathcal{H}^{\prime }$ such that $\left.
U^{\prime }\right| _{\mathcal{K}}$ decomposes into the direct sum of copies
of $U$.} space of $U$ in $L^{2}\left( G,\mu _{G}\right) $.
\end{enumerate}
If $G$ is unimodular, the operator $C$ is a multiple of the
identity.
\end{proposition}

If eq.~(\ref{galliani}) is satisfied, $U$ is called \textbf{square-integrable%
}. By eq.~(\ref{cheppalle}), the selfadjoint operator $C$ is
unbounded if $G$ is not unimodular.
The square of $C$ is called \textbf{formal degree} of $U$, and it is
uniquely determined up to a positive factor which depends on the choice of $%
\mu _{G}$.

\section{Characterisation of $E$ in the case $U$ unitary and $H=\left\{
e\right\} $}

We fix an irreducible representation $U$ of $G$ in the Hilbert space $%
\mathcal{H}$. The following theorem characterizes all the POM on $G$
covariant with respect to $U$ in terms of positive trace one operators on $%
\mathcal{H}$ \cite{CDT1}.

\begin{theorem}
\label{Teo. centr.}The irreducible representation $U$ admits a covariant POM
based on $G$ if and only if $U$ is square-integrable.

In this case, let $C$ be the square root of the formal degree of $U$. There
exists a one-to-one correspondence between the set of covariant POMs $E$ on $%
G$ and the set of positive trace one operators $T$ on $\mathcal{H}$. This
correspondence associates to each positive trace one operator $T$ the
covariant POM $E_{T}$ given by 
\begin{equation}
\left\langle u, E_{T}\left( X\right) v \right\rangle =\int_{X}\left\langle
CU\left( g^{-1}\right) u , TCU\left( g^{-1}\right) v \right\rangle \text{d}\mu
_{G}\left( g\right)  \label{La Povm}
\end{equation}
for all $u,v\in \mathrm{dom}\,C$ and $X\in \mathcal{B}\left( G\right) $.
\end{theorem}

\begin{proof}
Let $E$ be a $U$-covariant POM based on $G$. According to the generalized
imprimitivity theorem there exists a representation $\sigma $ of the trivial
subgroup $H=\left\{ e\right\} $ in a Hilbert space $\mathcal{K}$ and an
isometry $W:\mathcal{H}\longrightarrow \mathcal{H}^{\sigma }$ intertwining $%
U $ with $\lambda ^{\sigma }$ such that 
\begin{equation*}
E\left( X\right) =W^{\ast }P^{\sigma }\left( X\right) W
\end{equation*}
for all $X\in \mathcal{B}\left( G\right) $. Clearly, $\sigma $ is the
trivial representation in $\mathcal{K}$. We then have 
\begin{equation*}
\mathcal{H}^{\sigma }=L^{2}\left( G,\mu _{G}\right) \otimes \mathcal{K}\text{%
,\qquad }\lambda ^{\sigma }=\lambda \otimes I_{\mathcal{K}}\text{.}
\end{equation*}
In particular, $U$ is a subrepresentation of $\lambda $, hence it is
square-integrable.

Due to Prop.~\ref{rivaldo}, the operator $W^{\prime }=\left( \Sigma ^{\ast
}\otimes I_{\mathcal{K}}\right) W$ is an isometry from $\mathcal{H}$ to $%
\mathcal{H}\otimes \mathcal{H}^{\ast }\otimes \mathcal{K}$ such that 
\begin{eqnarray*}
W^{\prime }U(g) & = & \left( \Sigma ^{\ast
}\otimes I_{\mathcal{K}}\right) W U(g)  \\
&& = \left( \Sigma ^{\ast
}\otimes I_{\mathcal{K}}\right) \left( \lambda (g) \otimes I_{\mathcal{K}}
\right) W \\
&& = \left( U (g) \otimes I_{\mathcal{H}^{\ast}} \otimes I_{\mathcal{K}} \right)
\left( \Sigma ^{\ast
}\otimes I_{\mathcal{K}}\right) W \\
&& = U(g)\otimes I_{\mathcal{H}^{\ast }\otimes \mathcal{K}} W^{\prime}.
\end{eqnarray*}
Since $U$ is irreducible, by a standard result there is a unit vector $B\in 
\mathcal{H}^{\ast }\otimes \mathcal{K}$ such that 
\begin{equation*}
W^{\prime }u=u\otimes B\qquad \forall u\in \mathcal{H}\text{.}
\end{equation*}
(see for example \cite[p.~342, Proposition 14]{Gaal}). Let $(e_{i})_{i\geq 1}$ be an orthonormal
basis of $\mathcal{H}$ such that $e_{i}\in \mathop{\rm{dom}}\nolimits C$,
then 
\begin{equation*}
B=\sum e_{i}^{\ast }\otimes k_{i},
\end{equation*}
where $k_{i}\in \mathcal{K}$ and $\sum_{i}\left\| k_{i}\right\| _{\mathcal{K}%
}^{2}=1$.

If $u\in \mathrm{dom}\,C$, one has that 
\begin{eqnarray*}
Wu & = & (\Sigma \otimes I_{\mathcal{K}})\left( u\otimes B\right) \\
& = & \sum\nolimits_{i}\Sigma (u\otimes e_{i}^{\ast })\otimes k_{i} \\
& = & \sum\nolimits_{i}\left\langle U\left( \cdot \right) Ce_{i} , u\right\rangle
\otimes k_{i} \\
& = & \sum\nolimits_{i}\left\langle e_{i} ,
CU\left( \cdot ^{-1}\right)
u \right\rangle \otimes k_{i} \\
& = & \sum\nolimits_{i}(e_{i}^{\ast }\otimes k_{i})(CU\left( \cdot ^{-1}\right) u)%
\text{,}
\end{eqnarray*}
where the series converges in $\mathcal{H}^{\sigma }$. On the other hand,
for all $g\in G$ the series $\sum_{i}(e_{i}^{\ast }\otimes k_{i})(CU\left(
g^{-1}\right) u)$ converges to $BCU\left( g^{-1}\right) u$. Here and in
the following we
identify $\mathcal{H}^{\ast }\otimes \mathcal{K}$ with the space of
Hilbert-Schmidt operators mapping $\mathcal{H}$ into $\mathcal{K}$.
By uniqueness of the limit 
\begin{equation*}
(Wu)(g)=BCU\left( g^{-1}\right) u\qquad g\in G.
\end{equation*}

If $u,v\in \mathrm{dom}\,C$, the corresponding covariant POM is given by 
\begin{eqnarray*}
\left\langle u , E\left( X\right) v \right\rangle _{\mathcal{H}} & = & \left\langle
Wu , P^{\sigma }\left( X\right) Wv \right\rangle _{\mathcal{H}^{\sigma }} \\
& = & \int_{G}\chi _{X}\left( g\right) \left\langle BCU\left( g^{-1}\right) u ,
BCU\left( g^{-1}\right)
v \right\rangle _{\mathcal{K}}\text{d}\mu
_{G}\left( g\right) \\
& = & \int_{X}\left\langle CU\left( g^{-1}\right)
u , TCU\left( g^{-1}\right) v \right\rangle _{\mathcal{H}}\text{d}\mu _{G}\left( g\right) 
\text{,}
\end{eqnarray*}
where 
\begin{equation*}
T:=B^{\ast }B
\end{equation*}
is a positive trace class trace one operator on $\mathcal{H}$.

Conversely, assume that $U$ is square-integrable and let $T$ be a positive
trace class trace one operator on $\mathcal{H}$. Then 
\begin{equation*}
B:=\sqrt{T}
\end{equation*}
is a (positive) operator belonging to $\mathcal{H}^{\ast }\otimes \mathcal{H}
$ such that $B^{\ast }B=T$ and $\left\| B\right\| _{\mathcal{H}^{\ast
}\otimes \mathcal{H}}=1$. The operator $W$ defined by 
\begin{equation*}
Wv:=\left( \Sigma \otimes I_{\mathcal{H}}\right) \left( v\otimes B\right)
\qquad \forall v\in \mathcal{H}
\end{equation*}
is an isometry intertwining $U$ with the induced representation $\lambda
^{\sigma }$, where $\sigma $ is the trivial representation of $\left\{
e\right\} $ in $\mathcal{H}$. Define $E_{T}$ by 
\begin{equation*}
E_{T}\left( X\right) =W^{\ast }P^{\sigma }\left( X\right) W\qquad X\in 
\mathcal{B}(G)\text{.}
\end{equation*}
With the same computation as above, one has that 
\begin{equation*}
\left\langle v , E_{T}\left( X\right) u \right\rangle =\int_{X}\left\langle
CU\left( g^{-1}\right) u , TCU\left( g^{-1}\right) v \right\rangle \text{d}\mu
_{G}\left( g\right)
\end{equation*}
for all $u,v\in \mathrm{dom}\,C$.

Finally, we show that the correspondence $T\longmapsto E_{T}$ is injective.
Let $T_{1}$ and $T_{2}$ be positive trace one operators on $\mathcal{H}$,
with $E_{T_{1}}=E_{T_{2}}$. Set $T=T_{1}-T_{2}$. Since $U$ is strongly
continuous, for all $u,v\in \mathrm{dom}\,C$ the map {\setlength%
\arraycolsep{0pt} 
\begin{eqnarray*}
&& G\ni g\longmapsto \left\langle CU\left(
g^{-1}\right) u , TCU\left( g^{-1}\right) v \right\rangle \\
&& \quad \quad \quad \quad \quad =\Delta (g)^{-1}\left\langle 
U\left( g^{-1}\right) Cu ,
TU\left(g^{-1}\right) Cv \right\rangle \in \mathbb{C}
\end{eqnarray*}
}is continuous. Since 
\begin{equation*}
\int_{X}\left\langle CU\left( g^{-1}\right)
u , TCU\left( g^{-1}\right) v \right\rangle \text{d}\mu _{G}\left( g\right) =\left\langle 
\left[
u, E_{T_{1}}\left( X\right) -E_{T_{2}}\left( X\right) \right] v \right\rangle
=0
\end{equation*}
for all $X\in \mathcal{B}\left( G\right) $, we have 
\begin{equation*}
\left\langle CU\left( g^{-1}\right) u , TCU\left( g^{-1}\right) v \right\rangle
=0\qquad \forall g\in G\text{.}
\end{equation*}
In particular, 
\begin{equation*}
\left\langle Cu , TCv \right\rangle =0\text{,}
\end{equation*}
so that, since $C$ has dense range, $T=0$.
\end{proof}

\begin{remark}
If $G$ is unimodular, then $C=\lambda I$, with $\lambda >0$, and one can
normalize $\mu _{G}$ so that $\lambda =1$. Hence, 
\begin{equation*}
E_{T}\left( X\right) =\int_{X}U\left( g\right) TU\left( g^{-1}\right) \text{d%
}\mu _{G}\left( g\right) \qquad \forall X\in \mathcal{B}\left( G\right) 
\text{,}
\end{equation*}
the integral being understood in the weak sense.
\end{remark}

\section{An example: the $ax+b$ group}

The $ax+b$ group is the semidirect product $G=\mathbb{R}\times ^{\prime }%
\mathbb{R}_{+}$, where we regard $\mathbb{R}$ as additive group and $\mathbb{%
R}_{+}$ as multiplicative group. The composition law is 
\begin{equation*}
\left( b,a\right) \left( b^{\prime },a^{\prime }\right) =\left( b+ab^{\prime
},aa^{\prime }\right) \text{.}
\end{equation*}
The group $G$ is nonunimodular with left Haar measure 
\begin{equation*}
\text{d}\mu _{G}\left( b,a\right) =a^{-2}\text{d}b\text{d}a
\end{equation*}
and modular function 
\begin{equation*}
\Delta \left( b,a\right) =\frac{1}{a}\text{.}
\end{equation*}

Let $\mathcal{H}=L^{2}\left( \left( 0,+\infty \right) ,\text{d}x\right) $
and $(U,\mathcal{H})$ be the representation of $G$ in $\mathcal{H}$ given by 
\begin{equation*}
\left[ U^{+}\left( b,a\right) f\right] \left( x\right) =a^{\frac{1}{2}%
}e^{2\pi ibx}f\left( ax\right) \text{.}
\end{equation*}
It is known~(\cite{Foll}, \cite{Ali}) that $U^{+}$ is square-integrable, and
the square root of its formal degree is 
\begin{equation*}
\left( Cf\right) \left( x\right) =\Delta \left( 0,x\right) ^{\frac{1}{2}%
}f\left( x\right) =x^{-\frac{1}{2}}f\left( x\right) \qquad x\in \left(
0,+\infty \right)
\end{equation*}
acting on its natural domain.

By means of Theorem \ref{Teo. centr.} every POM based on $G$ and covariant
with respect to $U^{+}$ is described by a positive trace one operator $T$
according to eq.~(\ref{La Povm}). Explicitly, let $\left( e_{i}\right) _{i\geq
1}$ be an orthonormal basis of $\mathcal{H}$ such that $Te_{i} = \lambda_{i} e_{i}$
, $\lambda_{i} \geq 0$, for all $i$. If $u\in L^{2}\left( \left( 0,+\infty
\right) ,\text{d}x\right) $ is such that $x^{-\frac{1}{2}}u\in L^{2}\left(
\left( 0,+\infty \right) ,\text{d}x\right) $, the $U^{+}$-covariant POM
corresponding to $T$ is given by 
\begin{eqnarray*}
\left\langle  u,E_{T}\left( X\right) u\right\rangle &=&\int_{X}\left\langle
CU^{+}\left( g^{-1}\right) u,TCU^{+}\left( g^{-1}\right) u\right\rangle 
\text{d}\mu _{G}\left( g\right) \\
&=&\int_{X}\sum\nolimits_{i}\lambda _{i}\left| \left\langle e_{i} ,
CU^{+}\left( g^{-1}\right) u\right\rangle \right| ^{2}\text{d}\mu _{G}\left(
g\right) \\
&=&\sum\nolimits_{i}\lambda _{i}\int_{X}\left| \int_{\mathbb{R}_{+}}x^{-%
\frac{1}{2}}a^{-\frac{1}{2}}e^{-\frac{2\pi ibx}{a}}u\left( \frac{x}{a}%
\right) \overline{e_{i}\left( x\right) }\text{d}x\right| ^{2}a^{-2}\text{d}b%
\text{d}a\text{.}
\end{eqnarray*}

\section{\label{Caso H/Z compatto generico}Characterisation of $E$ for
projective representations in the case $H$ is compact}

We now suppose that $H$ is a compact subgroup of $G$ and $U$ is an
irreducible projective representation of $G$ with multiplier $m$.

We recall the standard construction that allows to extend $U$ to an
irreducible unitary representation $\widetilde{U}$ of the central extension $%
G_{m}$ of $G$ associated with the multiplier $m$. For more details about
multipliers and central extensions we refer to \cite{Var}. Let $\mathbb{T}%
=\left\{ z\in \mathbb{C}:\left| z\right| =1\right\} $ be the multiplicative
group of the torus. The group $G_{m}$ is the set $G\times \mathbb{T}$ with
the composition law 
\begin{equation*}
\left( g_{1},z_{1}\right) \left( g_{2},z_{2}\right) =\left(
g_{1}g_{2},z_{1}z_{2}m\left( g_{1},g_{2}\right) \right) \text{.}
\end{equation*}
The group $G_{m}$ can always be endowed with a locally compact second
countable Hausdorff topology in which it becomes a topological group. With
this topology, $T:=\left\{ e\right\} \times \mathbb{T}$ is a closed subgroup
in $G_{m}$, and $G_{m}/T$ is canonically isomorphic to $G$. Moreover, $%
G\times \left\{ 1\right\} $ is a Borel subset of $G_{m}$. The representation 
$\widetilde{U}$ is defined by 
\begin{equation*}
\widetilde{U}\left( g,z\right) :=z^{-1}U\left( g\right) \text{\quad }%
\forall z\in \mathbb{T},\,g\in G\text{.}
\end{equation*}
It is a strongly continuous unitary representation of $G_{m}$, with $%
\widetilde{U}\left( g,1\right) =U\left( g\right) $.

Let $\mu _{\mathbb{T}}$ be the Haar measure of $\mathbb{T}$ with $\mu _{%
\mathbb{T}}\left( \mathbb{T}\right) =1$. The measure $\mu _{G_{m}}$ of $%
G_{m} $, given by 
\begin{equation*}
\int_{G_{m}}\varphi \left( g,z\right) \text{d}\mu _{G_{m}}\left( g,z\right)
=\int_{G}\text{d}\mu _{G}\left( g\right) \int_{\mathbb{T}}\varphi \left(
g,z\right) \text{d}\mu _{\mathbb{T}}\left( z\right)
\end{equation*}
for all $\varphi \in C_{c}\left( G_{m}\right) $, is clearly a left Haar
measure of $G_{m}$. We have 
\begin{eqnarray*}
\int_{G_{m}}\left| \left\langle \widetilde{U}\left( g,z\right)
v, u\right\rangle \right| ^{2}\text{d}\mu _{G_{m}}\left( g,z\right) &=&\int_{G}%
\text{d}\mu _{G}\left( g\right) \left| \left\langle U\left( g\right)
v, u\right\rangle \right| ^{2}\int_{\mathbb{T}}\text{d}\mu _{\mathbb{T}}\left(
z\right) \\
&=&\int_{G}\left| \left\langle U\left( g\right) v, u\right\rangle \right| ^{2}%
\text{d}\mu _{G}\left( g \right) \text{.}
\end{eqnarray*}
It follows that $\widetilde{U}$ is square integrable if and only if there is
some vector $u\in \mathcal{H}$ such that 
\begin{equation}
\int_{G}\left| \left\langle U\left( g\right) u,u\right\rangle \right| ^{2}%
\text{d}\mu _{G}\left( g\right) <\infty \text{.}
\label{Square-integrabilita' proiettiva}
\end{equation}
If eq.~(\ref{Square-integrabilita' proiettiva}) is satisfied, we say that the
projective unitary representation $U$ is \textbf{square-integrable}.

Moreover, it is easily checked that a POM $E$ on the quotient $G/H=\left(
G_{m}/T\right) /\left( H_{m}/T\right) =G_{m}/H_{m}$ is covariant with
respect to $U$ if and only if it is covariant with respect to $\widetilde{U}$%
. We note that, since $H$ is compact, the subgroup $H_{m}$ is compact in $%
G_{m}$.

The above discussion is summarised in the following lemma.

\begin{lemma}
\label{Lemma su pi iff pi tilde}

\begin{enumerate}
\item  A POM based on $G/H=G_{m}/H_{m}$ is covariant with respect to the
projective unitary representation $U$ of $G$ if and only if it is covariant
with respect to the unitary representation $\widetilde{U}$ of $G_{m}$.

\item  The representation $\widetilde{U}$ is square integrable if and only
if $U$ is square integrable.
\end{enumerate}
\end{lemma}

The next theorem is the central result of this chapter. It is a
re-elaboration of \cite[Corollary 5]{Articolo con Bassano}. See also \cite[Theorem II.3.2]{Amann}.

\begin{theorem}
\label{Teo. centr.1}Suppose $H$ is a compact subgroup of $G$. Assume that $U$
is an irreducible projective unitary representation of $G$. Then $U$ admits
a covariant POM based on $G/H$ if and only if $U$ is square-integrable.

In this case, let $C$ be the square root of the formal degree of $\widetilde{%
U}$. There exists a one-to-one correspondence between covariant POMs $E$ on $%
G/H$ onto the set of positive trace one operators $T$ on $\mathcal{H}$ such
that 
\begin{equation}
TU\left( h\right) =U\left( h\right) T\quad \forall h\in H\text{.}
\label{commuta}
\end{equation}
This correspondence associates to each positive trace one operator $T$
satisfying eq.~$(\ref{commuta})$ the covariant POM $E_{T}$ given by 
\begin{equation}
\left\langle u, E_{T}\left( X\right) v \right\rangle =\int_{X}\left\langle
CU\left( g^{-1}\right) u, TCU\left( g^{-1}\right) v \right\rangle \text{d}\mu
_{G/H}\left( \dot{g}\right)  \label{La Povm1}
\end{equation}
for all $u,v\in \mathrm{dom}\,C$ and $X\in \mathcal{B}\left( G/H\right) $,
where $\mu _{G/H}$ is an invariant measure on $G/H$.
\end{theorem}

\begin{proof}
By Lemma \ref{Lemma su pi iff pi tilde} and compactness of $H_{m}$, possibly
switching from $G$ to $G_{m}$ and from $U$ to $\widetilde{U}$, we can assume
that $U$ itself is a unitary representation of $G$ (note that eqs.~(\ref
{commuta}) and (\ref{La Povm1}) are unaffected by this change). Let $\mu
_{H} $ be the invariant measure on $H$ with $\mu _{H}\left( H\right) =1$.
Due to the compactness of $H$, there exists a left $G$-invariant measure $%
\mu _{G/H} $ on $G/H$ such that the following measure decomposition holds 
\begin{equation}
\int_{G}f\left( g\right) \text{d}\mu _{G}\left( g\right) =\int_{G/H}\text{d}%
\mu _{G/H}\left( \dot{g}\right) \int_{H}f\left( gh\right) \text{d}\mu
_{H}\left( h\right) \text{.}  \label{Mackey-Bruhat}
\end{equation}
for all $f\in L^{1}\left( G,\mu _{G}\right) $ (see \cite{Foll}, \cite{Gaal}, 
\cite{Dieu2}).

Assume that $U$ is square-integrable and let $T$ be as in the statement of
the theorem. By means of eq.~(\ref{La Povm}) $T$ defines a POM $%
\widetilde{E}_{T}$ based on $G$ and covariant with respect to $U$. For all $%
X\in {\mathcal{B}}(G/H)$ let 
\begin{equation*}
E_{T}(X)=\widetilde{E}_{T}(\pi ^{-1}(X))\text{.}
\end{equation*}
Clearly, $E_{T}$ is a POM on $G/H$ covariant with respect to $U$. Moreover,
denoting with $\chi _{X}$ the characteristic function of $X$, if $u,v\in 
\mathrm{dom}\,C$, 
\begin{eqnarray*}
\left\langle u, E_{T}\left( X\right) v\right\rangle &=&\int_{G}\chi
_{X}\left( \pi \left( g\right) \right) \left\langle CU\left( g^{-1}\right) u ,
TCU\left( g^{-1}\right)
v \right\rangle \text{d}\mu _{G}\left( g\right) \\
&&\text{(by eq.~(\ref{Mackey-Bruhat}))} \\
&=&\int_{G/H}\text{d}\mu _{G/H}\left( \dot{g}\right) \int_{H}\chi _{X}\left(
\pi \left( gh\right) \right) \left\langle CU\left( gh\right) ^{-1}u ,
TCU\left( gh\right)
^{-1}v \right\rangle \text{d}\mu _{H}\left( h\right)
\\
&& \text{(by eq.~(\ref{commuta}) and since }\left. \Delta \right| _{H} =1%
\text{)} \\
&=&\int_{G/H}\chi _{X}\left( \pi \left( g\right) \right) \left\langle
CU\left( g\right) ^{-1}u, TCU\left( g\right) ^{-1}v\right\rangle \text{d}\mu
_{G/H}\left( \dot{g}\right)
\end{eqnarray*}
that is, equation~(\ref{La Povm1}) holds.

Conversely, let $E$ be a POM on $G/H$ which is covariant with respect to $U$%
. For all $Y\in {\mathcal{B}}(G)$, let $l_{Y}$ be the function on $G$ given
by 
\begin{equation*}
l_{Y}(g)=\mu _{H}(g^{-1}Y\cap H)=\int_{H}\chi _{Y}(gh)\text{d}\mu _{H}\left(
h\right) .
\end{equation*}
Clearly, $l_{Y}$ is a positive measurable function bounded by $1$ and, since 
$\mu _{H}$ is invariant, for all $h\in H$, $l_{Y}(gh)=l_{Y}(g)$. It follows
that there is a positive measurable bounded function $\ell _{Y}$ on $G/H$
such that $l_{Y}=\ell _{Y}\circ \pi $.

Define the operator $\widetilde{E}\left( Y\right) $ by means of 
\begin{equation*}
\widetilde{E}(Y)=\int_{G/H}\ell _{Y}(\dot{g})\text{d}E(\dot{g}),
\end{equation*}
which is well defined since $\ell _{Y}$ is bounded.

We claim that $Y\longmapsto \widetilde{E}\left( Y\right) $ is a POM on $G$
covariant with respect to $U$. Clearly, since $\ell _{Y}$ is positive, $%
\widetilde{E}\left( Y\right) $ is a positive operator. Recalling that $\ell
_{G}=1$, one has $\widetilde{E}\left( G\right) =I$. Let now $(Y_{i})_{i\geq
1}$ be a disjoint sequence of ${\mathcal{B}}(G)$ and $Y=\cup _{i}Y_{i}$.
Given $g\in G$, since $(g^{-1}Y_{i}\cap H)_{i\geq 1}$ is a disjoint sequence
of ${\mathcal{B}}(H)$ and $g^{-1}Y\cap H=\cup _{i}(g^{-1}Y_{i}\cap H)$, then 
$\ell _{Y}=\sum_{i}\ell _{Y_{i}}$, where the series converges pointwise. Let 
$u\in {\mathcal{H}}$, by monotone convergence theorem, one has that 
\begin{equation*}
\langle u,\widetilde{E}\left( Y\right) u\rangle =\sum_{i}\langle u,%
\widetilde{E}\left( Y_{i}\right) u\rangle \text{.}
\end{equation*}
Finally, let $g_{1}\in G$, then 
\begin{eqnarray*}
\widetilde{E}(g_{1}Y) &=&\int_{G/H}\mu _{H}(g^{-1}g_{1}Y\cap H)\text{d}E(%
\dot{g}) \\
&& (\dot{g} \longrightarrow g_{1}\dot{g}) \\
&=&\int_{G/H}\mu _{H}(g^{-1}Y\cap H)U\left( g_{1}\right) \text{d}E(\dot{g}%
)U\left( g_{1}\right) ^{\ast } \\
&=&U\left( g_{1}\right) \widetilde{E}(Y)U\left( g_{1}\right) ^{\ast }\text{,}
\end{eqnarray*}
where we used the fact that $E$ is covariant.

By means of Theorem \ref{Teo. centr.}, $U$ is square-integrable and there is
a positive trace one operator $T$ such that, for $u,v\in \mathrm{dom}\,C$, 
\begin{equation}
\left\langle u , \widetilde{E}\left( Y\right) v \right\rangle
=\int_{Y}\left\langle CU\left( g^{-1}\right)
u, TCU\left( g^{-1}\right) v \right\rangle \text{d}\mu \left( g\right) \text{.}  
\label{quasi}
\end{equation}

We now show that $T$ satisfies equation~(\ref{commuta}). First of all we
claim that, given $h\in H$ and $Y\in {\mathcal{B}}(G)$, 
\begin{equation}
\widetilde{E}\left( Yh\right) =\widetilde{E}\left( Y\right) \text{.}
\label{comm_H}
\end{equation}
Indeed, since $H$ is compact, $\mu _{H}$ is both left and right invariant,
so that 
\begin{equation*}
\mu _{H}(g^{-1}Yh\cap H)=\mu _{H}((g^{-1}Y\cap H)h)=\mu _{H}(g^{-1}Y\cap H)
\end{equation*}
and, hence, $\ell _{Y}=\ell _{Yh}$. By definition of $\widetilde{E}\left(
Y\right) $, equation (\ref{comm_H}) easily follows. Fixed $h\in H$, by means
of equation~(\ref{comm_H}) and equation~(\ref{quasi}) one has that
{\setlength\arraycolsep{0pt} 
\begin{eqnarray*}
&& \int_{Y}\left\langle CU\left( g^{-1}\right)
u , TCU\left( g^{-1}\right) v \right\rangle \text{d}\mu \left( g\right) \\
&& \quad \quad = \int_{Yh}\left\langle
CU\left( g^{-1}\right) u , TCU\left( g^{-1}\right) v \right\rangle \text{d}\mu
\left( g\right) \\
&& \quad \quad \quad (g \longrightarrow gh) \\
&& \quad \quad =\int_{Y}\left\langle CU\left( gh\right)
^{-1}u , TCU\left( gh\right) ^{-1}v \right\rangle \text{d}\mu \left( g\right) \text{,}
\end{eqnarray*}
}where we used the fact that $\left. \Delta \right| _{H}=1$. Since the
equality holds for all $Y\in {\mathcal{B}}(G)$, then, for a.a.$~g\in G$, 
\begin{eqnarray*}
\left\langle CU\left( g^{-1}\right) u ,
TCU\left( g^{-1}\right) v \right\rangle
&=&\left\langle CU\left( gh\right)
^{-1}u , TCU\left( gh\right) ^{-1}v \right\rangle \\
&=&\left\langle CU\left( g\right) ^{-1}u ,
U\left( h\right) TU\left( h\right) ^{-1}CU\left( g\right)
^{-1}v \right\rangle \text{.}
\end{eqnarray*}
Since both sides are continuous functions of $g$, the equality holds
everywhere and equation~(\ref{commuta}) follows by density of $\mathrm{ran}%
\,C$.

Let now $X\in {\mathcal{B}}(G/H)$. Since 
\begin{equation*}
g^{-1}\pi ^{-1}(X)\cap H=\left\{ 
\begin{array}{cc}
H & \text{if}\ gH\in X \\ 
\emptyset & \text{if}\ gH\not\in X
\end{array}
\right. \text{,}
\end{equation*}
then $\ell _{\pi ^{-1}(X)}=\chi _{X}$ and $\widetilde{E}\left( \pi
^{-1}(X)\right) =E(X)$. Reasoning as in the first part of the proof one has
that $E=E_{T}$.

The injectivity of the map $T\longmapsto E_{T}$ easily follows from the
injectivity of the map $T\longmapsto \widetilde{E}_{T}$ from the set of
positive trace one operators to the set of $U$-covariant POM based on $G$.
\end{proof}

\begin{remark}
Note that by eq.~$(\ref{Mackey-Bruhat})$\ the invariant measure $\mu _{G/H}$
in eq.~$(\ref{La Povm1})$ depends only on the normalisation of the Haar
measure $\mu _{G}$, hence on the choice of the operator $C$, which is
uniquely determined up to a positive factor.
\end{remark}

\begin{remark}
If $G$ is unimodular and $U$ is a projective square integrable
representation, it is known since $\cite{Holevo82}$ that every $U$-covariant
POM is given by eq.~$(\ref{La Povm1})$ (with $C=I$). But we actually have
more: if $U$ is not square integrable, then $U$ does not admit any covariant
POM.
\end{remark}

\begin{remark}
Scutaru shows in ref.~$\cite{scud}$ that there exists a one-to-one
correspondence between covariant POM's $E$ based on $G/H$ with the property 
\begin{equation}
\mathrm{tr}\,E\left( K\right) <+\infty  \label{inzaghi}
\end{equation}
for all compact sets $K\subset G/H$ and positive trace one operators on $%
\mathcal{H}$. Theorem $\ref{Teo. centr.1}$ shows that if $G$ is unimodular 
\emph{every} covariant POM $E$ based on $G/H$ shares property $(\ref{inzaghi}%
)$.
\end{remark}

\begin{remark}
Suppose that in Theorem $\ref{Teo. centr.1}$ $G$ is unimodular and $U$ is a
unitary representation. Then by eq.~$(\ref{La Povm1})$ 
\begin{equation*}
E_{T}\left( X\right) =\int_{X}U\left( g\right) TU\left( g\right) ^{-1}\text{d%
}\mu _{G/H}\left( \dot{g}\right)
\end{equation*}
for all $X\in \mathcal{B}\left( G/H\right) $ (the integral being understood
in the weak sense). In particular, each $U$-covariant POM based on $G/H$
admits the representation of eq.~$(\ref{(E)})$ in \S $\ref{Sez. sugli stati
coerenti}$, with operator valued density
$E\left( \dot{g}\right) =U\left( g\right) TU\left( g\right)
^{-1}$ and $\nu =\mu _{G/H}$.

Let $\sigma $ be a representation of $H$ in a Hilbert space $\mathcal{K}$
and $W:\mathcal{H}\longrightarrow \mathcal{H}^{\sigma }$ be an operator
intertwining $U$ with the representation induced from $\sigma $. By
irreducibility of $U$, $W$ is a multiple of an isometry. Then, by Theorem $%
\ref{Teo. sugli stati coerenti}$, there exists an operator $A:\mathcal{H}%
\longrightarrow \mathcal{K}$ such that $AU\left( h\right) =\sigma \left(
h\right) A$ for all $h\in H$, and 
\begin{equation*}
\left( Wv\right) \left( g\right) =AU\left( g\right) ^{-1}v\quad \forall v\in 
\mathcal{H}\text{.}
\end{equation*}
Since 
\begin{equation*}
E\left( X\right) :=W^{\ast }P^{\sigma }\left( X\right) W=\int_{X}U\left(
g\right) A^{\ast }AU\left( g\right) ^{-1}\text{d}\mu _{G/H}\left( \dot{g}%
\right)
\end{equation*}
is a multiple of a covariant POM, by Theorem $\ref{Teo. centr.1}$ $A^{\ast
}A $ is trace class, i.e.~$A$ is a Hilbert-Schmidt operator.
\end{remark}

\begin{remark}
Suppose that $H$ is a closed subgroup of $G$ such that

\begin{enumerate}
\item  $H$ contains a closed subgroup $Z$ which is central in $G$;

\item  $H/Z$ is compact.
\end{enumerate}

Let $U$ be an irreducible representation of $G$ in the Hilbert space $%
\mathcal{H}$ and $\gamma $ be the character of $Z$ such that $U\left(
z\right) =\gamma \left( z\right) I_{\mathcal{H}}$ for all $z\in Z$. Fix a
Borel section $s:G/Z\longrightarrow G$, and define 
\begin{equation*}
\widehat{U}\left( \dot{g}\right) :=U\left( s\left( \dot{g}\right) \right)
\quad \forall \dot{g}\in G/Z\text{.}
\end{equation*}
It is easily checked that $\widehat{U}$ is a projective unitary
representation of the quotient group $G/Z$ with multiplier 
\begin{equation*}
m\left( \dot{g}_{1},\dot{g}_{2}\right) =\gamma \left( s\left( \dot{g}_{1}%
\dot{g}_{2}\right) s\left( \dot{g}_{2}\right) ^{-1}s\left( \dot{g}%
_{1}\right) ^{-1}\right) \text{.}
\end{equation*}
Moreover, a POM based on $G/H=\left( G/Z\right) /\left( H/Z\right) $ is $U$%
-covariant iff it is $\widehat{U}$-covariant. By Theorem $\ref{Teo. centr.1}$%
, then $U$ admits covariant POM's iff there is some vector $u\in \mathcal{H}$
such that 
\begin{equation}
\int_{G/Z}\left| \left\langle U\left( g\right) u,u\right\rangle \right| ^{2}%
\text{d}\mu _{G/Z}\left( \dot{g}\right) <\infty \text{.}
\label{Square-integrability modulo Z}
\end{equation}
In this case, it is easily checked that the $U$-covariant POM's are given
again by formula $(\ref{La Povm1})$. An irreducible unitary representation $%
U $ of $G$ satisfying eq.~$(\ref{Square-integrability modulo Z})$ is called 
\textbf{square-integrable modulo }$Z$.
\end{remark}

\section{Two examples\label{subsec. Two examples}}

\subsection{\label{Esempio spazio fasi in 3 dim.}The isochronous Galilei
group}

The following example is taken from \cite{Articolo con Bassano}. Consider a
free nonrelativistic spin-$0$ particle with mass $m$ in the Euclidean space $
\mathbb{R}^{3}$. Its associated Hilbert space is $\mathcal{H}=L^{2}\left( 
\mathbb{R}^{3},\text{d}\vec{x}\right) $. The symmetry group of the system is
the isochronous Galilei group. We recall that this group is the topological
space $G=\mathbb{R}^{3}\times \mathbb{V}^{3}\times SO\left( 3\right) $,
where $\mathbb{R}^{3}$ is the $3$-dimensional vector group of space
translations, $\mathbb{V}^{3}\simeq \mathbb{R}^{3}$ is the $3$-dimensional
vector group of velocity boosts and $SO(3)$ is the group of rotations
(connected with the identity). The composition law of $G$ is 
\begin{equation*}
\left( \vec{a},\vec{v},R\right) \left( \vec{a}^{\prime },\vec{v}^{\prime
},R^{\prime }\right) =\left( \vec{a}+R\vec{a}^{\prime },\vec{v}+R\vec{v}%
^{\prime },RR^{\prime }\right) \text{.}
\end{equation*}
The group $G$ acts in $\mathcal{H}$ through the irreducible projective
unitary representation $U$ defined as follows 
\begin{equation}
\left[ U\left( \vec{a},\vec{v},R\right) \phi \right] \left( \vec{x}\right)
=e^{im\vec{v}\cdot \left( \vec{x}-\vec{a}\right) }\phi \left( R^{-1}\left( 
\vec{x}-\vec{a}\right) \right) \text{.} \label{rappr. U di Galilei in 3 dim.}
\end{equation}

The phase space of the system is $\Omega =\mathbb{R}^{3}\times \mathbb{P}%
^{3} $. The action of an element $g=\left( \vec{a},\vec{v},R\right) \in G$
on a point $x=\left( \vec{q},\vec{p}\right) \in \Omega $ is given by 
\begin{equation*}
g\left[ x\right] =\left( \vec{a}+R\vec{q},m\vec{v}+R\vec{p}\right) \text{.}
\end{equation*}
The stability subgroup at the point $\left( \vec{0},\vec{0}\right) $ is the
compact subgroup $H=SO\left( 3\right) $. In particular, $\Omega $ is
isomorphic to $G/H$ by means of 
\begin{equation}
\left( \vec{q},\vec{p}\right) \longmapsto \pi \left( \vec{q},\frac{\vec{p}}{m%
},I\right) \text{.}  \label{identificazione di Galilei}
\end{equation}

A \textbf{covariant phase space observable} is a POM $E$ based on $\Omega $
and covariant with respect to $U$. To apply Theorem \ref{Teo. centr.1} and
classify such POM's we need to check the square-integrability of $U$.
Denoting by d$R$ the normalised Haar measure of $SO\left( 3\right) $, we fix
in $G$ the Haar measure 
\begin{equation*}
\text{d}\mu _{G}\left( \vec{a},\vec{v},R\right) =\frac{m}{\left( 2\pi
\right) ^{3}}\text{d}\vec{a}\text{d}\vec{v}\text{d}R\text{.}
\end{equation*}
If $\psi \in L^{2}\left( \mathbb{R}^{3},\text{d}\vec{x}\right) $, we have 
\begin{eqnarray*}
&&\int_{G}\left| \left\langle U\left( \vec{a},\vec{v},R\right) \psi , \psi
\right\rangle \right| ^{2}\text{d}\mu _{G}\left( \vec{a},\vec{v},R\right) =
\\
&&\qquad =\int_{\mathbb{R}^{3}\times \mathbb{V}^{3}\times SO\left( 3\right)
}\left| \int_{\mathbb{R}^{3}}\psi \left( \vec{x}\right) e^{-im\vec{v}\cdot
\left( \vec{x}-\vec{a}\right) }\overline{\psi \left( R^{-1}\left( \vec{x}-%
\vec{a}\right) \right) }\text{d}\vec{x}\right| ^{2}\frac{m\text{d}\vec{a}%
\text{d}\vec{v}\text{d}R}{\left( 2\pi \right) ^{3}} \\
&&\qquad =\int_{\mathbb{R}^{3}\times SO\left( 3\right) }\left[ {\int_{%
\mathbb{V}^{3}}}\left| \mathcal{F}\left( \psi \left( \cdot \right) \overline{%
\psi \left( R^{-1}\left( \cdot -\vec{a}\right) \right) }\right) \left( m\vec{%
v}\right) \right| ^{2}m\text{d}\vec{v}\right] \text{d}\vec{a}\text{d}R \\
&&\qquad =\int_{\mathbb{R}^{3}\times SO\left( 3\right) }\left[ {\int_{%
\mathbb{V}^{3}}}\left| \psi \left( \vec{x}\right) \overline{\psi \left(
R^{-1}\left( \vec{x}-\vec{a}\right) \right) }\right| ^{2}\text{d}\vec{x}%
\right] \text{d}\vec{a}\text{d}R=\left\| \psi \right\| ^{4}\text{.}
\end{eqnarray*}
Thus, $U$ is a square-integrable representation with formal degree $C^{2}=I$%
. Choosing d$\mu _{G/H}\left( \vec{a},\vec{v}\right) =\frac{m}{\left( 2\pi
\right) ^{3}}$d$\vec{a}$d$\vec{v}$ and recalling identification (\ref
{identificazione di Galilei}), every $U$-covariant POM based on $\Omega $
has the form 
\begin{equation}
E_{T}\left( X\right) =\frac{1}{\left( 2\pi \right) ^{3}}\int_{X}U_{\left( 
\vec{q},\frac{\vec{p}}{m},I\right) }TU_{\left( \vec{q},\frac{\vec{p}}{m}%
,I\right) }^{\ast }\text{d}\vec{q}\text{d}\vec{p}  \label{L'altra}
\end{equation}
for all $X\in \mathcal{B}\left( \Omega \right) $, where $T$ is a positive
trace one operator commuting with $\left. U\right| _{SO\left( 3\right) }$.

We now characterize the positive trace one operators $T$ commuting with $%
\left. U\right| _{SO\left( 3\right) }$. We have the factorization 
\begin{equation*}
L^{2}\left( \mathbb{R}^{3},\text{d}\vec{x}\right) =L^{2}\left( S^{2},\text{d}%
\Omega \right) \otimes L^{2}\left( \mathbb{R}_{+},r^{2}\text{d}r\right) 
\text{.}
\end{equation*}
Denoting with $l$ the representation of $SO\left( 3\right) $ acting in $%
L^{2}\left( S^{2},\text{d}\Omega \right) $ by left translations, we have 
\begin{equation*}
\left. U\right| _{SO\left( 3\right) }=l\otimes I\text{.}
\end{equation*}
The representation $\left( l,L^{2}\left( S^{2},\text{d}\Omega \right)
\right) $ decomposes into 
\begin{equation*}
L^{2}\left( S^{2},\text{d}\Omega \right) =\bigoplus_{\ell \geq 0}M_{\ell }%
\text{,}
\end{equation*}
where each irreducible inequivalent subspace $M_{\ell }$ is generated by the
spherical harmonics $\left( Y_{\ell m}\right) _{-\ell \leq m\leq \ell }$. We
have 
\begin{equation*}
L^{2}\left( \mathbb{R}^{3},\text{d}\vec{x}\right) =\left( \bigoplus_{\ell
\geq 0}M_{\ell }\right) \otimes L^{2}\left( \mathbb{R}_{+},r^{2}\text{d}%
r\right) =\bigoplus_{\ell \geq 0}\left( M_{\ell }\otimes L^{2}\left( \mathbb{%
R}_{+},r^{2}\text{d}r\right) \right) \text{.}
\end{equation*}

Let $P_{\ell }:L^{2}\left( S^{2},\text{d}\Omega \right) \longrightarrow
L^{2}\left( S^{2},\text{d}\Omega \right) $ be the orthogonal projection onto
the subspace $M_{\ell }$. If $T$ commutes with $l\otimes I$, one has 
\begin{equation*}
T\left( P_{\ell }\otimes I\right) =\left( P_{\ell }\otimes I\right) T\text{,}
\end{equation*}
where $P_{\ell }\otimes I$ projects onto $M_{\ell }\otimes L^{2}\left( 
\mathbb{R}_{+},r^{2}\text{d}r\right) $. Given Hilbert spaces $\mathcal{H}%
_{1} $ and $\mathcal{H}_{2}$ and an irreducible representation $\pi $ acting
in a Hilbert space $\mathcal{K}$, a standard result asserts that the
operators interwining $\pi \otimes I_{\mathcal{H}_{1}}$ and $\pi \otimes I_{%
\mathcal{H}_{2}}$ are exactly the tensor product $I_{\mathcal{K}}\otimes 
\mathcal{L}\left( \mathcal{H}_{1},\mathcal{H}_{2}\right) $ (see \cite[p.~342, Proposition 14]{Gaal}%
). Since $M_{\ell }$ is irreducible, this implies 
\begin{equation*}
T\left( P_{\ell }\otimes I\right) =P_{\ell }\otimes T_{\ell }
\end{equation*}
with $T_{\ell }\in \mathcal{L}\left( L^{2}\left( \mathbb{R}_{+},r^{2}\text{d}%
r\right) \right) $. We then have 
\begin{equation*}
T=\sum_{\ell }T\left( P_{\ell }\otimes I\right) =\sum_{\ell }P_{\ell
}\otimes T_{\ell }\text{.}
\end{equation*}
Since $T$ is a positive trace one operator,
each $T_{\ell }$ is positive and 
\begin{equation}
1\equiv \sum_{\ell }\dim M_{\ell }\mathrm{tr}\,T_{\ell }=\sum_{\ell }\left(
2\ell +1\right) \mathrm{tr}\,T_{\ell }\text{.}  \label{quella}
\end{equation}
It follows that the operators $T$ associated to the $U$-covariant POM's $E$
by means of equation~(\ref{L'altra}) are all the operators of the form 
\begin{equation*}
T=\sum_{\ell }P_{\ell }\otimes T_{\ell }
\end{equation*}
with $T_{\ell }$ positive trace class operators satisfying equation~(\ref
{quella}).

\subsection{\label{Esempio spazio fasi in 1 dim.}Covariant phase space
observables in one dimension.}

In this subsection, the quantum system $\mathcal{S}$ is
a nonrelativistic
spin-$0$ particle with mass $m$ in the one dimensional space $\mathbb{R}$. The Hilbert 
space of $\mathcal{S}$ is $%
\mathcal{H}=L^{2}\left( \mathbb{R},\text{d}x\right) $ (d$x$ being the
Lebesgue measure of $\mathbb{R}$). The symmetry group of the system is
the isochronous Galilei group in one dimension. This is the additive abelian group 
$G=\mathbb{R}
\times \mathbb{V}$ ($\mathbb{V}\simeq \mathbb{R}$), which acts on $\mathcal{H}$ by
means of the projective unitary representation $U$ given by 
\begin{equation*}
\left[ U\left( a,v\right) \psi \right] \left( x\right) =e^{imv ( x-a ) }
\psi \left( x-q\right) \quad \forall \psi \in \mathcal{H}
\end{equation*}
(compare with eq.~(\ref{rappr. U di Galilei in 3 dim.})). The one dimensional phase space
is $\Omega = \mathbb{R} \times \mathbb{P}$, on which $G$ acts by
\begin{equation*}
\left( a,v \right) \left[ \left( q , p \right) \right]
=\left( q+a,p+mv \right) \text{.}
\end{equation*}
A \textbf{covariant phase space observable} is thus a POM based on $\Omega $
and covariant with respect to $U$.

To simplify our notations in the next chapters, we introduce the group $G^{\prime}
= \mathbb{R} \times \mathbb{P}$ of space translations and momentum boosts, also called
the group of phase space translations (see \cite
{Lahti95}, \cite{Holevo82}). We define the projective unitary representation of 
$G^{\prime}$ on $\mathcal{H}$, given by
\begin{equation}
\left[ W\left( q,p\right) \psi \right] \left( x\right) =e^{ip\left( x-\frac{q%
}{2}\right) }\psi \left( x-q\right) \quad \forall \psi \in \mathcal{H}\text{.%
}  \label{Def. della rappr. W}
\end{equation}
This can be written 
\begin{equation*}
W\left( q,p\right) =e^{i\left( pQ-qP\right) }\text{,}
\end{equation*}
where 
\begin{equation*}
\left[ Q\psi \right] \left( x\right) =x\psi \left( x\right) ,\qquad \left[
P\psi \right] \left( x\right) =-i\frac{\text{d}}{\text{d}t}\psi \left(
x\right)
\end{equation*}
defined on their natural domains are the usual selfadjoint generators of boosts and 
translations. The group $G^{\prime}$ acts on the phase space $\Omega$ by translations: 
\begin{equation*}
\left( q,p\right) \left[ \left( q^{\prime },p^{\prime }\right) \right]
=\left( q+q^{\prime },p+p^{\prime }\right) \text{.}
\end{equation*}
It is easily seen that a POM $E:\mathcal{B}(\Omega) \longrightarrow
\mathcal{L}(\mathcal{H})$ is a covariant phase space observable iff it is covariant with 
respect to the representation $W$ of $G^{\prime}$. In the following two chapters, we will 
work with $G^{\prime}$ and $W$ rather than $G$ and $U$.

We endow $G^{\prime}$ with the Haar measure 
\begin{equation*}
\text{d}\mu _{G^{\prime}}\left( p,q\right) =\frac{1}{2\pi }\text{d}p\text{d}q\text{.}
\end{equation*}
It is known~(\cite{Isot}, \cite{Foll}, \cite{Holevo82}) that the
representation $W$ is square-integrable, with $C=1$. It follows from Theorem 
\ref{Teo. centr.} that \emph{any} $W$-covariant POM $E$ based on $\Omega $
is of the form 
\begin{equation*}
E\left( X\right) =\frac{1}{2\pi }\int_{X}e^{i(pQ-qP)}Te^{-i(pQ-qP)}\text{d}p%
\text{d}q\qquad X\in \mathcal{B}\left( \mathbb{R}\times \mathbb{P}\right)
\end{equation*}
for some positive trace one operator $T$ on $\mathcal{H}$. This result was
known since \cite{Holevo82} and \cite{Art. di Warner sulle POM
Heisenberg-cov.} (although in the book of Holevo it follows from a more
general result, while the proof of Werner is specific for the case of the
Heisenberg group).
\
\newpage
\

\chapter{Covariant position and momentum observables\label%
{Capitolo sulle Covariant position and momentum observables}}

\section{Introduction}

In this chapter, we define the position and momentum observables for a
nonrelativistic quantum particle by means of their covariance property under the
transformations of the isochronous Galilei group.

We will show that these observables are a smeared or `fuzzy' version of the
canonical sharp position and momentum observables defined in eqs.~(\ref
{Posiz. canonica}) and (\ref{Mom. canonico}) below (for a detailed account
about fuzzy observables, we refer to \cite{Lahti95} and \cite{fuzzy}).

Moreover, we will characterise some operational properties of the position
and momentum observables, such as regularity and state distinction power. We
also introduce a variant of the concept of regularity which has a quite
transparent meaning for these observables. We call it $\alpha $\emph{%
-regularity}. $\alpha $-regularity allows one to characterise the \emph{%
limit of resolution} of a position or a momentum observable (for more
details, we refer to \S \ref{Resolution})

Most of our results will be given only in the one dimensional case, since
their extension to the particle in the Euclidean space is quite
straightforward. Moreover, we will work with the group of space translations and momentum 
boosts of \S \ref{Esempio spazio fasi in 1 dim.} (and with its three dimensional analogue) 
rather than with the Galilei group, since this will slightly simplify our notations 
(essentially, this drops the mass factor in our formulas).

Finally, in the next chapter we will conclude our discussion on position and
momentum observables treating the problem of their coexistence and joint
measurability.

Here we stress that, in different approaches, position and momentum observables can be 
defined in different ways, sometimes even without referring to their properties of 
covariance. We shall not enter into details about this. For more information, see 
\cite{Lahti95}, \cite{Werner04} and references therein.  

If not explicitely stated otherwise, the results of this chapter
are all taken from \cite{CHT04}.

We end this section introducing some notations that will be used in the
following. Since we will always be concerned with $\mathbb{R}^{n}$ endowed
with the Lebesgue measure d$x^{n}$, we will use the abbreviated notation $%
L^{p}\left( \mathbb{R}^{n}\right) $ for $L^{p}\left( \mathbb{R}^{n},\text{d}%
x^{n}\right) $. The Fourier transform of any $f\in L^{1}(\mathbb{R}^{n})$ is
denoted by $\hat{f}$. We set also $\hat{f}=\mathcal{F}(f)$ to denote the
Fourier-Plancherel transform of any $f\in L^{2}(\mathbb{R}^{n})$, and
similarly $\hat{\mu}=\mathcal{F}(\mu )$ is the Fourier-Stieltjes transform
of any complex Borel measure $\mu $ on $\mathbb{R}^{n}$.

If $E:\mathcal{A}\left( \Omega \right) \longrightarrow \mathcal{L}\left( 
\mathcal{H}\right) $ is a POM based on the measurable space $\left( \Omega ,%
\mathcal{A}\left( \Omega \right) \right) $, an element $E\left( X\right) \in 
\mathrm{ran}\,E\subset \mathcal{L}\left( \mathcal{H}\right) $ is called an 
\textbf{effect}. The effects $O$ and $I$ are called \textbf{trivial}.

\section{Definition of the observables of position and momentum on $\mathbb{R%
}$\label{Def1}}

Let us consider a nonrelativistic spin-$0$ particle living in the
one-dimensional space $\mathbb{R}$ and fix $\mathcal{H}=L^{2}(\mathbb{R})$.
Let $U$ and $V$ be the one-parameter unitary representations on $\mathcal{H}$
related to the groups of space translations and momentum boosts. They act on 
$\varphi \in \mathcal{H}$ as 
\begin{eqnarray*}
\left[ U(q)\varphi \right] (x) &=&\varphi (x-q)\text{,} \\
\left[ V(p)\varphi \right] (x) &=&e^{ipx}\varphi (x)\text{.}
\end{eqnarray*}
Let $P$ and $Q$ be the selfadjoint operators generating $U$ and $V$, that
is, $U(q)=e^{-iqP}$ and $V(p)=e^{ipQ}$ for every $q,p\in \mathbb{R}$. We
denote by $\Pi _{P}$ and $\Pi _{Q}$ the spectral decompositions of the
operators $P$ and $Q$, respectively. They have the form 
\begin{eqnarray}
&&\left[ \Pi _{Q}(X)\varphi \right] (x)=\chi _{X}(x)\varphi (x)\text{,}
\label{Posiz. canonica} \\
&&\Pi _{P}(X)=\mathcal{F}^{-1}\Pi _{Q}(X)\mathcal{F}\text{.}
\label{Mom. canonico}
\end{eqnarray}
The projection valued measure $\Pi _{Q}:\mathcal{B}(\mathbb{R}%
)\longrightarrow \mathcal{L}\left( \mathcal{H}\right) $ has the property
that, for all $q,p\in \mathbb{R}$ and $X\in \mathcal{B}(\mathbb{R})$, 
\begin{eqnarray}
U(q)\Pi _{Q}(X)U(q)^{\ast } &=&\Pi _{Q}(X+q)\text{,}  \label{covQ} \\
V(p)\Pi _{Q}(X)V(p)^{\ast } &=&\Pi _{Q}(X)\text{.}  \label{invQ}
\end{eqnarray}

More generally, the abelian group $G=\mathbb{R}\times \mathbb{P}$ of the
space translations and momentum boosts acts on the one dimensional space $%
\mathbb{R}$ by 
\begin{equation*}
\left( q,p\right) \left[ x\right] =x+q\quad \forall x\in \mathbb{R}\text{, }%
\left( q,p\right) \in \mathbb{R}\times \mathbb{P}\text{.}
\end{equation*}
On the other hand, its action on the Hilbert space $L^{2}\left( \mathbb{R}%
\right) $ of the quantum particle is given by the projective unitary
representation $W$ defined in eq.~(\ref{Def. della rappr. W}) (see \S \ref
{Esempio spazio fasi in 1 dim.}). So, a position observable $E:\mathcal{B}(%
\mathbb{R})\longrightarrow \mathcal{L}(\mathcal{H})$ must satisfy the
covariance condition 
\begin{equation*}
W\left( q,p\right) E\left( X\right) W\left( q,p\right) ^{\ast }=E\left(
X+q\right) \quad \forall X\in \mathcal{B}(\mathbb{R})\text{.}
\end{equation*}
Since $W\left( q,p\right) =e^{iqp/2}U\left( q\right) V\left( p\right) $, it
is easy to check that the last equation is equivalent to the analogues of
eqs.~(\ref{covQ}) and (\ref{invQ}) with the sharp observable $\Pi _{Q}$
replaced by the POM $E$. We thus take covariance under translations and
invariance under momentum boosts as the defining properties of a general
position observable.

\begin{definition}
An observable $E:\mathcal{B}(\mathbb{R})\longrightarrow \mathcal{L}(\mathcal{%
H})$ is a \textbf{position observable on }$\mathbb{R}$ if, for all $q,p\in \mathbb{R}$
and $X\in \mathcal{B}(\mathbb{R})$, 
\begin{eqnarray}
U(q)E(X)U(q)^{\ast } &=&E(X+q)\text{,}  \label{cov} \\
V(p)E(X)V(p)^{\ast } &=&E(X)\text{.}  \label{inv1}
\end{eqnarray}
We will denote by $\mathcal{POS}_{\mathbb{R}}$ the convex set of all
position observables on $\mathbb{R}$.
\end{definition}

The projection valued position observable $\Pi _{Q}$ is called the \textbf{%
canonical (sharp) position observable}.

In an analogous way we define a momentum observable to be an observable
which is covariant under momentum boosts and invariant under translations.

\begin{definition}
An observable $F:\mathcal{B}(\mathbb{R})\longrightarrow \mathcal{L}(\mathcal{%
H})$ is a \textbf{momentum observable on }$\mathbb{R}$ if, for all $q,p\in \mathbb{R}$
and $X\in \mathcal{B}(\mathbb{R})$, 
\begin{eqnarray}
U(q)F(X)U(q)^{\ast } &=&F(X),  \label{invF} \\
V(p)F(X)V(p)^{\ast } &=&F(X+p).  \label{covF}
\end{eqnarray}
\end{definition}

Since $\mathcal{F}U(q)=V(-q)\mathcal{F}$ and $\mathcal{F}V(p)=U(p)\mathcal{F}
$, the sharp observable $\Pi _{P}=\mathcal{F}^{-1}\Pi _{Q}\mathcal{F}$
satisfies (\ref{invF}) and (\ref{covF}). It is called the \textbf{canonical
(sharp) momentum observable}. Moreover, an observable $E$ is a position observable if and 
only if $%
\mathcal{F}^{-1}E\mathcal{F}$ is a momentum observable. Therefore, in the
following we will restrict ourselves to the study of position observables,
the results of Sections \ref{Structure}, \ref{State} and \ref{Resolution}
being easily converted to the case of momentum observables.

\begin{remark}
We classified in \S \ref{Esempio 2.5.1} the POM's $E:\mathcal{B}\left( 
\mathbb{R}\right) \longrightarrow \mathcal{L}\left( \mathcal{H}\right) $
which fulfill only eq.~(\ref{cov}). In some articles the name `position
observables' is used to denote these observables, without requiring the
additional property of invariance under boosts. In \S \ref{Esempio 2.5.1} we
say that such observables are `translation covariant observables', and we
reserve the name `position observables' only to the observables satisfying
eqs.~(\ref{cov}) and (\ref{inv1}). In \S \ref{Structure} it is shown, in
particular, that every position observable is commutative. However, using
the classification of translation covariant observables given in section \ref
{Esempio 2.5.1}, it is easy to check that there exist noncommutative
localization observables. Thus, the position observables are a strict subset
of the set of the translation covariant observables.
\end{remark}

\section{The structure of position observables\label{Structure}}

Let $\rho :\mathcal{B}(\mathbb{R})\rightarrow \lbrack 0,1]$ be a probability
measure. For any $X\in \mathcal{B}(\mathbb{R})$, the map $q\mapsto \rho
(X-q) $ is bounded and measurable, and hence the equation 
\begin{equation}
E_{\rho }(X)=\int \rho (X-q)\ \text{d}\Pi _{Q}(q)  \label{pos1}
\end{equation}
defines a bounded positive operator. The map 
\begin{equation*}
\mathcal{B}(\mathbb{R})\ni X\mapsto E_{\rho }(X)\in \mathcal{L}(\mathcal{H})
\end{equation*}
is an observable. It is straightforward to verify that the observable $%
E_{\rho }$ satisfies the covariance condition (\ref{cov}) and the invariance
condition (\ref{inv1}), hence it is a position observable on $\mathbb{R}$.
Denote by $\delta _{t}$ the Dirac measure concentrated at $t$. The
observable $E_{\delta _{0}}$ is the canonical position observable $\Pi _{Q}$%
. We may also write 
\begin{equation}
\Pi _{Q}(X)=\int \delta _{0}(X-q)\ \text{d}\Pi _{Q}(q)  \label{pos2}
\end{equation}
and comparing (\ref{pos1}) to (\ref{pos2}) we note that $E_{\rho }$ is
obtained when the sharply concentrated Dirac measure $\delta _{0}$ is
replaced by the probability measure $\rho $. The observable $E_{\rho }$
admits an interpretation as an imprecise, or fuzzy, version of the canonical
position observable $\Pi _{Q}$, unsharpness being characterised by the
probability measure $\rho $ (see \cite{AliDoe}, \cite{AliEmch}, \cite{AliPru}, 
\cite{fuzzy} for further
details).

The following is the central result of this chapter.

\begin{proposition}
\label{struc1} Any position observable $E$ on $\mathbb{R}$ is of the form $%
E=E_{\rho }$ for some probability measure $\rho :\mathcal{B}(\mathbb{R}%
)\rightarrow \lbrack 0,1]$.
\end{proposition}

Since the proof of Proposition \ref{struc1} is quite long and can be given
in a slightly greater generality which will be useful in the following, we
postpone it to the next section.

We denote by $M(\mathbb{R})$ the set of complex measures on $\mathbb{R}$. $%
M_{1}^{+}(\mathbb{R})$ is the convex set of probability measures.
Proposition \ref{struc1} thus estabilishes a map $\rho \longmapsto E_{\rho }$
from $M_{1}^{+}(\mathbb{R})$ onto the convex set $\mathcal{POS}_{\mathbb{R}}$
of the position observables on $\mathbb{R}$. It is immediately checked that
this map is convex. The next proposition (proved in 
\cite{Nuovo articolo con Teiko}) shows that it is an isomorphism of
convex sets.

\begin{proposition}
\label{inj}Let $\rho _{1},\rho _{2}\in M_{1}^{+}(\mathbb{R})$, $\rho
_{1}\neq \rho _{2}$. Then $E_{\rho _{1}}\neq E_{\rho _{2}}$.
\end{proposition}

\begin{proof}
For $\psi \in \mathcal{H}$, we define the real measure $\lambda _{\psi }$ by 
\begin{equation*}
\lambda _{\psi }\left( X\right) =\left\langle \psi ,
\left( E_{\rho _{1}}\left(
X\right) -E_{\rho _{2}}\left( X\right) \right) \psi \right\rangle =\mu
_{\psi }\ast \left( \rho _{1}-\rho _{2}\right) \left( X\right) ,
\end{equation*}
where $\ast $ is the convolution and d$\mu _{\psi }\left( x\right) =\left|
\psi \left( x\right) \right| ^{2}$d$x$. Taking the Fourier-Stieltjes
transform we get 
\begin{equation*}
\hat{\lambda}_{\psi }=\hat{\mu}_{\psi }\cdot \left( \hat{\rho}_{1}-\hat{\rho}%
_{2}\right) ,
\end{equation*}
where $\hat{\lambda}_{\psi }$, $\hat{\mu}_{\psi }$, $\hat{\rho}_{1}$ and $%
\hat{\rho}_{2}$ are continuous functions. By injectivity of the
Fourier-Stieltjes transform we have $\hat{\rho}_{1}\neq \hat{\rho}_{2}$.
Thus, choosing $\psi $ such that $\widehat{\left| \psi \right| ^{2}}(p)\neq 0
$ for every $p\in \mathbb{R}$, we have $\hat{\lambda}_{\psi }\neq 0$. This
means that $\lambda _{\psi }\neq 0$ and hence, $E_{\rho _{1}}\neq E_{\rho
_{2}}$.
\end{proof}

As we will see in Proposition \ref{dilat} below, a position observable $%
E_{\rho }$ is a sharp observable if and only if $\rho =\delta _{x}$ for some 
$x\in \mathbb{R}$, where $\delta _{x}$ is the Dirac measure concentrated at $%
x$. Since the Dirac measures are the extreme elements of the convex set $%
M_{1}^{+}(\mathbb{R})$, from the above discussion the following fact follows
\cite{Nuovo articolo con Teiko}.

\begin{proposition}
The sharp position observables are the extreme elements of the convex set $%
\mathcal{POS}_{\mathbb{R}}$.
\end{proposition}

The following useful property of a position observable is proved in \cite
{HLPPY03}. Here we give a slightly modified proof.

\begin{proposition}
If $E:\mathcal{B}(\mathbb{R})\longrightarrow \mathcal{L}(\mathcal{H})$
satisfies eq.~(\ref{cov}), then $E(X)=O$ if and only if the Borel set $X$
has zero Lebesgue measure.
\end{proposition}

\begin{proof}
This is exactly as in the first part of the proof of Lemma \ref{Lemma 1.3.1}%
. Let $\left( v_{n}\right) _{n\geq 1}$ be a countable dense subset in $%
\mathcal{H}$, and let $\mu $ be the bounded measure on $\mathbb{R}$ given by 
\begin{equation*}
\mu \left( X\right) =\sum\nolimits_{n}2^{-n}\left\| v_{n}\right\|
^{-2}\left\langle v_{n},E\left( X\right) v_{n}\right\rangle \text{.}
\end{equation*}
We have $\mu \left( X\right) =0$ iff $E\left( X\right) =0$. By this fact and
the translational covariance of $E$ we have $\mu \left( X+x\right) =0$ iff $%
\mu \left( X\right) =0$, i.e.~$\mu $ is a quasi invariant measure on $%
\mathbb{R}$. So, $\mu $ is equivalent to the Lebesgue measure, thus proving
our claim.
\end{proof}

We conclude this section with a standard example.

\begin{example}
\label{absc} Let $\rho \in M_{1}^{+}(\mathbb{R})$ be absolutely continuous
with respect to the Lebesgue measure and let $e\in L^{1}(\mathbb{R})$ be the
corresponding Radon-Nikod\'{y}m derivative. Then (\ref{pos1}) can be written
in the form 
\begin{equation}
E_{\rho }(X)=\left( \chi _{X}\ast \tilde{e}\right) \left( Q\right) , \label{convoluzione 
di Pekka}
\end{equation}
where $\tilde{e}(q)=e(-q)$ and $\ast $ denotes the convolution.
\end{example}

\section{Translation covariant and boost invariant observables in dimension $%
n$\label{proof1}}

Let $N=\mathbb{R}^{n+1}$ and $H=\mathbb{R}^{n}$, with the usual structure of
additive abelian groups. Denote with $(p,t)$, $p\in \mathbb{R}^{n}$, $t\in 
\mathbb{R}$, an element of $N$. Let $H$ act on $N$ as 
\begin{equation*}
\alpha _{q}\left( p,t\right) =\left( p,t+q\cdot p\right) \quad q\in H,\
\left( p,t\right) \in N\text{.}
\end{equation*}
The Heisenberg group\footnote{Usually, the Heisenberg group is defined as the topological 
set $\mathcal{R}\times \mathcal{R}^{n}\times \mathcal{R}^{n}$ with composition law
\begin{equation*}
(t,q,p)(t^{\prime},q^{\prime},p^{\prime}) =
\left( t+t^{\prime}+\frac{1}{2}(q\cdot p^{\prime} - p\cdot q^{\prime}), q+q^{\prime},
p+p^{\prime} \right).
\end{equation*} 
It is easy to show that this is isomorphic to our definition in eq.~(\ref{legge di comp. 
di Heisenberg}).} is the semidirect product $G=N\times _{\alpha }H$. We recall that such a 
group is the topological set $G=N\times H$ endowed with the composition law
\begin{equation}
((p,t), q) ((p^{\prime},t^{\prime}), q^{\prime}) =
((p+p^{\prime},t+t^{\prime}+q\cdot p^{\prime}), q+q^{\prime}). \label{legge di comp. di 
Heisenberg}
\end{equation}
Let $W$ be the following irreducible unitary representation of $G$ acting in 
$L^{2}\left( \mathbb{R}^{n}\right) $ 
\begin{equation*}
\left[ W\left( (p,t),q\right) f\right] \left( x\right) =e^{-i\left( t-p\cdot
x\right) }f\left( x-q\right) \text{.}
\end{equation*}
Clearly, $W\left( (0,0),q\right) =U\left( q\right) $, $W\left(
(p,0),0\right) =V\left( p\right) $, and $W\left( (0,t),0\right) =e^{-it}$.
The groups $H$ and $G/N$ are naturally identified. With such an
identification, the canonical projection $\pi :G\longrightarrow G/N$ is 
\begin{equation*}
\pi \left( (p,t),q\right) =q\text{,}
\end{equation*}
and an element $\left( (p,t),q\right) \in G$ acts on $q_{0}\in H$ as 
\begin{equation*}
\left( (p,t),q\right) \left[ q_{0}\right] =\pi \left( \left( (p,t),q\right)
\left( (0,0),q_{0}\right) \right) =q+q_{0}\text{.}
\end{equation*}

A POM $E$ based on $\mathbb{R}^{n}$ and acting in $L^{2}\left( \mathbb{R}%
^{n}\right) $ satisfies the analogues of eqs.~(\ref{cov}),~(\ref{inv1}) in
dimension $n$ if, and only if, for all $X\in \mathcal{B}\left( \mathbb{R}^{n}%
\right) $ and $\left( (p,t),q\right) \in G$, 
\begin{equation}
W\left( (p,t),q\right) E\left( X\right) W\left( (p,t),q\right) ^{\ast
}=E\left( X+q\right) \text{,}  \label{problema}
\end{equation}
i.e.~if and only if $E$ is a $W$-covariant POM based on $G/N$. By virtue of
the generalized imprimitivity theorem of \S \ref{subsec. 1.2}, $E$ is $W$%
-covariant if and only if there exists a representation $\sigma $ of $N$ and
an isometry $L$ intertwining $W$ with the induced representation $\mathrm{ind%
}_{N}^{G}\left( \sigma \right) $ such that 
\begin{equation*}
E\left( X\right) =L^{\ast }P^{\sigma }\left( X\right) L
\end{equation*}
for all $X\in \mathcal{B}\left( \mathbb{R}^{n}\right) $. Since $\mathrm{ind}%
_{N}^{G}\left( \sigma \right) \subset \mathrm{ind}_{N}^{G}\left( \sigma
^{\prime }\right) $ (as imprimitivity systems) if $\sigma \subset \sigma
^{\prime }$ (as representations), it is not restrictive to assume that such
a $\sigma $ has constant infinite multiplicity, so that there exists a
positive Borel measure $\mu _{\sigma }$ on $\widehat{N}=\mathbb{R}^{n+1}$
and an infinite dimensional Hilbert space $\mathcal{H}$ such that $\sigma $
is the diagonal representation acting in $L^{2}\left( \mathbb{R}^{n+1},\mu
_{\sigma };\mathcal{H}\right) $, i.e. 
\begin{equation*}
\left[ \sigma \left( p,t\right) \phi \right] \left( h,k\right) =e^{ih\cdot
p}e^{ikt}\phi \left( h,k\right) \text{.}
\end{equation*}

Denote with $\gamma _{h,k}$, $h\in \mathbb{R}^{n}$, $k\in \mathbb{R}$ the
following character of $N$ 
\begin{equation*}
\gamma _{h,k}(p,t)=e^{ih\cdot p}e^{ikt}.
\end{equation*}
The action of $H$ on $\widehat{N}$ is given by 
\begin{equation*}
\left( q\cdot \gamma _{h,k}\right) (p,t)=\gamma _{h,k}\left( \alpha
_{-q}\left( p,t\right) \right) =e^{i(h-kq)\cdot p}e^{ikt}\text{,}
\end{equation*}
or in other words 
\begin{equation*}
q\cdot \gamma _{h,k}=\gamma _{h-kq,k}.
\end{equation*}
If $k\neq 0$, the $H$-orbit passing through $\gamma _{h,k}$ is thus 
\begin{equation*}
\mathcal{O}_{\gamma _{h,k}}=\mathbb{R}^{n}\times \{k\}
\end{equation*}
and the corresponding stability subgroup is 
\begin{equation*}
H_{\gamma _{h,k}}=\{0\}.
\end{equation*}
From Mackey's theory (\cite{Foll}, \cite{Var}) it follows that the
representations 
\begin{equation*}
\rho _{h,k}:=\mathrm{ind}_{N}^{G}\left( \gamma _{h,k}\right)
\end{equation*}
are irreducible if $k\neq 0$, $\rho _{h,k}$ and $\rho _{h^{\prime
},k^{\prime }}$ are inequivalent if $k\neq k^{\prime }$ and, fixed $k\neq 0$%
, $\rho _{h,k}$ and $\rho _{h^{\prime },k}$ are equivalent.

The representation $\rho :=\mathrm{ind}_{N}^{G}(\sigma )$ acts on $%
L^{2}\left( \mathbb{R}^{n},\text{d}x;L^{2}\left( \mathbb{R}^{n+1},\mu
_{\sigma };\mathcal{H}\right) \right) $ according to 
\begin{equation*}
\left[ \rho \left( (p,t),q\right) f\right] \left( x\right) =\sigma \left(
p,t-p\cdot x\right) f\left( x-q\right)
\end{equation*}
(here we are using the second realisation of $\mathrm{ind}_{N}^{G}(\sigma )$
which we described in \S \ref{subsec. 1.2}). Using the fact that $\sigma $
acts diagonally in $L^{2}\left( \mathbb{R}^{n+1},\mu _{\sigma };\mathcal{H}%
\right) $ and the identification $L^{2}\left( \mathbb{R}^{n},\text{d}%
x;L^{2}\left( \mathbb{R}^{n+1},\mu _{\sigma };\mathcal{H}\right) \right)
\cong L^{2}\left( \mathbb{R}^{n}\times \mathbb{R}^{n+1},\text{d}x\otimes 
\text{d}\mu _{\sigma }(x);\mathcal{H}\right) $, we find that $\rho $ acts on 
$L^{2}\left( \mathbb{R}^{n}\times \mathbb{R}^{n+1},\text{d}x\otimes \text{d}%
\mu _{\sigma }(x);\mathcal{H}\right) $ as 
\begin{equation*}
\left[ \rho \left( (p,t),q\right) f\right] \left( x,h,k\right) =e^{ih\cdot
p}e^{ik\left( t-p\cdot x\right) }f\left( x-q,h,k\right) .
\end{equation*}
Write $\mu _{\sigma }=\mu _{\sigma _{1}}+\mu _{\sigma _{2}}$, where $\mu
_{\sigma _{1}}\perp \mu _{\sigma _{2}}$ and $\mu _{\sigma _{2}}\left( 
\mathcal{O}_{\gamma _{0,-1}}\right) =0$, and let $\sigma =\sigma _{1}\oplus
\sigma _{2}$ be the corresponding decomposition of $\sigma $. We then have 
\begin{equation*}
\mathrm{ind}_{N}^{G}(\sigma )=\mathrm{ind}_{N}^{G}(\sigma _{1})\oplus 
\mathrm{ind}_{N}^{G}(\sigma _{2}),
\end{equation*}
where the two representations in the sum are disjoint and the sum is a
direct sum of imprimitivity systems. So, since $W\simeq \mathrm{ind}%
_{N}^{G}\left( \gamma _{0,-1}\right) $, it is not restrictive to assume $%
\sigma =\sigma _{1}$, or, in other words, that $\mu _{\sigma }$ is
concentrated in the orbit $\mathcal{O}_{\gamma _{0,-1}}=\mathbb{R}^{n}\times
\{-1\}\cong \mathbb{R}^{n}$.

Let $T$ be the following unitary operator in $L^{2}(\mathbb{R}^{n}\times 
\mathbb{R}^{n},$d$x\otimes $d$\mu _{\sigma }(x);\mathcal{H})$: 
\begin{equation*}
\lbrack Tf]\left( x,h\right) =f\left( x+h,h\right) .
\end{equation*}
If we define the representation $\hat{\rho}$, given by 
\begin{equation*}
\left[ \hat{\rho}\left( (p,t),q\right) f\right] \left( x,h\right)
=e^{-i\left( t-p\cdot x\right) }f\left( x-q,h\right) ,
\end{equation*}
then $T$ intertwines $\hat{\rho}$ with $\rho $. Since $\hat{\rho}\simeq
W\otimes I_{L^{2}\left( \mathbb{R}^{n},\mu _{\sigma };\mathcal{H}\right) }$
and $W$ is irreducible, every isometry intertwining $W$ with $\hat{\rho}$
has the form 
\begin{equation*}
\lbrack \widetilde{L}f]\left( x,h\right) =f\left( x\right) \varphi (h)\quad
\forall f\in L^{2}(\mathbb{R}^{n})
\end{equation*}
for some $\varphi \in L^{2}(\mathbb{R}^{n},\mu _{\sigma };\mathcal{H})$ with 
$\left\| \varphi \right\| _{L^{2}}=1$. The most general isometry $L$
intertwining $W$ with $\rho $ has then the form $L=T\widetilde{L}$ for some
choice of $\varphi $, and the corresponding observable is given by 
\begin{eqnarray*}
\left\langle g , E(X)f \right\rangle &=&\left\langle g ,
L^{\ast }P^{\sigma
}(X)Lf
\right\rangle =\left\langle T\widetilde{L}g ,
P^{\sigma }(X)T\widetilde{L}f \right\rangle \\
&=&\int_{\mathbb{R}^{2n}}\chi _{X}(x)f(x+h)\overline{g(x+h)}\left\langle
\varphi (h),\varphi (h)\right\rangle \text{d}x\text{d}\mu _{\sigma }(h).
\end{eqnarray*}
It follows that 
\begin{eqnarray*}
\left[ E(X)f\right] (x) &=&f(x)\int_{\mathbb{R}^{n}}\chi _{X}(x-h)\left\|
\varphi (h)\right\| ^{2}\text{d}\mu _{\sigma }(h) \\
&=&f(x)\int_{\mathbb{R}^{n}}\chi _{X}(x-h)\text{d}\mu (h),
\end{eqnarray*}
where d$\mu (h)=\Vert \varphi (h)\Vert ^{2}$d$\mu _{\sigma }(h)$ is a
probability measure on $\mathbb{R}^{n}$.

The last formula can be rewritten as in eq.~(\ref{convoluzione di Pekka}) in terms of the 
selfadjoint operator $Q$
\begin{equation*}
E(X) = (\chi_{X} \ast \tilde{\mu}) (Q),
\end{equation*}
where $\chi_{X} \ast \tilde{\mu}$ is the convolution of the function $\chi_{X}$ with the 
measure defined by
\begin{equation*}
\int_{\mathbb{R}^{n}} \varphi (x) \text{d}\tilde{\mu} =
\int_{\mathbb{R}^{n}} \varphi (-x) \text{d}\mu \qquad \forall \varphi \in 
C_{c}(\mathbb{R}^{n}) . 
\end{equation*}

\section{Covariance under dilations}

Besides covariance (\ref{covQ}) and invariance (\ref{invQ}), the canonical
position observable $\Pi _{Q}$ has still more symmetry properties. Namely,
let $\mathbb{R}_{+}$ be the set of positive real numbers regarded as a
multiplicative group. It has a family of unitary representations $\left\{
A_{t}\mid t\in \mathbb{R}\right\} $ acting on $\mathcal{H}$, and given by 
\begin{equation*}
\lbrack A_{t}(a)f](x)=\frac{1}{\sqrt{a}}f\left( a^{-1}(x-t)+t\right) .
\end{equation*}
It is a direct calculation to verify that for all $a\in \mathbb{R}_{+},X\in 
\mathcal{B}(\mathbb{R})$, 
\begin{equation*}
A_{0}(a)\Pi _{Q}(X)A_{0}(a)^{\ast }=\Pi _{Q}(aX).
\end{equation*}
We adopt the following terminology, which we take from \cite{Cast}.

\begin{definition}
An observable $E:\mathcal{B}(\mathbb{R})\rightarrow \mathcal{L}(%
\mathcal{H})$ is \textbf{covariant under dilations} if there exists a
unitary representation $A$ of $\mathbb{R}_{+}$ such that for all $a\in 
\mathbb{R}_{+}$ and $X\in \mathcal{B}(\mathbb{R})$, 
\begin{equation}
A(a)E(X)A(a)^{\ast }=E(aX).  \label{e_dilat}
\end{equation}
\end{definition}

The canonical position observable $\Pi_Q$ is not the only position
observable which is covariant under dilations. An observable $E_{\delta_t},
t\in\mathbb{R},$ is a translated version of $\Pi_Q$, namely, for any $X\in%
\mathcal{B}(\mathbb{R})$, 
\begin{equation*}
E_{\delta_t}(X)=\Pi_Q(X-t)=U(t)^*\Pi_Q(X)U(t).
\end{equation*}
Since $A_{-t}(a) = U(t)^* A_{0}(a) U(t)$, the observable $E_{\delta_t}$ is
covariant under dilations, with, for example, $A = A_{-t}$.

\begin{proposition}
\label{dilat}Let $E$ be a position observable on $\mathbb{R}$. The following
conditions are equivalent:

\begin{itemize}
\item[(a)]  $E$ is covariant under dilations;

\item[(b)]  $\left\| E(U)\right\| =1$ for every nonempty open set $U\subset 
\mathbb{R}$;

\item[(c)]  $E=E_{\delta _{t}}$ for some $t\in \mathbb{R}$;

\item[(d)]  $E$ is a sharp observable.
\end{itemize}
\end{proposition}

\begin{proof}
Let $E$ be covariant under dilations. In a similar way as in \cite[Lemma 3]
{Cast} we can show that $\left\| E(U)\right\| =1$ for all nonempty open sets 
$U$. In fact, assuming the contrary, we can find a closed interval $I$ with
nonvoid interior such that $\left\| E(I)\right\| =1-\varepsilon <1$. By
translational covariance of $E$, it is not restrictive to assume that $I$ is
centered at the origin. If $f\in \mathcal{H}$ we then have 
\begin{equation*}
\mu _{f,f}\left( nI\right) =\left\langle f,E\left( nI\right) f\right\rangle
=\left\langle A\left( n\right) ^{\ast }f,E\left( I\right) A\left( n\right)
^{\ast }f\right\rangle \leq 1-\varepsilon 
\end{equation*}
for all $n\in \mathbb{N}$. By inner regularity of $\mu _{f,f}$, we have $\mu
_{f,f}\left( \mathbb{R}\right) \leq 1-\varepsilon $, hence $E\left( \mathbb{R%
}\right) \neq I$, which is absurd. So, (a) implies (b).

Assume then that (b) holds. For any nonempty open set $U$ we get 
\begin{equation}
1=\left\| E(U)\right\| =\mathrm{ess\,sup}_{x\in \mathbb{R}}\rho (x+U).
\label{norm1}
\end{equation}
It follows that $\mathrm{supp}(\rho )$ contains only one point. Indeed,
assume on the contrary that $\mathrm{supp}(\rho )$ contains two points $%
x_{1}\neq x_{2}$ and denote $U=\{x\in \mathbb{R}||x|<\frac{1}{4}%
|x_{1}-x_{2}|\}$. Since $x_{1}+U$ and $x_{2}+U$ are neighborhoods of $x_{1}$
and $x_{2}$, respectively, we have $m_{i}:=\rho (x_{i}+U)>0$ for $i=1,2$.
Then, for any $x\in \mathbb{R}$, $\rho (x+U)\leq 1-\mathrm{min}(m_{1},m_{2})$%
. This is in contradiction with (\ref{norm1}). Hence, (b) implies (c).

As previously mentioned, (c) implies (a). Clearly, (c) also implies (d).
Since (d) implies (b) the proof is complete.
\end{proof}

The dilation covariance means that the observable in question has no scale
dependence. A realistic position measurement apparatus has a limited
accuracy and hence it cannot define a position observable which is covariant
under dilations. Thus, sharp position observables are not suitable to
describe nonideal situations.

If $E=E_{\delta _{t}}$, one could ask what is the most general form of the
representation $A$ of $\mathbb{R}_{+}$ satisfying eq.~(\ref{e_dilat}). The
answer is given in the next proposition.

\begin{proposition}
\label{r_dilation}If $A$ is a unitary representation of $\mathbb{R}_{+}$
satisfying eq.~(\ref{e_dilat}) with $E=E_{\delta _{t}}$, then there exists a
measurable function $\beta :\mathbb{R}\longrightarrow \mathbb{T}$ such that 
\begin{equation*}
\left[ A(a)f\right] (x)=\frac{1}{\sqrt{a}}\beta (x+t)\overline{\beta \left(
a^{-1}(x+t)\right) }f\left( a^{-1}(x+t)-t\right) .
\end{equation*}
In particular, $A$ is equivalent to $A_{-t}$.
\end{proposition}

\begin{proof}
Let $A^{\prime }(a)=U(t)A(a)U(t)^{\ast }$. Then, $A^{\prime }(a)\Pi
_{Q}(X)A^{\prime }(a)^{\ast }=\Pi _{Q}(aX)$. Denote with $\Pi _{Q}^{+}$ the
restriction of $\Pi _{Q}$ to the Borel subsets of $\mathbb{R}_{+}$. Then, $%
S_{0}=\left( A_{0},\Pi _{Q}^{+},L^{2}\left( 0,+\infty \right) \right) $ and $%
S=\left( A^{\prime },\Pi _{Q}^{+},L^{2}\left( 0,+\infty \right) \right) $
are transitive imprimitivity systems of the group $\mathbb{R}_{+}$ based on $%
\mathbb{R}_{+}$. By Mackey imprimitivity theorem, there exists a Hilbert
space $\mathcal{K}$ such that $S=\mathrm{ind}_{\{1\}}^{\mathbb{R}_{+}}(I_{%
\mathcal{K}})$, where $I_{\mathcal{K}}$ is the trivial representation of ${%
\{1\}}$ acting in $\mathcal{K}$. Since $S_{0}=\mathrm{ind}_{\{1\}}^{\mathbb{R%
}_{+}}(1)$, we have the isomorphism of intertwining operators $\mathcal{C}%
\left( 1,I_{\mathcal{K}}\right) \simeq \mathcal{C}\left( S_{0},S\right) $,
and hence there exists an isometry $W_{1}:L^{2}\left( 0,+\infty \right)
\longrightarrow L^{2}\left( 0,+\infty \right) $ intertwining $S_{0}$ with $S$%
. In particular, $W_{1}\Pi _{Q}^{+}=\Pi _{Q}^{+}W_{1}$, and hence there
exists a measurable function $\beta _{1}:\mathbb{R}_{+}\longrightarrow 
\mathbb{T}$ such that 
\begin{equation*}
\lbrack W_{1}f](x)=\beta _{1}(x)f(x)\quad \forall f\in L^{2}\left( 0,+\infty
\right) .
\end{equation*}
It follows that $W_{1}$ is unitary.

Reasoning as above, one finds a unitary operator $W_{2}$ intertwining the
restrictions of $A_{0}$ and $A^{\prime }$ to $L^{2}\left( -\infty ,0\right) $%
, with 
\begin{equation*}
\lbrack W_{2}f](x)=\beta _{2}(x)f(x)\quad \forall f\in L^{2}\left( -\infty
,0\right) ,
\end{equation*}
for some measurable function $\beta _{2}:\mathbb{R}_{-}\longrightarrow 
\mathbb{T}$. Then, $W=W_{1}\oplus W_{2}$ is unitary on $L^{2}\left( -\infty
,+\infty \right) $, and $A(a)=U(t)^{\ast }WA_{0}(a)W^{\ast }U(t)$ has the
claimed form for all $a\in \mathbb{R}_{+}$.
\end{proof}

\section{State distinction power of a position observable\label{State}}

In this section, we investigate the ability of position observables to
discriminate between different states, that is we compare the state distinction power of 
two position observables (see also \cite{Lahti95}).

\begin{definition}
Let $E_{1}$ and $E_{2}$ be observables on $\mathbb{R}$. The \textbf{state
distinction power} of $E_{2}$ is greater than or equal to $E_{1}$ if for all 
$T,T^{\prime }\in \mathcal{S}(\mathcal{H})$, 
\begin{equation*}
p_{T}^{E_{2}}=p_{T^{\prime }}^{E_{2}}\ \Rightarrow
p_{T}^{E_{1}}=p_{T^{\prime }}^{E_{1}}.
\end{equation*}
In this case we denote $E_{1}\sqsubseteq E_{2}$. If $E_{1}\sqsubseteq
E_{2}\sqsubseteq E_{1}$ we say that $E_{1}$ and $E_{2}$ are \textbf{%
informationally equivalent} and denote $E_{1}\sim E_{2}$. If $%
E_{1}\sqsubseteq E_{2}$ and $E_{2}\not\sqsubseteq E_{1}$, we write $%
E_{1}\sqsubset E_{2}$.
\end{definition}

\begin{example}
\label{trivial1}An observable $E:\mathcal{B}(\mathbb{R})\rightarrow \mathcal{%
L}(\mathcal{H})$ is \textbf{trivial} if $p_{T}^{E}=p_{T^{\prime }}^{E}$ for
all states $T,T^{\prime }\in \mathcal{S}(\mathcal{H})$. A trivial observable 
$E$ is then of the form $E(X)=\lambda (X)I$, $X\in \mathcal{B}(\mathbb{R})$,
for some probability measure $\lambda $. The state distinction power of any
observable $E^{\prime }$ is greater than or equal to that of the trivial
observable $E$. Clearly there is no trivial position observable on $\mathbb{R%
}$.
\end{example}

\begin{example}
An observable $E:\mathcal{B}(\mathbb{R})\rightarrow \mathcal{L}(\mathcal{H})$
is called \textbf{informationally complete} if $p_{T}^{E}\neq p_{T^{\prime
}}^{E}$ whenever $T\neq T^{\prime }$. The state distinction power of an
informationally complete observable is greater than or equal to that of any
other observable $E_{1}$ on $\mathbb{R}$. It is easy to see that there is no
informationally complete position observable. Namely, let $\psi $ be a unit
vector, $p\neq 0$ a real number, and denote $\psi ^{\prime }=V(p)\psi $.
Then the states $T=\left\langle \psi , \cdot \right\rangle \psi $ and $%
T^{\prime }=\left\langle \psi^{\prime }, \cdot \right\rangle \psi ^{\prime }$
are different but for any position observable $E_{\rho }$, $p_{T}^{E_{\rho
}}=p_{T^{\prime }}^{E_{\rho }}$ since $V(p)$ commutes with all the effects $%
E_{\rho }(X)$, $X\in \mathcal{B}(\mathbb{R})$.
\end{example}

We will next think of $\sim$ as a relation on the set $\mathcal{POS}_{%
\mathbb{R}}$. The relation $\sim$ is clearly reflexive, symmetric and
transitive, and hence it is an equivalence relation. We denote the
equivalence class of a position observable $E$ by $[E]$ and the space of
equivalence classes as $\mathcal{POS}_{\mathbb{R}}/\sim$. The relation $%
\sqsubseteq$ induces a partial order in the set $\mathcal{POS}_{\mathbb{R}%
}/\sim$ in a natural way.

Let $E_{\rho}$ be a position observable and $T$ a state. The probability
measure $p^{E_{\rho}}_T$ is the convolution of the probability measures $%
p^{\Pi_Q}_T$ and $\rho$, 
\begin{equation}  \label{convol}
p^{E_{\rho}}_T=p^{\Pi_Q}_T\ast\rho.
\end{equation}
It is clear from (\ref{convol}) that for all $T,T^{\prime}\in\mathcal{S}(%
\mathcal{H})$, 
\begin{equation*}
p^{\Pi_Q}_T=p^{\Pi_Q}_{T^{\prime}}\ \Rightarrow
p^{E_{\rho}}_T=p^{E_{\rho}}_{T^{\prime}},
\end{equation*}
and hence $E_{\rho}\sqsubseteq \Pi_Q$. We conclude that $[\Pi_Q]$ is the
only maximal element of the partially ordered set $\mathcal{POS}_{\mathbb{R}%
}/\sim$.

It is shown in \cite[Prop. 5]{fuzzy} that a position observable $E_\rho$
belongs to the maximal equivalence class $[\Pi_Q]$ 
if and only if $\mathrm{supp}\left(\widehat{\rho}\right)=\mathbb{R}$. The
following proposition characterizes the equivalence classes completely.

\begin{proposition}
Let $\rho _{1},\rho _{2}$ be probability measures on $\mathbb{R}$ and $%
E_{\rho _{1}},E_{\rho _{2}}$ the corresponding position observables. Then 
\begin{equation*}
E_{\rho _{1}}\sqsubseteq E_{\rho _{2}}\Longleftrightarrow \mathrm{supp}%
\left( \widehat{\rho _{1}}\right) \subseteq \mathrm{supp}\left( \widehat{%
\rho _{2}}\right) .
\end{equation*}
\end{proposition}

\begin{proof}
Taking the Fourier transform of eq.~(\ref{convol}), we get 
\begin{equation}
\mathcal{F}(p_{T}^{E_{\rho }})=\mathcal{F}(p_{T}^{\Pi _{Q}})\mathcal{F}(\rho
).  \label{e_Fourier}
\end{equation}
Since the Fourier transform is injective, it is clear from the above
relation that $\mathrm{supp}\left( \widehat{\rho }_{1}\right) \subseteq 
\mathrm{supp}\left( \widehat{\rho }_{2}\right) $ implies $E_{\rho
_{1}}\sqsubseteq E_{\rho _{2}}$.

Conversely, suppose $\mathrm{supp}\left( \widehat{\rho }_{1}\right)
\nsubseteq \mathrm{supp}\left( \widehat{\rho }_{2}\right) $. As $\widehat{%
\rho }_{i}$, $i=1,2$, are continuous functions and $\widehat{\rho }%
_{i}\left( \xi \right) =\overline{\widehat{\rho }_{i}\left( -\xi \right) }$,
there exists a closed interval $[2a,2b]$, with $0\leq a<b$, such that $%
[2a,2b]\cup \lbrack -2b,-2a]\subseteq \mathrm{supp\left( \widehat{\rho }%
_{1}\right) }$ and $\left( [2a,2b]\cup \lbrack -2b,-2a]\right) \cap \mathrm{%
supp}\left( \widehat{\rho }_{2}\right) =\emptyset $. Define the functions 
\begin{eqnarray*}
h_{1} &=&\frac{1}{\sqrt{2(b-a)}}\left( \chi _{\lbrack a,b]}-\chi _{\lbrack
-b,-a]}\right) , \\
h_{2} &=&\frac{1}{\sqrt{2(b-a)}}\left( \chi _{\lbrack a,b]}+\chi _{\lbrack
-b,-a]}\right) ,
\end{eqnarray*}
and for $i=1,2$, denote 
\begin{equation*}
h_{i}^{\ast }(\xi ):=\overline{h_{i}(-\xi )}.
\end{equation*}
Define 
\begin{equation*}
f_{i}=\mathcal{F}^{-1}\left( h_{i}\right) ,
\end{equation*}
and let $T_{i}$ be the one-dimensional projection $\left| f_{i}\right\rangle
\left\langle f_{i}\right| $. We then have 
\begin{equation*}
\text{d}p_{T_{i}}^{\Pi _{Q}}(x)=\left| f_{i}(x)\right| ^{2}\text{d}x=\left|
\left( \mathcal{F}^{-1}h_{i}\right) (x)\right| ^{2}\text{d}x=\mathcal{F}%
^{-1}\left( h_{i}\ast h_{i}^{\ast }\right) (x)\text{d}x,
\end{equation*}
and 
\begin{equation*}
\begin{split}
\mathcal{F}(p_{T_{i}}^{\Pi _{Q}})& =\mathcal{F}\mathcal{F}^{-1}\left(
h_{i}\ast h_{i}^{\ast }\right) =h_{i}\ast h_{i}^{\ast } \\
& =\frac{1}{2(b-a)}\Bigl(2\chi _{\lbrack a,b]}\ast \chi _{\lbrack
-b,-a]}+(-1)^{i}\chi _{\lbrack -b,-a]}\ast \chi _{\lbrack -b,-a]} \\
& \quad +(-1)^{i}\chi _{\lbrack a,b]}\ast \chi _{\lbrack a,b]}\Bigr).
\end{split}
\end{equation*}
Since 
\begin{eqnarray*}
\mathrm{supp}\left( \chi _{\lbrack a,b]}\ast \chi _{\lbrack -b,-a]}\right) 
&=&[a-b,b-a], \\
\mathrm{supp}\left( \chi _{\lbrack a,b]}\ast \chi _{\lbrack a,b]}\right) 
&=&[2a,2b], \\
\mathrm{supp}\left( \chi _{\lbrack -b,-a]}\ast \chi _{\lbrack -b,-a]}\right)
&=&[-2b,-2a],
\end{eqnarray*}
an application of (\ref{e_Fourier}) shows that 
\begin{eqnarray*}
\mathcal{F}(p_{T_{1}}^{E_{\rho _{1}}}) &\neq &\mathcal{F}(p_{T_{2}}^{E_{\rho
_{1}}}), \\
\mathcal{F}(p_{T_{1}}^{E_{\rho _{2}}}) &=&\mathcal{F}(p_{T_{2}}^{E_{\rho
_{2}}}),
\end{eqnarray*}
or in other words, $E_{\rho _{1}}\not\sqsubseteq E_{\rho _{2}}$.
\end{proof}

\begin{remark}
It follows from the above proposition that $E_{1}\sqsubset
E_{2}\Longleftrightarrow \mathrm{supp}\left( \widehat{\rho _{1}}\right)
\subset \mathrm{supp}\left( \widehat{\rho _{2}}\right) $, and hence the set $%
\mathcal{POS}_{\mathbb{R}}/\sim $ has no minimal element. Indeed, if $\rho
_{2}$ is a probability measure, there always exists a probability measure $%
\rho _{1}$ such that $\mathrm{supp}\left( \widehat{\rho _{1}}\right) \subset 
\mathrm{supp}\left( \widehat{\rho _{2}}\right) $. In fact, since $\widehat{%
\rho _{2}}$ is continuous, $\widehat{\rho _{2}}(\xi )=\overline{\widehat{%
\rho _{2}}(-\xi )}$ and $\widehat{\rho _{2}}(0)=\rho _{2}(\mathbb{R})\neq 0$%
, there exists $a>0$ such that the closed interval $[-a,a]$ is strictly
contained in $\mathrm{supp}\left( \widehat{\rho _{2}}\right) $. If we define 
$h=\frac{1}{\sqrt{a}}\chi _{\left[ -\frac{a}{2},\frac{a}{2}\right] }$, $f=%
\mathcal{F}^{-1}h$, then d$\rho _{1}(x):=|f(x)|^{2}$d$x$ is a probability
measure, and $\mathrm{supp}\left( \widehat{\rho _{1}}\right) =\mathrm{supp}%
\left( h\ast h\right) =[-a,a]$.
\end{remark}

\section{Limit of resolution of a position observable\label{Resolution}}

Let $\Pi :\mathcal{B}(\mathbb{R})\rightarrow \mathcal{L}(\mathcal{H})$ be a
sharp observable. For any nontrivial projection $\Pi (X)$, there exist
states $T,T^{\prime }\in \mathcal{S}(\mathcal{H})$ such that $p_{T}^{\Pi
}(X)=1$ and $p_{T^{\prime }}^{\Pi }(X)=0$. Using the terminology of 
\cite[II.2.1]{Lahti95}, we may say that $\Pi (X)$ is a \emph{sharp property}
and it is \emph{real} in the state $T$.

In general, an observable $E$ has effects as its values which are not
projections and, hence, not sharp properties. An effect $E\left( X\right)
\in \mathrm{ran}\,E$ is called \textbf{regular} if its spectrum extends both
above and below $1/2$. This means that there exist states $T,T^{\prime }\in 
\mathcal{S}(\mathcal{H})$ such that $\mathrm{tr}[TE\left( X\right) ]>1/2$
and $\mathrm{tr}[T^{\prime }E\left( X\right) ]<1/2$. In this sense regular
effects can be seen as \emph{approximately realizable properties} (see again 
\cite[II.2.1]{Lahti95}). The observable $E$ is called \textbf{regular} if
all the nontrivial effects $E\left( X\right) $ are regular.

It is shown in \cite[Prop. 4]{fuzzy} that if a probability measure $\rho$ is
absolutely continuous with respect to the Lebesgue measure, then the
position observable $E_{\rho}$ is not regular. Here we modify the notion of
regularity to get a quantification of sharpness, or resolution, of position
observables.

For any $x\in \mathbb{R},r\in \mathbb{R}_{+}$, we denote the interval $%
[x-r/2,x+r/2]$ by $I_{x;r}$. We also denote $I_{r}=I_{0;r}$. Let $E:\mathcal{%
B}(\mathbb{R})\rightarrow \mathcal{L}(\mathcal{H})$ be an observable and $%
\alpha >0$. We say that $E$ is $\alpha $\textbf{-regular} if all the
nontrivial effects $E(I_{x;r})$, $x\in \mathbb{R}$, $r\geq \alpha $, are
regular.


\begin{definition}
Let $E:\mathcal{B}(\mathbb{R})\rightarrow \mathcal{L}(\mathcal{H})$ be an
observable. We denote 
\begin{equation*}
\gamma _{E}=\inf \left\{ \alpha >0\mid \text{$E$ is $\alpha $-regular}%
\right\} 
\end{equation*}
and say that $\gamma _{E}$ is the \textbf{limit of resolution of }$E$.
\end{definition}

It follows directly from definitions that the limit of resolution of a
regular observable is 0. Especially, the limit of resolution of canonical
position observables is 0.

\begin{example}
\label{trivial2}Let $E$ be a trivial observable (see Example \ref{trivial1}%
). For any $X\in \mathcal{B}(\mathbb{R})$, we have either $E(X)\geq \frac{1}{%
2}I$ or $E(X)\leq \frac{1}{2}I$. Hence, $\gamma _{E}=\infty $.
\end{example}

\begin{proposition}
\label{reg2}A position observable $E_{\rho }$ is $\alpha $-regular if and
only if 
\begin{equation}
\mathrm{ess\,sup}_{x\in \mathbb{R}}\rho (I_{x,\alpha })>1/2\text{,}
\label{areg}
\end{equation}
where the essential supremum is taken with respect to the Lebesgue measure
of $\mathbb{R}$.
\end{proposition}

\begin{proof}
An effect $E_{\rho }(X)$ is regular if and only if $\left\| E_{\rho
}(X)\right\| >1/2$ and $\left\| E_{\rho }(\mathbb{R}\setminus X)\right\| >1/2
$. Since the norm of the multiplicative operator $E_{\rho }(X)$ is $\mathrm{%
ess\,sup}_{x\in \mathbb{R}}\rho (X-x)$, we conclude that $E_{\rho }(X)$ is
regular if and only if 
\begin{equation*}
\mathrm{ess\,sup}_{x\in \mathbb{R}}\rho (X-x)>1/2\quad \text{ and}\quad 
\mathrm{ess\,inf}_{x\in \mathbb{R}}\rho (X-x)<1/2\text{.}
\end{equation*}
Thus, $E_{\rho }$ is $\alpha $-regular if and only if, for all $r\geq \alpha 
$, 
\begin{equation*}
\mathrm{ess\,sup}_{x\in \mathbb{R}}\rho (I_{x;r})>1/2\quad \text{ and}\quad 
\mathrm{ess\,inf}_{x\in \mathbb{R}}\rho (I_{x;r})<1/2\text{.}
\end{equation*}
The second condition is always satisfied and from the first eq.~(\ref{areg})
follows.
\end{proof}

\begin{corollary}
A position observable $E_{\rho }$ has a finite limit of resolution and 
\begin{equation*}
\gamma _{E_{\rho }}=\inf \left\{ \alpha >0\mid \mathrm{ess\,sup}_{x\in 
\mathbb{R}}\rho (I_{x;\alpha })>1/2\right\} .
\end{equation*}
\end{corollary}

In the next chapter, we will prove an Heisenberg-like uncertainty relation
involving the limits of resolution of coexistent position and momentum
observables. For the moment, we end this section with a result that
characterises the position observables which are regular. It is taken
from \cite{Articolo con Teiko in preparazione}.

\begin{proposition}
\label{limit0} Let $E_{\rho }$ be a position observable. The following
conditions are equivalent:

\begin{itemize}
\item[(i)]  $E_{\rho }$ is regular;

\item[(ii)]  $\gamma _{E_{\rho }}=0$;

\item[(iii)]  there exists $\overline{x}\in \mathbb{R}$ and $\lambda \in
M_{1}^{+}(\mathbb{R})$ with $\overline{x}\in \mathrm{supp}\lambda $ such
that $\rho =\frac{1}{2}\delta _{\overline{x}}+\frac{1}{2}\lambda $.
\end{itemize}
\end{proposition}

\begin{proof}
It is clear that (i) implies (ii).

Let $\gamma _{E_{\rho }}=0$. This means that 
\begin{equation}
\inf \{\alpha >0\mid \mathrm{ess\,sup}_{x\in \mathbb{R}}\rho (I_{x;\alpha })>%
\frac{1}{2}\}=0.  \label{reg10}
\end{equation}
For every $\alpha >0$, denote 
\begin{equation*}
A_{\alpha }=\{x\in \mathbb{R}\mid \rho (I_{x;\alpha })>\frac{1}{2}\}.
\end{equation*}
Since $\alpha _{1}>\alpha _{2}$ implies $A_{\alpha _{1}}\supset A_{\alpha
_{2}}$, it follows from (\ref{reg10}) that $A_{\alpha }\neq \emptyset $ for
all $\alpha >0$. For each $n=1,2\dots $, choose an element $x_{n}\in A_{1/n}$%
. We have $\rho (I_{x_{n},1/n})>\frac{1}{2}$. Since $\rho $ is a finite
measure, the sequence $(x_{n})_{n\geq 1}$ is bounded. Hence, there exists a
subsequence $(x_{n_{k}})_{k\geq 1}$ converging to some $\overline{x}\in 
\mathbb{R}$. For each $\beta >0$, there exists $k$ such that $I_{x_{n_{k}},1/%
{n_{k}}}\subset I_{\overline{x},\beta }$, so that $\rho (I_{\overline{x}%
,\beta })>\frac{1}{2}$. Thus, 
\begin{equation*}
\rho (\{\overline{x}\})=\rho (\cap _{\beta >0}I_{\overline{x},\beta
})=\lim_{\beta \rightarrow 0}\rho (I_{\overline{x},\beta })\geq \frac{1}{2}.
\end{equation*}
It follows that $\lambda =2\rho -\delta _{\overline{x}}$ is a probability
measure. For all $\beta >0$, we have $\lambda (I_{\overline{x},\beta
})=2\rho (I_{\overline{x},\beta })-1>0$, which implies $\overline{x}\in 
\mathrm{supp}\lambda $. Thus, (ii) implies (iii).

Assume that (iii) holds. We start by noticing that $E_{\rho }$ is regular if
and only if $\left\| E_{\rho }(X)\right| =\frac{1}{2}\mathrm{ess\,sup}_{q\in 
\mathbb{R}}\left\{ \delta _{x_{0}}(X-q)+\lambda (X-q)\right\} >\frac{1}{2}$
for every Borel set in which $E_{\rho }\neq 0,I$. Since we know that $%
E_{\rho }(X)=0$ is equivalent to $\mu (X)=0$ we can assume that $X$ is not a
null or co-null set. Let us begin by evaluating 
\begin{equation*}
\mathrm{ess\,sup}_{q\in \mathbb{R}}\left\{ \delta _{x_{0}}(X-q)+\lambda
(X-q)\right\} =\mathrm{ess\,sup}_{q\in \mathbb{R}}\left\{ \chi
_{X-x_{0}}(q)+\lambda (X-q)\right\} 
\end{equation*}
in the case $X$ is a finite measure set. It is a matter of proving that $%
\lambda (X-(\cdot ))$ has non void essential image on the support of the
characteristic function $\chi _{X-x_{0}}$, that is:

\begin{equation*}
\mathrm{ess\,sup}_{q\in X-x_{0}}\lambda (X-(\cdot ))>0
\end{equation*}
This is also equivalent to prove that 
\begin{equation*}
\int_{X-x_{0}}\lambda (X-q)d\mu (q)>0
\end{equation*}
but an easy calculation shows that the last integral is equal to 
\begin{equation*}
\int_{\mathbb{R}}\chi _{X-x_{0}}\ast \check{\chi}_{X-x_{0}}(x)d\lambda (x)
\end{equation*}
where $\check{\chi}_{X}(x):=\chi _{X}(-x)$. It is a well known fact that the
integrand is a continuous function (since it is the convolution of two
continuous functions). Observing that $x_{0}\in \mathrm{supp}\left( \chi
_{X-x_{0}}\ast \check{\chi}_{X-x_{0}}\right) $, we get immediately that the
integral is not zero. So we have the thesis in the case of finite measure
sets.

Let us now consider the case $\mu (X)=\infty $. We must show that either $%
E_{\rho }(X)=I$ or $E_{\rho }(X)$ has spectrum below $\frac{1}{2}$. By a
previous observation we can limit ourselves to the case $\mu (X^{\prime
})\neq 0$ since in the case of co-null set $E_{\rho }(X)=I$. Hence we can
assume that $X^{\prime }$ cointains a finite measure set $Y$ to which the
above result apply. So we have 
\begin{equation*}
\left\| E_{\rho }(X^{\prime })\right\| \geq \left\| E_{\rho }(Y)\right\| >%
\frac{1}{2}
\end{equation*}
Since $E_{\rho }(X^{\prime })=I-E_{\rho }(X)$, denoting with $\sigma (A)$
the spectrum of the operator $A$, we have $\sigma (E_{\rho }(X))=1-\sigma
(E_{\rho }(X^{\prime }))$ and we are done.
\end{proof}

\section{Position and momentum observables on $\mathbb{R}^{3}$\label{R3}}

\subsection{Definitions\label{Def3}}

In this section, $\mathcal{H}=L^{2}(\mathbb{R}^{3};\mathbb{C}^{2j+1})$ is
the Hilbert space of a nonrelativistic particle with spin $j$ in the
Euclidean space. Let $Q_{i},i=1,2,3$, denote the multiplicative operators on 
$\mathcal{H}$ given by $\left[ Q_{i}f\right] (\vec{x})=x_{i}f(\vec{x})$,
where $x_{i}$ is the $i$th component of $\vec{x}$. By $P_{i}$ we mean the
operator $\mathcal{F}^{-1}Q_{i}\mathcal{F}$ and we denote $\vec{Q}%
=(Q_{1},Q_{2},Q_{3})$, $\vec{P}=(P_{1},P_{2},P_{3})$. The space translation
group $\mathbb{R}^{3}$ has a unitary representation $U(\vec{q})=e^{-i\vec{q}%
\cdot \vec{P}}$ and similarly, the momentum boost group has a representation 
$V(\vec{p})=e^{i\vec{p}\cdot \vec{Q}}$. It is an immediate observation that
the sharp observables $\Pi _{\vec{Q}}$ and $\Pi _{\vec{P}}$ on $\mathbb{R}%
^{3}$, respectively associated to the representations $V$ and $U$ by
Stone-Naimark-Ambrose-Godement theorem, satisfy the obvious covariance and
invariance conditions, analogous to (\ref{cov})-(\ref{covF}). The rotation
group $SO(3)$ acts in $\mathcal{H}$ according to the projective
representation $D$, given by
\begin{equation*}
\left[ D(R)f\right] (\vec{x})=D^{j}\left( R^{\prime }\right) f(R^{-1}\vec{x})%
\text{.}
\end{equation*}
Here, $D^{j}$ is the irreducible representation of $SU\left( 2\right) $ in $%
\mathbb{C}^{2j+1}$, and $R^{\prime }$ is an element of $SU\left( 2\right) $
lying above $R$. It is straightforward to verify that the sharp observables $%
\Pi _{\vec{Q}}$ and $\Pi _{\vec{P}}$ are covariant under rotations, that is,
for all $R\in SO(3)$ and $X\in \mathcal{B}(\mathbb{R}^{3})$, 
\begin{eqnarray*}
D(R)\Pi _{\vec{Q}}(X)D(R)^{\ast } &=&\Pi _{\vec{Q}}(RX), \\
D(R)\Pi _{\vec{P}}(X)D(R)^{\ast } &=&\Pi _{\vec{P}}(RX).
\end{eqnarray*}

These observations motivate to the following definitions.

\begin{definition}
An observable $E:\mathcal{B}(\mathbb{R}^{3})\rightarrow \mathcal{L}(\mathcal{%
H})$ is a \textbf{position observable on }$\mathbb{R}^{3}$ if, for all $\vec{q},\vec{p%
}\in \mathbb{R}^{3}$, $R\in SO(3)$ and $X\in \mathcal{B}(\mathbb{R}^{3})$, 
\begin{eqnarray}
U(\vec{q})E(X)U(\vec{q})^{\ast } &=&E(X+\vec{q}),  \label{cov3} \\
V(\vec{p})E(X)V(\vec{p})^{\ast } &=&E(X),  \label{inv3} \\
D(R)E(X)D(R)^{\ast } &=&E(RX).  \label{rot3}
\end{eqnarray}
We will denote by $\mathcal{POS}_{\mathbb{R}^{3}}$ the set of all position
observables on $\mathbb{R}^{3}$.
\end{definition}

\begin{definition}
An observable $F:\mathcal{B}(\mathbb{R}^{3})\rightarrow \mathcal{L}(\mathcal{%
H})$ is a \textbf{momentum observable on }$\mathbb{R}^{3}$ if, for all $\vec{q},\vec{p%
}\in \mathbb{R}^{3}$, $R\in SO(3)$ and $X\in \mathcal{B}(\mathbb{R}^{3})$, 
\begin{eqnarray*}
U(\vec{q})F(X)U(\vec{q})^{\ast } &=&F(X), \\
V(\vec{p})F(X)V(\vec{p})^{\ast } &=&F(X+\vec{p}), \\
D(R)F(X)D(R)^{\ast } &=&F(RX).
\end{eqnarray*}
\end{definition}

\subsection{Structure of position observables on $\mathbb{R}^{3}$\label%
{Structure3}}

We say that a probability measure $\rho $ on $\mathbb{R}^{3}$ is rotation
invariant if for all $X\in \mathcal{B}(\mathbb{R}^{3})$ and $R\in SO(3)$, 
\begin{equation*}
(R\cdot \rho )(X):=\rho (R^{-1}X)\equiv \rho (X).
\end{equation*}
The set of rotation invariant probability measures on $\mathbb{R}^{3}$ is
denoted by $M(\mathbb{R}^{3})_{inv}^{1,+}$. Using the isomorphism $\mathbb{R}%
^{3}\setminus \{0\}\simeq \mathbb{R}_{+}\times S^{2}$ and the disintegration
of measures, the restriction of any measure $\rho \in M(\mathbb{R}%
^{3})_{inv}^{1,+}$ to the subset $\mathbb{R}^{3}\setminus \{0\}$ can be
written in the form 
\begin{equation*}
\left. \text{d}\rho \right| _{\mathbb{R}^{3}\setminus \{0\}}\left( \vec{r}%
\right) =\text{d}\rho _{\mathrm{rad}}\left( r\right) \text{d}\rho _{\mathrm{%
ang}}\left( \Omega \right) ,
\end{equation*}
where $\rho _{\mathrm{rad}}$ is a finite measure on $\mathbb{R}_{+}$ with $%
\rho _{\mathrm{rad}}(\mathbb{R}_{+})=1-\rho (\{0\})$, and $\rho _{\mathrm{ang%
}}$ is the $SO(3)$-invariant measure on the sphere $S^{2}$ normalized to $1$.

Given a rotation invariant probability measure $\rho $, the formula 
\begin{equation}
E_{\rho }(X)=\int \rho (X-\vec{q})\ \text{d}\Pi _{\vec{Q}}(\vec{q}),\quad
X\in \mathcal{B}(\mathbb{R}^{3}),  \label{pos3}
\end{equation}
defines a position observable on $\mathbb{R}^{3}$.

\begin{proposition}
\label{struc3}Any position observable $E$ on $\mathbb{R}^{3}$ is of the form 
$E=E_{\rho }$ for some $\rho \in M(\mathbb{R}^{3})_{inv}^{1,+}$.
\end{proposition}

\begin{proof}
It is shown in \S \ref{proof1} that if $E$ satisfies eqs.~(\ref{cov3}), (\ref
{inv3}), then $E$ is given by eq.~(\ref{pos3}) for some probability measure $%
\rho $ in $\mathbb{R}^{3}$. If $\varphi \in C_{c}\left( \mathbb{R}%
^{3}\right) $, we define $E\left( \varphi \right) $ as in Remark \ref{Rem.
sulla rappr. di E come funz. su Cc}. For all $f\in L^{2}\left( \mathbb{R}%
^{3};\mathbb{C}^{2j+1}\right) $, define the measure 
\begin{equation*}
\text{d}\mu _{f}(\vec{x})=\left| f(\vec{x})\right| ^{2}\text{d}\vec{x}.
\end{equation*}
We then have 
\begin{equation*}
\left\langle f,E(\varphi ) f\right\rangle =\left( \mu _{f}\ast \rho \right)
(\varphi ).
\end{equation*}
From (\ref{rot3}) it then follows 
\begin{equation}
\left( \mu _{D(R)f}\ast \rho \right) (\varphi )=\left( \mu _{f}\ast \rho
\right) \left( \varphi ^{R^{-1}}\right) ,  \label{covmis}
\end{equation}
where $\varphi ^{R^{-1}}(\vec{x})=\varphi (R\vec{x})$. Rewriting explicitly (%
\ref{covmis}), we then find (since $D^{j}$ is unitary on $\mathbb{C}^{2j+1}$)
\begin{eqnarray*}
&&\int_{\mathbb{R}^{3}\times \mathbb{R}^{3}}\varphi (\vec{x}+\vec{y})\left|
f\left( R^{-1}\vec{x}\right) \right| ^{2}\text{d}\vec{x}\text{d}\rho (\vec{y}%
) \\
&& \qquad \quad =\int_{\mathbb{R}^{3}\times \mathbb{R}^{3}}\varphi ^{R^{-1}}\left( 
\vec{x}+\vec{y}\right) \left| f\left( \vec{x}\right) \right| ^{2}\text{d}%
\vec{x}\text{d}\rho (\vec{y}),
\end{eqnarray*}
With some computations, setting $\psi \left( \vec{x}\right) =\varphi \left(
-R\vec{x}\right) $, this gives 
\begin{equation}
\int_{\mathbb{R}^{3}}\left( \psi \ast \left| f\right| ^{2}\right) (-\vec{y})%
\text{d}\left( R^{-1}\cdot \rho \right) (\vec{y})=\int_{\mathbb{R}%
^{3}}\left( \psi \ast \left| f\right| ^{2}\right) (-\vec{y})\text{d}\rho (%
\vec{y})  \label{questa}
\end{equation}
(here $R^{-1}\cdot \rho $ denotes the measure on $\mathbb{R}^{3}$ such that $%
\left( R^{-1}\cdot \rho \right) \left( \phi \right) =\rho \left( \phi
^{R}\right) $ for all $\phi \in C_{c}\left( \mathbb{R}^{3}\right) $).
Letting $\psi $ and $\left| f\right| $ vary in $C_{c}\left( \mathbb{R}%
^{3}\right) $, the functions $\psi \ast \left| f\right| ^{2}$ span a dense
subset of $C_{0}\left( \mathbb{R}^{3}\right) $. From eq.~(\ref{questa}), it
then follows that $R^{-1}\cdot \rho =\rho $.
\end{proof}

\begin{proposition}
Let $E$ be a position observable on $\mathbb{R}^{3}$. The following facts
are equivalent:

\begin{itemize}
\item[(a)]  $\left\| E(U)\right\| =1$ for every nonempty open set $U\subset 
\mathbb{R}$;

\item[(b)]  $E$ is a sharp observablen ;

\item[(c)]  $E=\Pi _{\vec{Q}}$.
\end{itemize}
\end{proposition}

\begin{proof}
It is clear that (c) $\Longrightarrow $ (b) $\Longrightarrow $ (a). Hence,
it is enough to show that (a) implies (c). As in the proof of Proposition 
\ref{dilat}, it follows from (a) that $\rho =\delta _{\vec{t}}$ for some $%
\vec{t}\in \mathbb{R}^{3}$. However, the probability measure $\delta _{\vec{t%
}}$ is rotation invariant if and only if $\vec{t}=\vec{0}$. This means that $%
E=\Pi _{\vec{Q}}$.
\end{proof}

\chapter{Coexistence of position and momentum observables\label%
{Cap. sulla coesistenza}}

We now conclude the discussion we have begun in the previous chapter, and we
finally consider the joint observables of position and momentum, i.e.~those
observables defined in the phase space of the quantum system whose margins
are observables of position and momentum (see definition \ref{Def. di joint
obs.} below). The covariant phase space observables described in \S \ref
{Esempio spazio fasi in 1 dim.} are just a subclass of the whole set of
joint observables of position and momentum.

The problem of joint measurability of position and momentum observables has
a long history in quantum mechanics and different viewpoints have been
presented (for an overview of the subject, see e.g. \cite{BL84}). In this
chapter, we will show that the following remarkable fact holds: if a
position observable and a momentum observable admit a joint observable, then
they also admit a covariant phase space joint observable (Proposition \ref
{fcoex1}). From this fact, one can derive many properties of coexistent
position and momentum observables. For example, they must satisfy
Heisenberg's uncertainty relation, which can be restated in different forms
(see Proposition \ref{up} and Corollary \ref{fcoex2}).

Here is a brief synopsis of this chapter. In section \ref{Concepts} we
recall some concepts which are essential for our investigation. In section 
\ref{Joint1} we characterize those pairs of position and momentum
observables which are functionally coexistent and can thus be measured
jointly. Also some consequences on the properties of joint observables are
investigated. In Section \ref{Coexistence} we present an observation on the
general problem of coexistence of position and momentum observables.

The material in this chapter is contained in
\cite{Nuovo articolo con Teiko}.

\section{Coexistence and joint observables\label{Concepts}}

The notions of coexistence, functional coexistence and joint observables are
essential when joint measurability of quantum observables is analyzed. We
next shortly recall the definitions of these concepts. For further details
we refer to a convenient survey \cite{Lahti03} and to the references given
there.

\begin{definition}
Let $\left( \Omega _{i},\mathcal{A}\left( \Omega _{i}\right) \right) $, $%
i=1,2$, be measurable spaces and let $E_{i}:\mathcal{A}\left( \Omega
_{i}\right) \longrightarrow \mathcal{L}(\mathcal{H})$ be observables.

\begin{itemize}
\item[(i)]  $E_{1}$ and $E_{2}$ are \textbf{coexistent} if there is a
measurable space $\left( \Omega ,\mathcal{A}\left( \Omega \right) \right) $
and an observable $G:\mathcal{A}\left( \Omega \right) \longrightarrow 
\mathcal{L}(\mathcal{H})$ such that 
\begin{equation*}
\mathrm{ran}\,(E_{1})\cup \mathrm{ran}\,(E_{2})\subseteq \mathrm{ran}\,(G)%
\text{.}
\end{equation*}

\item[(ii)]  $E_{1}$ and $E_{2}$ are \textbf{functionally coexistent} if
there is a measurable space $\left( \Omega ,\mathcal{A}\left( \Omega \right)
\right) $, an observable $G:\mathcal{A}\left( \Omega \right) \longrightarrow 
\mathcal{L}(\mathcal{H})$ and measurable functions $f_{1}:\Omega
\longrightarrow \Omega _{1}$, $f_{2}:\Omega \longrightarrow \Omega _{2}$,
such that, for any $X_{1}\in \mathcal{A}\left( \Omega _{1}\right) $, $%
X_{2}\in \mathcal{A}\left( \Omega _{2}\right) $, 
\begin{equation*}
E_{1}(X_{1})=G(f_{1}^{-1}(X_{1}))\text{,\qquad }%
E_{2}(X_{2})=G(f_{2}^{-1}(X_{2}))\text{.}
\end{equation*}
\end{itemize}
\end{definition}

Functionally coexistent observables are clearly coexistent, but it is an
open question if the reverse holds.

We now confine our discussion to observables on $\mathbb{R}$.

\begin{definition}
\label{Def. di joint obs.}Let $E_{1},E_{2}:\mathcal{B}(\mathbb{R}%
)\rightarrow \mathcal{L}(\mathcal{H})$ be observables. An observable $G:%
\mathcal{B}(\mathbb{R}^{2})\rightarrow \mathcal{L}(\mathcal{H})$ is their 
\textbf{joint observable} if for all $X,Y\in \mathcal{B}(\mathbb{R})$, 
\begin{eqnarray*}
E_{1}(X) &=&G(X\times \mathbb{R}), \\
E_{2}(Y) &=&G(\mathbb{R}\times Y).
\end{eqnarray*}
In this case $E_{1}$ and $E_{2}$ are the \textbf{margins} of $G$.
\end{definition}

For observables $E_1$ and $E_2$ defined on $\mathcal{B}(\mathbb{R})$ the
existence of a joint observable is equivalent to their functional
coexistence. These conditions are also equivalent to the \emph{joint
measurability} of $E_1$ and $E_2$ in the sense of the quantum measurement
theory (see Section 7 in \cite{Lahti03}).

The \textbf{commutation domain} of observables $E_{1}$ and $E_{2}$, denoted
by $\mathrm{com}\,(E_{1},E_{2})$, is the closed subspace of $\mathcal{H}$
defined as 
\begin{equation*}
\mathrm{com}\,(E_{1},E_{2})=\left\{ \psi \in \mathcal{H}\mid
E_{1}(X)E_{2}(Y)\psi -E_{2}(Y)E_{1}(X)\psi =0\ \forall X,Y\in \mathcal{B}(%
\mathbb{R})\right\} \text{.}
\end{equation*}
If $E_{1}$ and $E_{2}$ are sharp observables, then $E_{1}$ and $E_{2}$ are
coexistent if and only if they are functionally coexistent and this is the
case exactly when $\mathrm{com}\,(E_{1},E_{2})=\mathcal{H}$. In general, for
two observables $E_{1}$ and $E_{2}$ the condition $\mathrm{com}%
\,(E_{1},E_{2})=\mathcal{H}$ is sufficient but not necessary for the
functional coexistence of $E_{1}$ and $E_{2}$.

In conclusion, given a pair of observables on $\mathbb{R}$ one may pose the
questions of their commutativity, functional coexistence, and coexistence,
in the order of increasing generality.

For position and momentum observables, the following known fact holds \cite
{BL89}, which will be needed later. For completeness we give a proof of it.

\begin{proposition}
\label{noncommu}A position observable $E_{\rho }$ and a momentum observable $%
F_{\nu }$ are totally noncommutative, that is, $\mathrm{com}\,(E_{\rho
},F_{\nu })=\{0\}$.
\end{proposition}

\begin{proof}
It is shown in \cite{BSS87} and \cite{Ylinen89} that for functions $f,g\in
L^{\infty }(\mathbb{R})$ the equation 
\begin{equation*}
f(Q)g(P)-g(P)f(Q)=O
\end{equation*}
holds if and only if one of the following is satisfied: (i) either $f(Q)$ or 
$g(P)$ is a multiple of the identity operator, (ii) $f$ and $g$ are both
periodic with minimal periods $a,b$ satisfying $2\pi /ab\in \mathbb{Z}%
\setminus \{0\}$.

Let $X\subset \mathbb{R}$ be a bounded interval. Then the operators $E_{\rho
}(X)$ and $F_{\nu }(X)$ are not multiples of the identity operator. Indeed,
let us assume, in contrary, that $E_{\rho }(X)=cI$ for some $c\in \mathbb{R}$%
. Denote $a=2\mu (X)$. Then the sets $X+na$, $n\in \mathbb{Z}$, are
pairwisely disjoint and 
\begin{eqnarray*}
I &\geq &E_{\rho }(\cup _{n\in \mathbb{Z}}(X+na))=\sum_{n\in \mathbb{Z}%
}E_{\rho }(X+na) \\
&=&\sum_{n\in \mathbb{Z}}U(na)E_{\rho }(X)U(na)^{\ast }=\sum_{n\in \mathbb{Z}%
}cI.
\end{eqnarray*}
This means that $c=0$. However, since $\mu (X)>0$, we have 
\begin{equation*}
O\neq E_{\rho }(X)=cI=O.
\end{equation*}
Thus, $E_{\rho }(X)$ is not a multiple of the identity operator. Moreover,
since $\rho (\mathbb{R})=1$, the function $q\mapsto \rho (X-q)$ is not
periodic. We conclude that, by the above mentioned result, the operators $%
E_{\rho }(X)$ and $F_{\nu }(X)$ do not commute and hence, $\mathrm{com}%
\,(E_{\rho },F_{\nu })\neq \mathcal{H}$.

Assume then that there exists $\psi \neq 0$, $\psi \in \mathrm{com}%
\,(E_{\rho },F_{\nu })$. Using the symmetry properties (\ref{cov}), (\ref
{inv1}), (\ref{invF}) and (\ref{covF}), a short calculation shows that for
any $q,p\in \mathbb{R}$, $U(q)V(p)\psi \in \mathrm{com}\,(E_{\rho },F_{\nu
}) $. This implies that $\mathrm{com}\,(E_{\rho },F_{\nu })$ is invariant
under the irreducible projective representation $W$ defined in (\ref{Def.
della rappr. W}) (recall that $W\left( q,p\right) =e^{iqp/2}U\left( q\right)
V\left( p\right) $). It follows that we have either $\mathrm{com}\,(E_{\rho },F_{\nu
})=\{0\}$ or $\mathrm{com}\,(E_{\rho },F_{\nu })=\mathcal{H}$. Since the
latter possibility is ruled out, this completes the proof.
\end{proof}

\section{Joint observables of position and momentum\label{Joint1}}

Looking at the symmetry conditions (\ref{cov}), (\ref{inv1}), (\ref{invF}), (%
\ref{covF}), and recalling the definition (\ref{Def. della rappr. W}) of the
projective unitary representation $W$ of the group of phase space
translations, it is clear that an observable $G:\mathcal{B}(\mathbb{R}%
^{2})\rightarrow \mathcal{L}(\mathcal{H})$ has a position observable and a
momentum observable as its margins if and only if, for all $q,p\in \mathbb{R}
$ and $X,Y\in \mathcal{B}(\mathbb{R})$, the following conditions hold: 
\begin{eqnarray}
W(q,p)G(X\times \mathbb{R})W(q,p)^{\ast } &=&G(X\times \mathbb{R}+(q,p)),
\label{WW1} \\
W(q,p)G(\mathbb{R}\times Y)W(q,p)^{\ast } &=&G(\mathbb{R}\times Y+(q,p)).
\label{WW2}
\end{eqnarray}
(recall that $W\left( q,p\right) =$ $e^{iqp/2}U\left( q\right) V\left(
p\right) $).

On the other hand, a covariant phase space observable is a POM $G:\mathcal{B}%
(\mathbb{R}^{2})\rightarrow \mathcal{L}(\mathcal{H})$ such that, for all $%
q,p\in \mathbb{R}$ and $Z\in \mathcal{B}(\mathbb{R}^{2})$, 
\begin{equation}
W(q,p)G(Z)W(q,p)^{\ast }=G(Z+(q,p))\text{,}  \label{W}
\end{equation}
and it is trivial that (\ref{W}) implies (\ref{WW1}) and (\ref{WW2}). Hence,
a covariant phase space observable is a joint observable of some position
and momentum observables. To our knowledge, it is an open question whether (%
\ref{WW1}) and (\ref{WW2}) imply (\ref{W}).

As proved in \S\ref{Esempio spazio fasi in 1 dim.}, each covariant phase
space observable is of the form $G=G_{T}$, with 
\begin{equation*}
G_{T}(Z)=\frac{1}{2\pi }\int_{Z}W(q,p)TW(q,p)^{\ast }\ \text{d}q\text{d}%
p\quad Z\in \mathcal{B}(\mathbb{R}^{2})\text{,}
\end{equation*}
for some $T\in \mathcal{S}(\mathcal{H})$. Moreover, the correspondence $%
T\longmapsto G_{T}$ is injective from $\mathcal{S}(\mathcal{H})$ onto the
set of the covariant phase space observables (see Theorem \ref{Teo. centr.1}%
).

Let $G_{T}$ be a covariant phase space observable and let $\sum_{n}\lambda
_{n}\langle \cdot ,\varphi _{n}\rangle \varphi _{n}$ be the spectral
decomposition of the state $T$. With an easy computation, one finds that the
margins of $G_{T}$ are the position observable $E_{\rho }$ and a the
momentum observable $F_{\nu }$ with 
\begin{eqnarray}
&&\text{d}\rho (q)=e(q)\text{d}q,\quad e(q)=\sum\nolimits_{n}\lambda
_{n}|\varphi _{n}(-q)|^{2},  \label{marginrho} \\
&&\text{d}\nu (p)=f(p)\text{d}p,\quad f(p)=\sum\nolimits_{n}\lambda _{n}|%
\widehat{\varphi _{n}}(-p)|^{2}.  \label{marginnu}
\end{eqnarray}

The form of $\rho $ and $\nu $ in (\ref{marginrho}) and (\ref{marginnu})
implies that, in general, the margins $E_{\rho }$ and $F_{\nu }$ do not
determine $G_{T}$, that is, another covariant phase space observable $%
G_{T^{\prime }}$ may have the same margins. Indeed, the functions $|\varphi
(\cdot )|$ and $|\hat{\varphi}(\cdot )|$ do not define the vector $\varphi $
uniquely up to a phase factor. Here we provide an example in which this
phenomenon occurs.

\begin{example}
\label{sopra1}Consider the functions 
\begin{equation*}
\varphi _{a,b}\left( q\right) =\left( \frac{2a}{\pi }\right)
^{1/4}e^{-\left( a+ib\right) q^{2}},
\end{equation*}
with $a,b\in \mathbb{R}$ and $a>0$. The Fourier transform of $\varphi _{a,b}$
is 
\begin{eqnarray*}
\hat{\varphi}_{a,b}\left( p\right)  &=&\left( \frac{a}{2\pi \left(
a^{2}+b^{2}\right) }\right) ^{1/4}\exp \left( -\frac{ap^{2}}{4\left(
a^{2}+b^{2}\right) }\right)  \\
&&\times \exp \left( \frac{ibp^{2}}{4\left( a^{2}+b^{2}\right) }-\frac{i}{2}%
\arctan \frac{b}{a}\right) .
\end{eqnarray*}
For $b\neq 0$, we see that $T_{1}=\langle \varphi _{a,b}, \cdot \rangle
\varphi _{a,b}$ and $T_{2}=\langle \varphi _{a,-b},\cdot \rangle \varphi
_{a,-b}$ are different, but the margins of $G_{T_{1}}$ and $G_{T_{2}}$ are
the same position and momentum observables $E_{\rho }$ and $F_{\nu }$, with 
\begin{eqnarray*}
\text{d}\rho (q) &=&\left( \frac{2a}{\pi }\right) ^{1/2}e^{-2aq^{2}}\text{d}q
\\
\text{d}\nu (p) &=&\left( \frac{a}{2\pi \left( a^{2}+b^{2}\right) }\right)
^{1/2}\exp \left( -\frac{ap^{2}}{2\left( a^{2}+b^{2}\right) }\right) \text{d}%
p
\end{eqnarray*}
\end{example}

As $\rho $ and $\nu $ in (\ref{marginrho}) and (\ref{marginnu}) arise from
the same state $T$, a multitude of uncertainty relations can be derived for
the observables $E_{\rho }$ and $F_{\nu }$. One of the most common
uncertainty relation is in terms of variances. Namely, let $\mathrm{Var}%
\,(p) $ denote the variance of a probability measure $p$, 
\begin{equation*}
\mathrm{Var}\,(p)=\int \left( x-\int x\text{d}p(x)\right) ^{2}\text{d}p(x).
\end{equation*}
Then for any state $S$, 
\begin{equation}
\mathrm{Var}\,(p_{S}^{E_{\rho }})\mathrm{Var}\,(p_{S}^{F_{\nu }})\geq 1
\label{ur}
\end{equation}
(see e.g. \cite{Lahti95}, Section III.2.4 or \cite{CRQM}, Section 5.4.) The
lower bound in (\ref{ur}) can be achieved only if 
\begin{equation}
\mathrm{Var}\,(\rho )\mathrm{Var}\,(\nu )=\frac{1}{4},  \label{min}
\end{equation}
and it is well known (\cite{Heisenberg}, \cite{Il libro che ci aveva detto
Teiko}) that (\ref{min}) holds if and only if $T=\langle \cdot ,\varphi
\rangle \varphi $ and $\varphi $ is a function of the form 
\begin{equation*}
\varphi \left( q\right) =\left( 2a/\pi \right) ^{1/4}e^{ibq}e^{-a\left(
q-c\right) ^{2}},\quad a>0,\ b,c\in \mathbb{R}.
\end{equation*}
It is also easily verified that, if $T$ is as above, choosing $S=T$ the
equality in (\ref{ur}) is indeed obtained.

The following proposition is the main result in this chapter, since it gives
a complete characterisation of jointly measurable position and momentum
observables.

\begin{proposition}
\label{fcoex1}Let $E_{\rho }$ be a position observable and $F_{\nu }$ a
momentum observable. If $E_{\rho }$ and $F_{\nu }$ have a joint observable,
then they also have a joint observable which is a covariant phase space
observable.
\end{proposition}

The proof of Proposition \ref{fcoex1} is given in section \ref{proof2},
since it needs some notations and the introduction of some additional
mathematical concepts. It is a rearrangement of a similar proof given by
Werner in \cite{Werner04}.

We now describe some consequences of Proposition \ref{fcoex1}. Corollaries 
\ref{fcoex2} and \ref{up} are two different restatements of Heisenberg
uncertainty relation. Corollary \ref{up} is Proposition 6 of \cite{CHT04}
slightly rearranged.

\begin{corollary}
\label{fcoex2} A position observable $E_{\rho }$ and a momentum observable $%
F_{\nu }$ are functionally coexistent if and only if there is a state $T\in 
\mathcal{S}(\mathcal{H})$ such that $\rho $ and $\nu $ are given by (\ref
{marginrho}) and (\ref{marginnu}). Especially, the uncertainty relation (\ref
{ur}) is a necessary condition for the functional coexistence, and thus for
the joint measurability of $E_{\rho }$ and $E_{\nu }$.
\end{corollary}

\begin{corollary}
\label{up} Let $E_{\rho }$ be a position observable and $F_{\nu }$ a
momentum observable. If $E_{\rho }$ and $F_{\nu }$ are functionally
coexistent, then the product of the respective limits of resolutions
satisfies the inequality 
\begin{equation}
\gamma _{E_{\rho }}\cdot \gamma _{F_{\nu }}\geq 3-2\sqrt{2}.  \label{ineq_up}
\end{equation}
\end{corollary}

\begin{proof}
Since $E_{\rho }$ and $F_{\nu }$ have a covariant phase space observable as
a joint observable, there is a vector valued function $\theta \in L^{2}(%
\mathbb{R},\mathcal{H})$ such that d$\rho (q)=\Vert \theta (q)\Vert _{%
\mathcal{H}}^{2}$d$q$ and d$\nu (p)=\Vert \hat{\theta}(p)\Vert _{\mathcal{H}%
}^{2}$d$p$. In fact, let $\left( f_{n}\right) _{n\geq 1}$ be an orthonormal
basis of $\mathcal{H}$. Then, with the notations of eqs.~(\ref{marginrho})
and (\ref{marginnu}), the function $\theta $ is 
\begin{equation*}
\theta \left( q\right) =\sum\nolimits_{n}\lambda _{n}^{1/2}\varphi
_{n}(-q)f_{n}\text{.}
\end{equation*}

By Proposition \ref{reg2} the observable $E_{\rho }$ is $\alpha $-regular if
and only if 
\begin{equation*}
\mathrm{ess\,sup}_{x\in \mathbb{R}}\rho \left( I_{x;\alpha }\right) >1/2.
\end{equation*}
Since the map $x\longmapsto \rho \left( I_{x;\alpha }\right) $ is
continuous, this is equivalent to 
\begin{equation*}
\mathrm{sup}_{x\in \mathbb{R}}\rho \left( I_{x;\alpha }\right) =\mathrm{sup}%
_{x\in \mathbb{R}}\int_{I_{x;\alpha }}\left\| \theta (x)\right\| _{\mathcal{H%
}}^{2}\text{d}x>1/2.
\end{equation*}
By the same argument, $F_{\nu }$ is $\beta $-regular if and only if 
\begin{equation*}
\mathrm{sup}_{\xi \in \mathbb{R}}\nu \left( I_{\xi ;\beta }\right) =\mathrm{%
sup}_{\xi \in \mathbb{R}}\int_{I_{\xi ;\beta }}\Vert \hat{\theta}(\xi )\Vert
_{\mathcal{H}}^{2}\text{d}\xi >1/2.
\end{equation*}
Using \cite[Theorem 2]{DonSta} extended to the case of vector valued
functions, we find 
\begin{equation*}
\alpha \cdot \beta \geq 3-2\sqrt{2},
\end{equation*}
and hence (\ref{ineq_up}) follows.
\end{proof}

\begin{corollary}
If a position observable $E_{\rho }$ and a momentum observable $F_{\nu }$
are functionally coexistent, then neither $E_{\rho }$ nor $F_{\nu }$ are
regular.
\end{corollary}

\begin{proof}
This is an immediate consequence of Corollary \ref{fcoex2} and of item (iii)
in Proposition \ref{limit0}.
\end{proof}

We end this section with an observation about a (lacking) localization
property of a joint observable of position and momentum observables. We wish
to emphasize again that $G$ in Proposition \ref{Gprop} is not assumed to be
a covariant phase space observable.

\begin{proposition}
\label{Gprop}Let $G$ be a joint observable of a position observable $E_{\rho
}$ and a momentum observable $F_{\nu }$ and let $Z\in \mathcal{B}(\mathbb{R}%
^{2})$ be a bounded set. Then

\begin{itemize}
\item[(i)]  $\left\| G(Z)\right\| \neq 1$;

\item[(ii)]  there exists a number $k_{Z}<1$ such that for any $T\in 
\mathcal{S}(\mathcal{H})$, $p_{T}^{G}(Z)\leq k_{Z}$.
\end{itemize}
\end{proposition}

\begin{proof}
(i) It follows from Proposition \ref{fcoex1} and the Paley-Wiener Theorem
that either $\rho $ or $\nu $ has unbounded support. Let us assume that, for
instance, $\rho $ has unbounded support.

Let $Z\in \mathcal{B}(\mathbb{R}^{2})$ be a bounded set. Then the closure $%
\bar{Z}$ is compact and also the set 
\begin{equation*}
X:=\{x\in \mathbb{R}\mid \exists y\in \mathbb{R}:(x,y)\in \bar{Z}\}\subset 
\mathbb{R}
\end{equation*}
is compact. Since 
\begin{equation*}
\left\| G(Z)\right\| \leq \left\| G(X\times \mathbb{R})\right\| =\left\|
E_{\rho }(X)\right\|
\end{equation*}
and 
\begin{equation*}
\left\| E_{\rho }(X)\right\| =\mathrm{ess\,sup}_{x\in \mathbb{R}}\rho
(X-x)\leq \mathrm{sup}_{x\in \mathbb{R}}\rho (X-x),
\end{equation*}
it is enough to show that 
\begin{equation}
\mathrm{sup}_{x\in \mathbb{R}}\rho (X-x)<1.  \label{sup1}
\end{equation}
Let us suppose, in contrary, that 
\begin{equation}
\mathrm{sup}_{x\in \mathbb{R}}\rho (X-x)=1  \label{sup2}
\end{equation}
This means that there exists a sequence $(x_{n})_{n\geq 1}\subset \mathbb{R}$
such that 
\begin{equation}
\lim_{n\rightarrow \infty }\rho (X-x_{n})=1.  \label{lim}
\end{equation}
Since $\rho (\mathbb{R})=1$ and $X$ is a bounded set, the sequence $%
(x_{n})_{n\geq 1}$ is also bounded. It follows that $B:=\bigcup_{n=1}^{%
\infty }\ X-x_{n}$ is a bounded set and by (\ref{lim}) we have $\rho (B)=1$.
This is in contradiction with the assumption that $\rho $ has an unbounded
support. Hence, (\ref{sup2}) is false and (\ref{sup1}) follows.

(ii) From (i) it follows that 
\begin{equation*}
1>k_{Z}:=\left\| G(Z)\right\| =\sup \{\left\langle \psi ,G(Z)\psi
\right\rangle \mid \psi \in \mathcal{H},\left\| \psi \right\| =1\}.
\end{equation*}
Let $T\in \mathcal{S}(\mathcal{H})$ and let $\sum_{i}\lambda _{i}\langle
\psi _{i},\cdot \rangle \psi _{i}$ be the spectral decomposition of $T$.
Then 
\begin{equation*}
p_{T}^{G}(Z)=\mathrm{tr}\,[TG(Z)]=\sum_{i}\lambda _{i}\left\langle \psi
_{i},G(Z)\psi _{i}\right\rangle \leq k_{Z}.
\end{equation*}
\end{proof}

\section{Coexistence of position and momentum observables\label{Coexistence}}

Since coexistence is, a priori, a more general concept than functional
coexistence, we are still left with the problem of characterizing coexistent
pairs of position and momentum observables. In lack of a general result we
close our investigation with some observations on this problem.

\begin{proposition}
\label{projection} Let $E_{\rho }$ be a position observable and $F_{\nu }$ a
momentum observable. If $\mathrm{com}\,(E_{\rho })\cup \mathrm{com}\,(F_{\nu
})$ contains a nontrivial projection (not equal to $O$ or $I$), then $%
E_{\rho }$ and $F_{\nu }$ are not coexistent.
\end{proposition}

\begin{proof}
Let us assume, in contrary, that there exists an observable $G$ such that $%
\mathrm{com}\,(E_{\rho })\cup \mathrm{com}\,(F_{\nu })\subseteq \mathrm{com}%
\,(G)$. Suppose, for instance, that $E_{\rho }(X)$ is a nontrivial
projection. Then $E_{\rho }(X)$ commutes with all operators in the range of $%
G$ (see e.g. \cite{LP01}). In particular, $E_{\rho }(X)$ commutes with all $%
F_{\nu }(X)$, $X\in \mathcal{B}(\mathbb{R})$. However, this is impossible by
the result proved in \cite{BSS87} and \cite{Ylinen89} (see also the
beginning of the proof of Proposition \ref{noncommu}).
\end{proof}

\begin{corollary}
If $E$ and $F$ are position and momentum observables, and either one of them
is sharp, then $E$ and $F$ are not coexistent.
\end{corollary}

\section*{Proof of Proposition \ref{fcoex1}\label{proof2}}

In order to prove Proposition \ref{fcoex1} we need some general results
about means on topological spaces, and for readers convenience they are
briefly reviewed. The following material is based on \cite{AHAI},
Chapter~IV, \S 17, and \cite{Werner04}.

Let $\Omega$ be a locally compact separable metric space with a metric $d$.
By $BC\left( \Omega\right) $ we denote the Banach space of complex valued
bounded continuous functions on $\Omega$, with the uniform norm $\left\|
f\right\|_{\infty }=\sup_{x\in \Omega}\left| f\left( x\right) \right|$. The
linear subspace of continuous functions with compact support is denoted by $%
C_{c}\left( \Omega\right)$. Adding the index $^{r}$ we denote the subsets of
real functions in $BC\left(\Omega\right) $ or in $C_{c}\left( \Omega\right) $%
. With the index $^{+}$ we denote the subsets of positive functions. 

\begin{definition}
A \emph{mean} on $\Omega $ is a linear functional 
\begin{equation*}
m:BC\left( \Omega \right) \longrightarrow \mathbb{C}
\end{equation*}
such that:

\begin{itemize}
\item[(i)]  $m\left( f\right) \geq 0$ if $f\in BC^{+}\left( \Omega \right) $;

\item[(ii)]  $m\left( 1\right) =1$.
\end{itemize}

For a mean $m$ on $\Omega $ we denote 
\begin{equation*}
m\left( \infty \right) =1-\sup \left\{ m\left( f\right) \mid f\in
C_{c}^{+}\left( \Omega \right) ,\ f\leq 1\right\} .
\end{equation*}
\end{definition}

Let $m$ be a mean on $\Omega $. By the Riesz Representation Theorem, there
exists a unique positive Borel measure $m_{0}$ on $\Omega $ such that 
\begin{equation*}
m\left( f\right) =\int_{\Omega }f\left( x\right) \text{d}m_{0}\left(
x\right) \quad \forall f\in C_{c}\left( \Omega \right) .
\end{equation*}
By inner regularity of $m_{0}$ we have 
\begin{equation*}
m_{0}\left( \Omega \right) =\sup \left\{ m\left( f\right) \mid f\in
C_{c}^{+}\left( \Omega \right) ,\ f\leq 1\right\} =1-m\left( \infty \right)
\leq 1.
\end{equation*}
Especially, any function in $BC\left( \Omega \right) $ is integrable with
respect to $m_{0}$. For any $f\in BC\left( \Omega \right) $, we use the
abbreviated notation 
\begin{equation*}
m_{0}\left( f\right) :=\int_{\Omega }f\left( x\right) \text{d}m_{0}\left(
x\right) .
\end{equation*}

\begin{proposition}
\label{minfty} If $m\left( \infty \right) =0$, then 
\begin{equation*}
m\left( f\right) =m_{0}\left( f\right) \quad \forall f\in BC\left( \Omega
\right) .
\end{equation*}
\end{proposition}

\begin{proof}
We fix a point $x_{0}\in \Omega $. For all $R>0$ we define 
\begin{equation*}
g_{R}\left( x\right) =\left\{ 
\begin{array}{ccc}
1 & \mbox{if} & d\left( x_{0},x\right) \leq R/2, \\ 
3/2-d\left( x_{0},x\right) /R & \mbox{if} & R/2<d\left( x_{0},x\right) \leq
3R/2, \\ 
0 & \mbox{if} & d\left( x_{0},x\right) >3R/2.
\end{array}
\right.
\end{equation*}
Then $g_{R}\in C_{c}^{+}\left( \Omega \right) $ and $g_{R}\leq 1$. Moreover,
for any $f\in C_{c}^{+}\left( \Omega \right) $ such that $f\leq 1$ there
exists $R>0$ such that $f\leq g_{R}$, and hence 
\begin{equation*}
1=\sup \left\{ m\left( f\right) \mid f\in C_{c}^{+}\left( \Omega \right) ,\
f\leq 1\right\} =\lim_{R\rightarrow \infty }m\left( g_{R}\right) .
\end{equation*}

Let $f\in BC^{+}\left( \Omega \right) $ and $R>0$. Since $g_{R}f\in
C_{c}\left( \Omega \right) $, we have 
\begin{equation}
m\left( f\right) =m_{0}\left( g_{R}f\right) +m\left( \left( 1-g_{R}\right)
f\right) .  \label{uno}
\end{equation}
We have $0\leq g_{R}f\leq f$, $f$ is $m_{0}$-integrable and $%
\lim_{R\rightarrow \infty }g_{R}\left( x\right) f\left( x\right) =f\left(
x\right) $ for all $x\in \Omega $. Therefore, by the dominated convergence
theorem we have 
\begin{equation*}
\lim_{R\rightarrow \infty }\int_{\Omega }g_{R}\left( x\right) f\left(
x\right) \text{d}m_{0}\left( x\right) =\int_{\Omega }f\left( x\right) \text{d%
}m_{0}\left( x\right) .
\end{equation*}
For the other term in the sum (\ref{uno}), we have 
\begin{equation*}
m\left( \left( 1-g_{R}\right) f\right) \leq \left\| f\right\| _{\infty
}m\left( 1-g_{R}\right) \underset{R\rightarrow \infty }{\longrightarrow }%
\left\| f\right\| _{\infty }m\left( \infty \right) =0.
\end{equation*}
Taking the limit $R\rightarrow \infty $ in (\ref{uno}) we then get 
\begin{equation*}
m\left( f\right) =m_{0}\left( f\right) .
\end{equation*}

If $f\in BC\left( \Omega \right) $, we write $f=f_{1}+if_{2}$ with $%
f_{1},f_{2}\in BC^{r}\left( \Omega \right) $, and $f_{i}=f_{i}^{+}-f_{i}^{-}$
with $f_{i}^{\pm }=\frac{1}{2}\left( \left| f_{i}\right| \pm f_{i}\right)
\in BC^{+}\left( \Omega \right) $, and we use the previous result to obtain
the conclusion.
\end{proof}

Let $i\in \left\{ 1,2\right\}$. For $f\in BC\left( \Omega\right)$ we define 
\begin{equation*}
\widetilde{f}_{i}\left( x_{1},x_{2}\right) := f\left( x_{i}\right)\quad
\forall x_{1},x_{2}\in \Omega.
\end{equation*}
Clearly, $\widetilde{f}_{i}\in BC\left( \Omega\times \Omega\right) $. For a
mean $m : BC\left( \Omega\times \Omega\right) \longrightarrow \mathbb{C}$,
we then define 
\begin{equation*}
m _{i}\left( f\right) :=m \left( \widetilde{f}_{i}\right) \quad \forall f\in
BC\left( \Omega\right) .
\end{equation*}
The linear functional $m_{i}:$ $BC\left( \Omega\right) \longrightarrow 
\mathbb{C}$ is a mean on $\Omega$, which we call the $i$th \emph{margin} of $%
m$.

\begin{proposition}
\label{mminfty} Let $m$ be a mean on $\Omega \times \Omega $. If $%
m_{1}\left( \infty \right) =m_{2}\left( \infty \right) =0$, then $m\left(
\infty \right) =0$.
\end{proposition}

\begin{proof}
For all $R>0$, we define the function $g_{R}\in C_{c}\left( \Omega \right) $
as in the proof of Proposition~\ref{minfty}. We set 
\begin{equation*}
h_{R}\left( x_{1},x_{2}\right) =g_{R}\left( x_{1}\right) g_{R}\left(
x_{2}\right) .
\end{equation*}
Clearly, $h_{R}\in C_{c}^{+}\left( \Omega \times \Omega \right) $, and, if $%
h\in C_{c}^{+}\left( \Omega \times \Omega \right) $ and $h\leq 1$, there
exists $R>0$ such that $h\leq h_{R}$. Since 
\begin{eqnarray*}
1-h_{R}\left( x_{1},x_{2}\right) &=&\left( 1-g_{R}\left( x_{1}\right)
\right) +g_{R}\left( x_{1}\right) \left( 1-g_{R}\left( x_{2}\right) \right)
\\
&\leq &\left( 1-g_{R}\left( x_{1}\right) \right) +\left( 1-g_{R}\left(
x_{2}\right) \right) ,
\end{eqnarray*}
we have 
\begin{equation*}
m\left( 1-h_{R}\right) \leq m_{1}\left( 1-g_{R}\right) +m_{2}\left(
1-g_{R}\right) ,
\end{equation*}
and the thesis follows from 
\begin{eqnarray*}
m\left( \infty \right) &=&1-\lim_{R\rightarrow \infty }m\left( h_{R}\right)
\leq \lim_{R\rightarrow \infty }m_{1}\left( 1-g_{R}\right)
+\lim_{R\rightarrow \infty }m_{2}\left( 1-g_{R}\right) \\
&=&m_{1}\left( \infty \right) +m_{2}\left( \infty \right) =0.
\end{eqnarray*}
\end{proof}

For a positive Borel measure $m_{0}$ on $\Omega\times \Omega$, we denote by $%
\left( m_{0} \right)_{i}$, $i=1,2$, the two measures on $\Omega$ which are
margins of $m_{0}$.

\begin{proposition}
\label{m0i} Let $m$ be a mean on $\Omega \times \Omega $. If $m\left( \infty
\right) =0$, then $\left( m_{0}\right) _{i}=\left( m_{i}\right) _{0}$.
\end{proposition}

\begin{proof}
Let $f\in BC\left( \Omega \right) $. By Proposition~\ref{minfty} we have 
\begin{equation*}
m_{0}\left( \widetilde{f}_{i}\right) =m\left( \widetilde{f}_{i}\right) .
\end{equation*}
Using this equality and the definitions of $(m_{0})_{i}$ and $(m_{i})_{0}$
we get 
\begin{equation*}
\left( m_{0}\right) _{i}\left( f\right) =m_{0}\left( \widetilde{f}%
_{i}\right) =m\left( \widetilde{f}_{i}\right) =m_{i}\left( f\right) =\left(
m_{i}\right) _{0}\left( f\right) .
\end{equation*}
\end{proof}

\begin{definition}
\label{ovm} An \emph{operator valued mean} on $\Omega $ is a linear
functional 
\begin{equation*}
M:BC\left( \Omega \right) \longrightarrow \mathcal{L}\left( \mathcal{H}%
\right)
\end{equation*}
such that:

\begin{itemize}
\item[(i)]  $M\left( f\right) \geq O$ if $f\in BC^{+}\left( \Omega \right) $;

\item[(ii)]  $M\left( 1\right) =I$.
\end{itemize}

For an operator valued mean $M$ on $\Omega $ we denote 
\begin{equation*}
M\left( \infty \right) =I-\mathrm{LUB}\,\left\{ M\left( f\right) \mid f\in
C_{c}^{+}\left( \Omega \right) ,\ f\leq 1\right\} .
\end{equation*}
\end{definition}

The least upper bound in Definition \ref{ovm} exists by virtue of
Proposition~1 in \cite{NST}.

Let $M$ be an operator valued mean on $\Omega $. For each $f\in BC^{r}\left(
\Omega \right) $, we have 
\begin{equation*}
M\left( f-\left\| f\right\| _{\infty }1\right) \leq O,\quad M\left(
f+\left\| f\right\| _{\infty }1\right) \geq O.
\end{equation*}
It follows that 
\begin{equation*}
\left\| M\left( f\right) \right\| \leq \left\| f\right\| _{\infty }.
\end{equation*}
By Theorem~19 in \cite{NST}, there exists a unique positive operator measure 
$M_{0}$ on $\Omega $ such that 
\begin{equation*}
M\left( f\right) =\int_{\Omega }f\left( x\right) \text{d}M_{0}\left(
x\right) \quad \forall f\in C_{c}\left( \Omega \right) ,
\end{equation*}
where the integral is understood in the weak sense. Similarly to the scalar
case we have 
\begin{equation}
M_{0}\left( \Omega \right) =I-M\left( \infty \right) \leq I,  \label{M0}
\end{equation}
and, for any $f\in BC\left( \Omega \right) $ we define 
\begin{equation*}
M_{0}\left( f\right) :=\int_{\Omega }f\left( x\right) \text{d}M_{0}\left(
x\right) .
\end{equation*}

Given an operator valued mean $M$ on $\Omega$ and a unit vector $\psi\in 
\mathcal{H}$, we set 
\begin{equation*}
m_{\psi}\left( f\right) := \left\langle \psi, M\left( f\right) \psi
\right\rangle \quad \forall f\in BC\left( \Omega\right) .
\end{equation*}
It is clear that $m_{\psi}$ is a mean on $\Omega$. By Proposition~1 in \cite
{NST}, 
\begin{equation*}
m_{\psi}\left( \infty \right) = \left\langle \psi , M\left( \infty \right) 
\psi \right\rangle .
\end{equation*}

\begin{proposition}
\label{Minfty} If $M\left( \infty \right) =O$, then 
\begin{equation*}
M\left( f\right) =M_{0}\left( f\right) \quad \forall f\in BC\left( \Omega
\right) .
\end{equation*}
\end{proposition}

\begin{proof}
For a unit vector $\psi \in \mathcal{H}$ and a function $f\in C_{c}\left(
\Omega \right) $, we have by definitions 
\begin{equation*}
\left( m_{\psi }\right) _{0}\left( f\right) =\left\langle \psi ,M_{0}\left(
f\right) \psi \right\rangle ,
\end{equation*}
and this equality is valid also for any $f\in BC\left( \Omega \right) $.
Since 
\begin{equation*}
m_{\psi }\left( \infty \right) =\left\langle \psi ,M\left( \infty \right)
\psi \right\rangle =0,
\end{equation*}
it follows from Proposition \ref{minfty} that the functional $m_{\psi }$ on $%
BC\left( \Omega \right) $ coincides with integration with respect to the
measure $\left( m_{\psi }\right) _{0}$. If $f\in BC\left( \Omega \right) $,
we then have 
\begin{equation*}
\left\langle \psi ,M_{0}\left( f\right) \psi \right\rangle =\left( m_{\psi
}\right) _{0}\left( f\right) =m_{\psi }\left( f\right) =\left\langle \psi ,
M\left(f\right) \psi \right\rangle ,
\end{equation*}
and the thesis follows.
\end{proof}

The margins $M_1$ and $M_2$ of an operator valued mean $M$ on $%
\Omega\times\Omega$ are defined in an analogous way as in the case of scalar
means.

\begin{proposition}
\label{MM} Let $M$ be an operator valued mean on $\Omega \times \Omega $.

\begin{itemize}
\item[(i)]  If $M_{1}\left( \infty \right) =M_{2}\left( \infty \right) =O$,
then $M\left( \infty \right) =O$;

\item[(ii)]  If $M\left( \infty \right) =O$, then $\left( M_{0}\right)
_{i}=\left( M_{i}\right) _{0}$.
\end{itemize}
\end{proposition}

\begin{proof}
(i) Let $\psi \in \mathcal{H}$ be a unit vector. We have, by definitions, $%
\left( m_{\psi }\right) _{i}(f)=\left\langle \psi , M_{i}(f) \psi
\right\rangle $ $\forall f\in BC(\Omega )$ and $\left( m_{\psi }\right)
_{i}(\infty )=\left\langle \psi , M_{i}(\infty ) \psi \right\rangle $. It
follows from Proposition \ref{mminfty} that $m_{\psi }(\infty )=0$. Since
this is true for any unit vector, $M(\infty )=O$. The proof of (ii) is
similar.
\end{proof}

With these results we are ready to prove Proposition \ref{fcoex1}.

\begin{proof}[Proof of Proposition~\ref{fcoex1}]
Given a function $f:\mathbb{R}\times \mathbb{R}\longrightarrow \mathbb{C}$
and $\left( q,p\right) \in \mathbb{R}\times \mathbb{R}$, we denote by $%
f^{\left( q,p\right) }$ the translate of $f$, 
\begin{equation*}
f^{\left( q,p\right) }\left( x,y\right) :=f\left( x+q,y+p\right) \quad
\forall x,y\in \mathbb{R}.
\end{equation*}
Since $\mathbb{R}\times \mathbb{R}$ (with addition) is an Abelian group,
there exists a mean $m$ on $\mathbb{R}\times \mathbb{R}$ such that 
\begin{equation*}
m\left( f^{\left( q,p\right) }\right) =m\left( f\right)
\end{equation*}
for all $f\in BC\left( \mathbb{R}\times \mathbb{R}\right) $ and $(q,p)\in 
\mathbb{R}\times \mathbb{R}$, (see \cite{AHAI}, Theorem~IV.17.5).

Let $M_{0}$ be a joint observable of $E_{\rho }$ and $F_{\nu }$. For each $%
f\in BC\left( \mathbb{R}\times \mathbb{R}\right) $, for all $\varphi ,\psi
\in \mathcal{H}$ and $q,p\in \mathbb{R}$ we define 
\begin{equation*}
\Theta \left[ f;\varphi ,\psi \right] \left( q,p\right) :=\left\langle
W\left( q,p\right) ^{\ast }\varphi
,M_{0}\left( f^{\left( q,p\right) }\right) W\left( q,p\right) ^{\ast } \psi \right\rangle 
.
\end{equation*}
Since 
\begin{equation*}
\left\| M_{0}\left( f^{\left( q,p\right) }\right) \right\| \leq \left\|
f^{\left( q,p\right) }\right\| _{\infty }=\left\| f\right\| _{\infty }
\end{equation*}
and $W\left( q,p\right) $ is a unitary operator, we have 
\begin{equation*}
\left| \Theta \left[ f;\varphi ,\psi \right] \left( q,p\right) \right| \leq
\left\| f\right\| _{\infty }\left\| \varphi \right\| \left\| \psi \right\|
\end{equation*}
and hence, $\Theta \left[ f;\varphi ,\psi \right] $ is a bounded function.
We claim that $\Theta \left[ f;\varphi ,\psi \right] $ is continuous. Since 
\begin{equation*}
\Theta \left[ f;\varphi ,\psi \right] \left( x+q,y+p\right) =\Theta \left[
f^{(q,p)};W\left( q,p\right) ^{\ast }\varphi ,W\left( q,p\right) ^{\ast
}\psi \right] \left( x,y\right) ,
\end{equation*}
it is sufficient to check continuity at $\left( 0,0\right) $. We have
{\setlength\arraycolsep{0pt} 
\begin{eqnarray*}
&&\left| \Theta \left[ f;\varphi ,\psi \right] \left( q,p\right) -\Theta %
\left[ f;\varphi ,\psi \right] \left( 0,0\right) \right| \\
&& \qquad \leq \left| \left\langle W\left( q,p\right)
^{\ast }\varphi ,
M_{0}\left( f^{\left( q,p\right) }\right)
\left( W\left( q,p\right) ^{\ast }\psi -\psi \right) 
\right\rangle \right| \\
&&\qquad \quad +\left| \left\langle \left( W\left( q,p\right) ^{\ast }\varphi -\varphi 
\right) , M_{0}\left( f^{\left( q,p\right)
}\right) \psi 
\right\rangle \right| \\
&&\qquad \quad +\left| \left\langle \varphi ,
M_{0}\left( f^{\left( q,p\right)
}-f\right) \psi \right\rangle \right| \\
&& \qquad \leq \left\| f\right\| _{\infty }\left( \left\| \varphi \right\|
\left\| W\left( q,p\right) ^{\ast }\psi -\psi \right\| +\left\| W\left(
q,p\right) ^{\ast }\varphi -\varphi \right\| \left\| \psi \right\| \right) \\
&&\qquad \quad +\left| \left\langle \varphi ,
M_{0}\left( f^{\left( q,p\right)
}-f\right) \psi 
\right\rangle \right| .
\end{eqnarray*}}

As $\left( q,p\right) \rightarrow \left( 0,0\right) $, the first two terms
go to $0$ by strong continuity of $W$, and the third by the dominated
convergence theorem. We have thus shown that $\Theta \left[ f;\varphi ,\psi %
\right] \in BC\left( \mathbb{R}\times \mathbb{R}\right) $.

For each $f\in BC\left( \mathbb{R}\times \mathbb{R}\right) $ we can then
define a linear bounded operator $M^{av}\left( f\right) $ by 
\begin{equation*}
\left\langle \varphi , M^{av}\left( f\right) \psi \right\rangle :=m\left(
\Theta \left[ f;\varphi ,\psi \right] \right) .
\end{equation*}
It is also immediately verified that the correspondence $M^{av}:BC\left( 
\mathbb{R}\times \mathbb{R}\right) \longrightarrow \mathcal{L}(\mathcal{H})$
is an operator valued mean on $\mathbb{R}\times \mathbb{R}$, and a short
calculation shows that 
\begin{equation}
M^{av}\left( f^{\left( q,p\right) }\right) =W\left( q,p\right) ^{\ast
}M^{av}\left( f\right) W\left( q,p\right) .  \label{due}
\end{equation}
If $f\in BC\left( \mathbb{R}\right) $ and $(q,p)\in \mathbb{R}\times \mathbb{%
R}$, we have 
\begin{eqnarray*}
\Theta \left[ \widetilde{f}_{1};\varphi ,\psi \right] \left( q,p\right)
&=&\left\langle W\left( q,p\right) ^{\ast }\varphi ,
M_{0}\left( \widetilde{f}_{1}^{\left( q,p\right) }\right)
W\left( q,p\right) ^{\ast }\psi 
\right\rangle \\
&=&\left\langle W\left( q,p\right) ^{\ast
}\varphi ,
W\left( q,p\right) ^{\ast }E_{\rho }\left( f\right) W\left(
q,p\right) W\left( q,p\right) ^{\ast }\psi 
\right\rangle \\
&=&\left\langle \varphi , E_{\rho }\left( f\right) \psi 
\right\rangle .
\end{eqnarray*}
(Especially, $\Theta \left[ \widetilde{f}_{1};\varphi ,\psi \right] $ is a
constant function.) Similarly, 
\begin{equation*}
\Theta \left[ \widetilde{f}_{2};\varphi ,\psi \right] \left( q,p\right)
=\left\langle \varphi , F_{\nu }\left( f\right) \psi \right\rangle .
\end{equation*}
It follows that 
\begin{eqnarray*}
M_{1}^{av}\left( f\right) &=&E_{\rho }\left( f\right) , \\
M_{2}^{av}\left( f\right) &=&F_{\nu }\left( f\right) .
\end{eqnarray*}
Since $E_{\rho }(\mathbb{R})=F_{\nu }(\mathbb{R})=I$, (\ref{M0}) shows that 
\begin{equation*}
M_{1}^{av}\left( \infty \right) =M_{2}^{av}\left( \infty \right) =O.
\end{equation*}
This together with Proposition~\ref{MM} implies that $M_{0}^{av}\left( 
\mathbb{R}\times \mathbb{R}\right) =I$ and 
\begin{eqnarray*}
\left( M_{0}^{av}\right) _{1} &=&E_{\rho }, \\
\left( M_{0}^{av}\right) _{2} &=&F_{\nu }.
\end{eqnarray*}
By (\ref{due}) the observable $M_{0}^{av}$ satisfies covariance condition (%
\ref{W}).
\end{proof}

\chapter{Conclusions}

As we have seen in chapters \ref{cap. 1} and \ref{cap. 2}, if $G$ is a
topological group and $H$ is a closed subgroup of $G$, by generalised
imprimitivity theorem the classification of the POM's based on $G/H$ and
covariant with respect to a fixed representation of $G$ is strictly related
to the problem of diagonalising the representations of $G$ induced from $H$.
This is an highly nontrivial problem, and a satisfactory solution can be
achieved only assuming that $G$ and $H$ are of particular type, as we have
in fact done in chapters \ref{cap. 1} and \ref{cap. 2}.

For example, if $G$ is abelian, things are enormously simplified by the fact
that the dual $\widehat{G}$ of $G$ is itself an abelian group having the
same topological properties of $G$. In addition, $\widehat{G/H}$ and $%
\widehat{H}$ are identified as subgroups or quotients of $\widehat{G}$.
These are the essential features which allow to construct the map $\Sigma $
and the measure $\widetilde{\nu }$ on $\widehat{G}$ diagonalising $\mathrm{%
ind}_{H}^{G}\left( \sigma \right) $ as in section \ref{subsec. 2.3}.

If $G$ is generic and $H$ is compact, one can quite easily prove that $%
\mathrm{ind}_{H}^{G}\left( \sigma \right) $ is contained in the regular
representation of $G$, thus showing that the diagonalisation of $\mathrm{ind}%
_{H}^{G}\left( \sigma \right) $ follows from Plancherel theory applied to $G$
(actually, in \S \ref{Caso H/Z compatto generico} we followed a different
and quicker approach, but Theorem \ref{Teo. centr.1} follows from Theorem 
\ref{Teo. centr.}, and the latter is an application of Plancherel theory
applied to $G$).

There are few other cases in which $\mathrm{ind}_{H}^{G}\left( \sigma
\right) $ can be diagonalised by known methods. One of them has been worked
out by Kirillov \cite{Ref. di Kirillov} in the case $G$ is a nilpotent Lie
group and $H$ is a generic closed subgroup of $G$. But the practical
application of the method of Kirillov is rather complicated, and does not
yeld to a compact form for covariant POM's like the quite simple expressions
we obtained in eqs.~(\ref{eq. di M buona}) and (\ref{La Povm1}). Moreover,
the only nilpotent Lie group of interest in physics is the Heisenberg group,
and, as we have seen in \S \ref{Esempio spazio fasi in 1 dim.}, its
covariant POM's can be classified using the theory exposed in chapter \ref
{cap. 2}.

A method for diagonalising $\mathrm{ind}_{H}^{G}\left( \sigma \right) $ can
also be elaborated when $G$ is the semidirect product $N\times ^{\prime }H$,
with $N$ normal abelian factor. We do not enter into details, but we only
remark that this method can be used to describe the localisation observables
for a relativistic quantum particle. For some hints about these facts we refer
to the work of Castrigiano \cite{Cast}, \cite{Cast2} and to the last part of
the book of Varadarajan \cite{Var}.

Finally, the results in chapters \ref{Capitolo sulle Covariant position and
momentum observables} and \ref{Cap. sulla coesistenza} give an answer to
some questions raised in \cite{Lahti95}. In particular, Proposition \ref
{fcoex1} shows that in order that a position observable $E_{\rho }$ and a
momentum observable $F_{\nu }$ are functionally coexistent, the probability measures $%
\rho $ and $\nu $ must form a Fourier couple in the sense of \cite[sec.
III.2.4]{Lahti95}. Up to our knowledge, no example is known of a joint
observable of a position and a momentum observable which is \emph{not} a
covariant phase space observable.

The last paragraph is of course intended according to our definition of position and 
momentum observables. We stress again that in the literature one often encounters many 
different approaches to the problem of position and momentum measurements in quantum 
mechanics, and that our results only refer to the covariant case. 

\chapter{Acknowledgements}

This work is the result of the collaboration of many people, and in these last lines
I would like to thank at least some of them.
Gianni is of course the first in the list, since he accepted me in this workgroup three
years ago, and he supported me with his guidance and patience in all this time. I am 
profoundly
grateful to him for all he have done for me.
Many thanks go to Pekka for his hospitality and
collaboration during my stays in Turku. I am also in debt of his most valuable suggestions 
and of
his great work to correct and improve my thesis.
I would like to thank Ernesto, Claudio and Teiko, not only for their precious 
collaboration in the
work of these years, but also for the lot of good time we have spent together.
Last, but not least, many thanks to all my friends in the Physics Department for having 
made of these years a wonderful experience for me.
\newpage
\

\end{document}